\documentclass[twoside,12pt]{article}
\usepackage{epsfig,amssymb,amsmath,eufrak,cite,pstricks,array,ragged2e,hyperref}
\usepackage[nottoc]{tocbibind}

\newcommand{\SLASH}[2]{\makebox[#2ex][l]{$#1$}/}

\newcommand{\pslash}{\SLASH{p}{.2}}
\newcommand{\qslash}{\SLASH{q}{.08}}

\newcommand{\MZ}{M_{\rm Z}}
\newcommand{\GZ}{\Gamma_{\rm Z}}
\newcommand{\mz}{\overline{M}_{\rm Z}}
\newcommand{\gz}{\overline{\Gamma}_{\rm Z}}
\newcommand{\MW}{M_{\rm W}}
\newcommand{\GW}{\Gamma_{\rm W}}
\newcommand{\mw}{\overline{M}_{\rm W}}
\newcommand{\gw}{\overline{\Gamma}_{\rm W}}
\newcommand{\MH}{M_{\rm H}}
\newcommand{\hMH}{\hat M_{\rm H}}
\newcommand{\mt}{m_{\rm t}}
\newcommand{\hmt}{\hat m_{\rm t}}
\newcommand{\sw}{s_{\rm W}}
\newcommand{\cw}{c_{\rm W}}
\newcommand{\gev}{\,\, \mathrm{GeV}}
\newcommand{\mev}{\,\, \mathrm{MeV}}
\newcommand{\seff}[1]{\sin^2\theta_{\rm eff}^{#1}}
\newcommand{\as}{\alpha_{\rm s}}
\newcommand{\at}{\alpha_{\rm t}}

\newcommand{\msbar}{{\ensuremath{\overline{\text{MS}}} }}

\begin{document}

\title{ \vspace{1cm} Numerical multi-loop integrals and applications}

\date{\vspace{2em}}

\author{A.\ Freitas \\ 
\\
\small Pittsburgh Particle-physics Astro-physics \& Cosmology Center
(PITT-PACC),\\[-.5ex]
\small Deptartment of Physics \& Astronomy, University of Pittsburgh,
Pittsburgh, PA 15260, USA}

\maketitle
\begin{abstract} 
Higher-order radiative corrections play an important role in precision studies
of the electroweak and Higgs sector, as well as for the detailed understanding
of large backgrounds to new physics searches. For corrections beyond the
one-loop level and involving many independent mass and momentum scales, it is in
general not possible to find analytic results, so that one needs to resort to
numerical methods instead. This article presents an overview of a variety of
numerical loop integration techniques, highlighting their range of
applicability, suitability for automatization, and numerical precision and
stability. 

In a second part of this article, the application of numerical loop integration
methods in the area of electroweak precision tests is illustrated. Numerical
methods were essential for obtaining full two-loop predictions for the most important
precision observables within the Standard Model. The
theoretical foundations for these corrections will be described in some detail,
including aspects of the renormalization, resummation of leading log
contributions, and the evaluation of the theory uncertainty from missing higher
orders.
\end{abstract}

\newpage

\tableofcontents

\newpage


\section{Introduction}
\label{}

With high-statistics data from LEP, SLC, Tevatron, LHC, B factories, and other
experiments, particle physics has forcefully moved into the precision realm
during the last few decades. Due to the small uncertainties of many experimental
results, it is possible to test the Standard Model at the quantum level and to
put stringent indirect constraints on new physics beyond the Standard Model. In
particular, electroweak precision tests put lower bounds on generic heavy new
physics of several TeV \cite{pdg}. Due to the level of precision, the inclusion
of radiative corrections has become an integral part in the analysis and
interpretation of experimental results.

In many cases, the dominant corrections arise from QED or QCD contributions due
to the radiation of photons, gluons or other massless partons from the initial
or final legs of a scattering or decay process. These can amount to tens of
percent or more in some situations. However, for observables with percent-level
or better precision, electroweak corrections also become relevant.  With
increasing  order in perturbation theory and increasing number of independent
mass and momentum scales, it becomes more difficult to find analytical
solutions to the virtual radiative corrections. This problem is particularly
acute for electroweak corrections, which involve many particles with
different, non-negligible masses. Consequently, in these situations it becomes
more convenient or even necessary to consider numerical techniques.

This review presents an
overview of numerical integration techniques, which are primarily used for the
calculations of electroweak loop corrections. The advantages and disadvantages
of the different methods are outlined, elucidating that there is no single
technique that works best in all circumstances.
In a second part, the application
of these methods for the calculation of two-loop corrections to electroweak
precision observables is discussed, and the phenomenological role of these
corrections in precision tests of the Standard Model is elucidated.

\vspace{\medskipamount}
A ``Feynman integral'' is obtained from diagrams or amplitudes that contain one
or more closed loops in the corresponding Feynman graph. The momentum flowing
through the loop $i$, denoted $q_i$, is not constrained by energy-momentum conservation and thus
must be integrated over, $\int d^4q_i/(i\pi^2) \dots$ 
In general, these loop integrals are divergent. One class of divergences, called
ultraviolet (UV) singularities, are
associated with the limit $|q_i| \to \infty$. These are removed through the
renormalization of
couplings, masses and the wave function of the incoming and outgoing fields of a
given physical process. A second class of divergences, called infrared
singularities (IR), can be divided into two types: soft and collinear
singularities. The former can occur when the momentum of a massless propagator
inside the loop tends to zero, while the latter can appear
if the momentum of a massless loop propagator becomes collinear with an external
light-like momentum that connects to one end of this propagator.  Soft and
final-state collinear singularities
cancel when the virtual loop corrections are combined with real emission
contributions.

Nevertheless, UV and IR singularities in individual loop amplitudes must be
regulated before they can be canceled. Throughout this article, dimensional
regularization \cite{dimreg} is employed. Within dimensional regularization, the
space-dimension is analytically continued from 4 to an arbitrary number
$D=4-2\epsilon$. The UV and IR singularities are then manifested 
terms with $1/\epsilon^n$ poles when taking the limit $\epsilon \to 0$.

For a physical observable, after summing over unobserved polarizations, a $L$-loop
integral in dimensional regularization can be written in the following generic
form:
\begin{align}
&I = \int {\mathfrak D}q_1 \cdots {\mathfrak D}q_L \;
\frac{(q_{i_1}\cdot p_{j_1})(q_{i_2}\cdot p_{j_2})\cdots}
{D_1^{\nu_1} \cdots D_n^{\nu_n}} \label{eq:genint}\,, \\
&{\mathfrak D}q_i = \frac{d^{D}q_i}{i\pi^{D/2}}, 
\qquad\qquad
D_j = k_j^2-m_j^2,
\end{align}
where the $k_j$ are linear combinations of one or more loop momenta $q_1,\,q_2,\,\dots$
and zero or more external momenta $p_1,\,p_2,\,\dots$, while $m_j$ are the masses of the
internal propagators, and $\nu_j$ are integer numbers. Note that
here and for the rest of this article, the Feynman $i\varepsilon$ prescription
is implicitly assumed, $D_j = k_j^2-m_j^2 \to k_j^2-m_j^2 + i\varepsilon$. In
other words, the propagator momentum $k_j$ is supposed to be endowed
with an infinitesimal imaginary part. Integrals with non-trivial numerator terms
in eq.~\eqref{eq:genint} are often called ``tensor integrals'' since they can
be written as
\begin{align}
I &= p_{j_1}^{\mu_1} p_{j_2}^{\mu_2} \cdots \int {\mathfrak D}q_1 \cdots {\mathfrak D}q_L \;
\frac{q_{i_1,\mu_1} q_{i_2,\mu_2}\cdots}{D_1^{\nu_1} \cdots D_n^{\nu_n}}\,.
\end{align}
A helpful and widely used tool for the analysis of Feynman integrals is the
Feynman parametrization. With its help, an integral of the form in
eq.~\eqref{eq:genint} without numerator terms turns into
\begin{align}
I = \int \frac{{\mathfrak D}q_1 \cdots {\mathfrak D}q_L}{D_1^{\nu_1} \cdots
 D_n^{\nu_n}} 
= \frac{\Gamma(N)}{\Gamma(\nu_1) \cdots \Gamma(\nu_n)} 
 \int_0^1 dx_1 \cdots dx_n \; \delta\Bigl(1-\sum_{i=1}^n x_i\Bigr) 
 \int {\mathfrak D}q_1 \cdots {\mathfrak D}q_L \;
 \frac{x_1^{\nu_1-1} \cdots x_n^{\nu_n-1}}{[x_1D_1 + \ldots + x_nD_n]^N},
\label{eq:feynp}
\end{align}
where $N = \nu_1 + ... + \nu_n$. The denominator sum can be written as
\begin{align}
x_1D_1 + \ldots + x_nD_n = \sum_{i,j=1}^L A_{ij} q_i\cdot q_j
 - 2\sum_{i=1}^L q_i\cdot P_i - M,
\end{align}
where the $(L{\times}L)$-matrix $A$, the $L$-column vector $P$, and the scalar
function $M$ depend on the Feynman parameters $x_i$. Upon shifting the loop
momenta to remove the linear term and carrying out the loop integration, the integral
becomes
\begin{align}
I &= \frac{(-1)^N\,\Gamma(N-LD/2)}{\Gamma(\nu_1) \cdots \Gamma(\nu_n)} 
 \int_0^1 dx_1 \cdots dx_n \; \delta\Bigl(1-\sum_{i=1}^n x_i\Bigr) \;
 x_1^{\nu_1-1} \cdots x_n^{\nu_n-1}\,\frac{{\cal U}^{N-(L+1)D/2}}{{\cal
 V}^{N-LD/2}}
\label{eq:feynp2},
\intertext{with} 
{\cal U} &= \det(A), \qquad\qquad 
{\cal V} = \det(A)\,\Bigl [M + \sum_{i,j=1}^L A^{-1}_{ij}\, P_i\cdot P_j \Bigr ].
\label{eq:feynpuv}
\end{align}
Instead of introducing Feynman parameters for all $L$ loop integrations at once,
as in \eqref{eq:feynp}, one can alternatively introduce them loop by loop, which
is advantageous for some applications discussed in this review.

Besides using Feynman parameters, another useful representation of Feynman
integrals is obtained from the use of so-called Schwinger or alpha parameters.
In fact, the alpha parametrization is closely related to the Feynman
parametrization; see chapter 2.3 of Ref.~\cite{Smirnov:2006ry} for more
information.


\subsection{Analytic methods}

From a historical perspective, the default approach to loop calculations is the
use of analytical methods. This procedure can be divided into two steps:
\emph{(a)} reduction of the complete lists of loop integrals for a given
physical process to a small set of scalar ``master integrals''; and \emph{(b)}
evaluation of the master integrals in terms of analytical functions that depend
on the propagator masses, invariants of the external momenta, and the
integration dimension $D$. In practice, it is usually sufficient to carry out
the last step as an expansion in $\epsilon = 2/(4-D)$, dropping all terms
$\epsilon^n$ with powers $n>0$\footnote{Occasionally, it may be necessary to retain higher powers
in $\epsilon$ if the coefficient in front of a certain master integral diverges
in the limit $\epsilon\to 0$.}.

The reduction to master integrals can be achieved through a number of different
methods:
\begin{itemize}

\item
The Passarino-Veltman approach \cite{pasvelt}, which is based on the 
decomposition of integrals with different terms in the numerator of
\eqref{eq:genint} into Lorentz covariant monomials: This technique is applicable
to generic one-loop integrals, and it has been extended for some classes of
two-loop integrals \cite{weiglein1}, but it does not work for arbitrary
multi-loop integrals.

\item
Integration-by-parts relations \cite{ibp}: In dimensional regularization, these
take the form
\begin{align}
\int \prod_{l=1}^L{\mathfrak D}q_l \,  \partial q_i^\mu \, k_{j,\mu} \,F(\{q_k\},\{p_m\},\{m_n\})
= \oint dS^\mu_{q_i} \int \prod_{l\neq i} {\mathfrak D}q_l \,k_{j,\mu} \,F(\{q_k\},\{p_m\},\{m_n\}) = 0, \label{eq:ibp}
\end{align}
where $k_j$ may be loop or external momentum and $F$ is an expression containing
propagators and dot products of momenta, but no free Lorentz indices. The
surface integral on the right-hand side of \eqref{eq:ibp} extends over the boundary
of the $D$-dimensional integration volume of the loop momentum $q_i$, and it
vanishes in dimensional regularization. On the other hand, when evaluating the
derivative on the left-hand side, one obtains a linear relation between
different integrals of the form \eqref{eq:genint}. By considering different
choices for $q_i$, $k_j$ and $F$ one can generate a large, overconstrained
linear equation system, which can be solved to find how a complicated loop
integral can be expressed as a linear combination of simpler master integrals.

This approach is very suitable for the automated implementation in computer
algebra systems. The first such computer program was MINCER \cite{mincer}, which
was developed for the reduction of 3-loop massless propagator diagrams. A
systematic prescription for building and solving such linear equation systems
for arbitrary multi-loop integrals was presented in Refs.~\cite{laporta} and is
usually referred to as the ``Laporta algorithm''. The integration-by-parts
identities can be supplemented by Lorentz invariance identities \cite{li} to
arrive at a more economical system of equations. Today, several public codes are
available that can perform the reduction of general multi-loop integrals, such
as AIR \cite{air}, FIRE \cite{fire}, {\sc Reduze} \cite{reduze}, {\sc LiteRED}
\cite{litered}. These programs  take advantage of integration-by-parts and
Lorentz invariance identities together with symmetry properties of the integrals
and advanced linear reduction algorithms.

Instead of solving the system of integration-by-parts identities through the
``Laporta algorithm'', an alternative method for vacuum integrals has been
presented by Baikov \cite{baikov}. This technique allows one to directly
determine the coefficients $c_k$ in the reduction of a Feynman integral $I$,
$
I = \sum_k c_k \, I^0_k,
$
where $I^0_k$ are the master integrals. It was shown in Ref.~\cite{baikov} that
the integration-by-parts identities can be transformed into differential
equations, for which an explicit solution in terms of the masses $m_{ij}$ and
propagator indices $\nu_j$ of the original integral can be found. This method
can also be used for integrals with external momenta by relating them to vacuum
integrals with additional propagators. The idea's from Baikov's method can also
be used to find a suitable basis of master integrals \cite{baikov2}.

\item
Tensor reduction through tensor operators \cite{tarasov,box2}: In
Ref.~\cite{tarasov} it was shown that integrals with non-trivial numerator terms
can be written as a tensor operator acting on a scalar integral
with numerator 1. The proof follows from
a careful examination of the Schwinger parametrization of the tensor integrals.
The result of the action of the tensor operator are scalar integrals with
higher powers of propagators in the denominator and/or 
shifted space-time dimension $D+2, \, D+4\,
\dots$ 
Integrals with shifted dimension can be related to $D$-dimensional scalar integrals by using a variant of
the tensor operator mentioned above. Scalar integrals with higher powers of
propagators can then be reduced to the master integrals by the
integration-by-parts identities as described in the previous bullet \cite{tarasov,box2}.

\item
Unitarity-based methods, which are founded on the basic tenet of the optical
theorem that the sum of all diagrams contributing to a certain process is
related to the discontinuities of the amplitude across its branch cuts:  By
introducing the concepts of generalized cuts and the on-shell singularity
structure of loop amplitudes \cite{unitarity1}, it has become feasible to reduce
a general one-loop diagram to master integrals by analyzing the residues of
these singularities \cite{opp1}.

These methods are very powerful for the computation of multi-leg one-loop
processes, and the implementation of automated algorithms into computer codes
has been achieved by several groups \cite{oppx}. For a recent review on this
topic, see $e.\,g.$ Ref.~\cite{Ellis:2011cr}. The extension of the
unitarity-based approach to higher loop orders is more difficult, but notable
advances have been made (see $e.\,g.$ Ref.~\cite{unitarity2}).

\end{itemize}
For the analytic calculation of multi-loop master integrals, a variety of
different techniques have been developed:

\begin{itemize}

\item
Direct integration over Feynman parameters, starting from the expression
\eqref{eq:feynp2}, is the most straightforward method for evaluating master
integrals. For more complicated Feynman integrals, the last few Feynman
parameter integrations are typically only possible after expanding in powers of
$\epsilon = 2/(4-D)$.
This method has been used, for example, for two-loop
vertex corrections with massless propagators \cite{ver2a}, massive two-loop
diagrams at threshold \cite{ver2b}, and massive three-loop vacuum integrals
\cite{master3}. By integrating over Schwinger instead of Feynman parameters, it is possible
to evaluate certain more difficult cases with more than three loops and
multiple independent scales, see $e.\,g.$ Ref.~\cite{panzer}.
However, the parametric integration approach 
reaches a limit for general multi-loop diagrams with several independent scales. 

\item
Mellin-Barnes representations are very useful for further processing of
difficult Feynman parameter or alpha parameter integrals. The key idea is to
replace the sum in the denominator of a Feynman parameter or alpha parameter
integral by the Mellin-Barnes integral
\begin{equation}
\begin{aligned}
\frac{1}{(A_0+\ldots+A_m)^Z} = \frac{1}{(2\pi i)^m} \int_{{\cal C}_1} dz_1
\cdots &\int_{{\cal C}_m} dz_m \; A_1^{z_1} \cdots A_m^{z_m}
A_0^{-Z-z_1-\ldots-z_m} \\
&\times \frac{\Gamma(-z_1) \cdots \Gamma(-z_m)\Gamma(Z+z_1+\ldots+z_m)}{\Gamma(Z)}
\end{aligned}
\label{eq:mb}
\end{equation}
where the integration contours ${\cal C}_i$ for $z_i$ are straight lines parallel to
the imaginary axis chosen such that all arguments of the gamma functions have
positive real parts. This representation is very convenient for the isolation of
singularities in $\epsilon$ (see section~\ref{sec:mb} for more details). After
carrying out the Feynman or alpha parameter integrals, the Mellin-Barnes
integration can be performed by closing the integration contours in the complex
plane and summing up the residues. This technique has been used for the
calculation of various two- and three-loop master integrals, see for instance
Refs.~\cite{2lvac,2lbox,3lbox}.

\item
The differential equation method \cite{deq} has been used widely for the
evaluation of master integrals beyond one-loop order. The basic idea is to take
derivatives of a given master integral $I_k^0$ with respect to a kinematical
invariant or mass. The result of this differentiation on the integrand of
$I_k^0$ produces a different Feynman integral, which in general is non-minimal
but can be reduced to a linear combination of master integrals by using, for
example, integration-by-parts identities. Thus one obtains a differential
equation of the form
\begin{align}
\frac{\partial}{\partial(m_i^2)}I_k^0 &= 
 \sum_l f_{kl}(\{m_n\}, \{s_{pr}\}, D) \, I^0_l  \label{eq:deq1}
\intertext{or}
\frac{\partial}{\partial s_{ij}}I_k^0 &= 
 \sum_l f'_{kl}(\{m_n\}, \{s_{pr}\}, D) \, I^0_l, \label{eq:deq2}
\end{align}
where the coefficient functions $f_{kl}$ depend on the masses $m_n$ and kinematical
invariants $s_{pr} = (p_p+p_r)^2$ of the integrals, as well as the dimension
$D$. By organizing the differential equations appropriately, one can make sure
that the right-hand side of \eqref{eq:deq1},\eqref{eq:deq2} contains only
$I_k^0$ itself and simpler master integrals, whose solution is assumed to be
known already. A good choice of the basis of master integrals $\{I_k^0\}$ is
essential in this context \cite{henn}.
Then the system of differential equations can be solved
sequentially to obtain solutions for all masters.
In practice, it is often difficult to find analytical solutions for arbitrary
dimension $D$, in which case one can instead expand both sides of the
differential equation in powers of $\epsilon = 2/(4-D)$ and then construct the
solution order by order in $\epsilon$.

\end{itemize}
A more detailed exposition of analytic loop calculation methods can be found
$e.\,g.$ in Ref.~\cite{Smirnov:2006ry}.

Analytical methods work well for problems with few independent momentum and mass
scales, leading to compact results suitable for fast evaluation in Monte-Carlo
event generators.
However, for multi-scale problems, both of the main steps of the analytical
approach run into difficulties. The reduction to master integrals requires
significantly larger computing resources and leads to large and unwieldy
expressions. Furthermore, master integrals with many independent momentum and mass
scales can typically not be solved in terms of known elementary functions. This
can be circumvented to some extent by introducing new types of functions, such
as harmonic polylogarithms \cite{hpl} and Goncharov polylogarithms \cite{gpl}, 
but it is unclear if these concepts can be extended to arbitrary Feynman integrals.

In the following two subsections, the two main philosophies for moving beyond the
limitations of the analytical approach are outlined.


\subsection{Asymptotic expansions}

In the presence of a suitable small expansion parameter, for example the ratio
$m/M$ of a small mass $m$ and a large mass $M$, a difficult multi-scale Feynman
diagram may be expanded in powers of this small parameter. The coefficients of
this expansion are diagrams with fewer scales and/or fewer loops, which are
simpler to evaluate analytically. This procedure, called ``asymptotic
expansion'', is an extension of Taylor expansions which can contain
non-analytical functions of the small parameter, such a logarithms.

Typically, a small expansion parameter is obtained if one of the masses or
momenta in a given problem are significantly larger ($e.\,g.$ the top-quark mass
in the Standard Model or the beam energy of a high-energy collider) or smaller
($e.\,g.$ the charm-quark mass in B-meson decays) than other relevant scales. In
some cases, the small parameter could also be the difference between two masses
and/or momenta. In some cases, even an expansion parameter of magnitude close to
1 leads to satisfactory results (see $e.\,g.$ Ref.~\cite{2lvac}). For many
applications, the first few terms in an asymptotic expansion are sufficient to
achieve the desired precision of the result.

The general prescription for the asymptotic expansion in the presence of a large
scale $\Lambda$ is \cite{asymp1}
\begin{align}
\Gamma(\Lambda, \{m_i\}, \{p_j\}) = \sum_\gamma \Gamma{/}\gamma(\{m_i\}, \{p_j\})
\star T_{\{m_i\}, \{p_j\}}\,\gamma(\Lambda,\{m_i\}, \{p_j\}).
\end{align}
Here $\Lambda \gg m_i, p_j$ is a momentum or mass that is significantly larger
than the collection of other masses, $\{m_i\}$, and momenta, $\{p_j\}$. $\Gamma$
is the Feynman diagram under consideration, and the sum runs over all subgraphs
$\gamma$ of $\Gamma$
that contain all vertices and propagators where $\Lambda$ appears. The subgraphs
also must be one-particle irreducible in their connected parts.
$\Gamma{/}\gamma\star T_{\{m_i\}, \{p_j\}}\gamma$ denotes that the integrand of
the subgraph 
$\gamma$ of $\Gamma$ is replaced by its Taylor expansion with respect to
all small masses and external momenta. In particular, the loop momenta of
$\Gamma$ that are external to $\gamma$ also have to be treated as small.

The major advantage of this method is that the expansion in the small parameter
is carried out in the integrand, before any loop integral is evaluated, and thus
leads to simpler integrals than the original problem. It can be shown that this
produces correct results by using the ``strategy of regions''
\cite{regions,threx}.
This technique subdivides the integration space into regions where the different
loop momenta are large or small. The ``strategy of regions'' also allows one to
tackle other cases than simple large or small mass and momentum expansions, such
as threshold expansions \cite{threx} or mass difference expansions
\cite{mdiff}.

The method of asymptotic expansions fails if an internal threshold of the loop
diagram is crossed when taking the limit of the small expansion parameter.
Furthermore, while asymptotic expansions typically lead to fairly compact final results, one has
to deal with large and unwieldy expressions during intermediate steps,
especially in the case of multiple expansions. Also, in the case of multiple
expansions, more terms may be needed to achieve satisfactory precision.


\subsection{Numerical integration}
\label{sec:numi}

Instead of trying to obtain a final result in terms of an analytical formula,
one can alternatively perform at least some of the integrations for a loop diagram
numerically. While in principle the numerical integrations can be carried out
directly in the space of the loop momenta $q_i$ in \eqref{eq:genint}, it is
typically more convenient to switch to different variables, such as Feynman
parameters, Mellin-Barnes integrals, and other options that will be discussed in
the following chapter.

The advantage of the numerical integration approach is that, at least
conceptually, it poses no limit to the number of different propagators and mass
and momentum scales present in a loop diagram. Thus it is particularly suitable
for the calculation of multi-loop corrections in the full Standard Model.
However, there are other difficulties encountered by numerical loop integration
techniques:

\paragraph{Isolation of UV and IR singularities:} 
Physical amplitudes can exhibit UV and IR (soft and collinear) divergences, which
appear as $1/\epsilon$ poles in dimensional regularization. These need to be
identified and extracted before the numerical integration can be performed.
Ideally one would like an algorithmic prescription for this step, which can be
implemented in a computer algebra system and works automatically for a large
class of Feynman diagrams. For the case of QED (as opposed to non-Abelian
theories like QCD), the IR divergences may also be regulated by a small photon
mass, which does not require any specific treatment before the numerical
evaluation. On the other hand, a photon mass that is much smaller than other
mass and momentum scales may pose a challenge to the
convergence and precision of the numerical integrator.

\paragraph{Stability and convergence:} For a numerical integration technique to
be practical, it must be able to produce a sufficiently precise result with a
reasonable number of integrand evaluation points. The precision should improve
in a predictable manner when the number of integration points is increased.
Typically, these requirements can be satisfied if the numerical integration
volume is of relatively low dimension, or if the integrand is ensured to be
relatively smooth, without large peaks or oscillatory behavior. In the former
case, it is typically advantageous to use standard discrete integration
algorithms, whereas the latter case is suitable for Monte-Carlo and
Quasi-Monte-Carlo integration routines.

A particular difficult situation is the occurrence of local 
singularities that are formally integrable, but which cannot be
handled by standard numerical integration routines. Such singularities usually
originate from internal thresholds of a loop diagram, $i.\,e.$ if a Feynman
diagram has physical cuts that meet the condition $0 < \sum_{i \in \rm cut} m_i
< \sqrt{p^2_{\rm tot,cut}}$, where the sum runs over the masses of the cut propagators
and $p_{\rm tot,cut}$ is the total momentum flowing through the cut.
These singularities are typically located in the inner part of the integration
domain ($i.\,e.$ not on its boundary), and their impact needs to be mitigated
through a suitable change of integration variables or some manipulation of the
integrand.

\paragraph{Generality:}
A numerical loop integration method should preferably be applicable to a large
class of Feynman diagrams, without special techniques for each different diagram
topology. Obviously, this is already a problem for analytical methods, and in
fact numerical integration approaches have the potential to be superior in this
aspect.

\vspace{1.5em}\noindent
In the following, some of the most commonly used and powerful
numerical integration techniques will be discussed in more detail. At the end of
the next chapter, their strengths and weaknesses with respect to the
aforementioned three
criteria will be summarized.


\section{Numerical integration techniques}
\label{sec:num}


\subsection{Feynman parameter integration of massive two-loop integrals}
\label{sec:ghin}

A general method for the numerical evaluation of massive two-loop diagrams was
introduced in Refs.~\cite{ghin1,ghin2}. Let us consider an arbitrary
$N$-propagator two-loop integral\footnote{Note that Refs.~\cite{ghin1,ghin2} use
a different convention for the $q_{1,2}$ integration measure.}
\begin{equation}
I_{(2)} = \int {\mathfrak D}q_1 {\mathfrak D}q_2 \; N(q_1,q_2,\{p_i\})
\begin{aligned}[t]
&\times \frac{1}{[(q_1+p_1)^2-m_1^2] \cdots [(q_1+p_n)^2-m_n^2]} \\
&\times \frac{1}{[(q_2+p_{n+1})^2-m_{n+1}^2] \cdots [(q_1+p_{n+m})^2-m_{n+m}^2]}
 \\
&\times \frac{1}{[(q_1+q_2+p_{n+m+1})^2-m_{n+m+1}^2] \cdots 
 [(q_1+q_2+p_N)^2-m_N^2]},
\end{aligned}
\label{eq:twol1}
\end{equation}
where $p_1, \dots, p_N$ are external momenta, some of which may be zero or
linearly dependent on other momenta, and $N$ is a polynomial function of loop
and external momenta. As a first step, three sets of Feynman parameters are
introduced, one set
each for all propagators with loop momentum $q_1$, $q_2$ and $q_1+q_2$,
respectively. After shifting the loop momenta, one thus 
can write \eqref{eq:twol1} in the form
\begin{align}
I_{(2)} = \int_0^1 dx_1 \cdots dx_{N-3} \int {\mathfrak D}q_1 {\mathfrak D}q_2 \; 
\frac{\tilde N(q_1,q_2,\{p_i\})}{[q_1^2-\tilde{m}_1^2]^{\nu_1}
[q_2^2-\tilde{m}_2^2]^{\nu_2}
[(q_1+q_2+\tilde{p})^2-\tilde{m}_3^2]^{\nu_3}}.
\label{eq:twol_fp}
\end{align}
Here $\tilde{p}$ is a linear combination of the external momenta $p_i$, which
depends on the Feynman parameters $x_j$, and $\tilde{m}_{1,2,3}$ are functions
of the masses and momenta in \eqref{eq:twol1} and of the Feynman parameters.

Using a decomposition of the $q_{1,2}$-dependent terms in the numerator into
parts that are transverse and longitudinal with respect to $\tilde{p}$
\cite{ghin2}, one finds that all such integrals can be reduced to
scalar integrals of the form
\begin{align}
I_{(2)} &= \int_0^1 dx_1 \cdots dx_{N-3} \; 
{\cal P}^{ab}_{\nu_1\nu_2\nu_3}(\tilde{m}_1,\tilde{m}_2,\tilde{m}_3;\tilde{p}^2), 
\\ 
{\cal P}^{ab}_{\nu_1\nu_2\nu_3}(\tilde{m}_1,\tilde{m}_2,\tilde{m}_3;\tilde{p}^2) &=
\int {\mathfrak D}q_1 {\mathfrak D}q_2 \; 
\frac{(q_1\cdot p)^a(q_2\cdot p)^b}{[q_1^2-\tilde{m}_1^2]^{\nu_1}
[q_2^2-\tilde{m}_2^2]^{\nu_2}
[(q_1+q_2+p)^2-\tilde{m}_3^2]^{\nu_3}}.
\end{align}
Scalar integrals ${\cal P}^{ab}_{\nu_1\nu_2\nu_3}$ with different indices can be
related by considering derivatives with respect to its mass and momentum
arguments, leading to
\begin{align}
{\cal P}^{ab}_{\nu_1+1,\nu_2,\nu_3} &=
 -\frac{1}{\nu_1} \, \frac{\partial}{\partial (\tilde{m}_1^2)}
 {\cal P}^{ab}_{\nu_1\nu_2\nu_3}, \\ 
{\cal P}^{a+1,b}_{\nu_1+1,\nu_2,\nu_3} &=
 \frac{1}{2\nu_1}\Bigl [ 2\tilde{p}^2 \frac{\partial}{\partial(\tilde{p}^2)} - (a+b) \Bigr ]
  {\cal P}^{ab}_{\nu_1\nu_2\nu_3} + \frac{a \tilde{p}^2}{2\nu_1} 
 {\cal P}^{a-1,b}_{\nu_1\nu_2\nu_3}, \\ 
{\cal P}^{a,b+1}_{\nu_1,\nu_2+1,\nu_3} &= 
 \frac{1}{2\nu_2}\Bigl [ 2\tilde{p}^2 \frac{\partial}{\partial(\tilde{p}^2)} - (a+b) \Bigr ]
  {\cal P}^{ab}_{\nu_1\nu_2\nu_3} + \frac{b \tilde{p}^2}{2\nu_2} 
 {\cal P}^{a,b-1}_{\nu_1\nu_2\nu_3}.
\end{align}
Using the relations, one can express all loop functions ${\cal
P}^{ab}_{\nu_1\nu_2\nu_3}$ in terms of a minimal set, for which the 
authors of Ref.~\cite{ghin2} chose the following ten: 
${\cal P}^{00}_{211}$, 
${\cal P}^{10}_{211}$, 
${\cal P}^{01}_{211}$, 
${\cal P}^{20}_{211}$, 
${\cal P}^{11}_{211}$, 
${\cal P}^{02}_{211}$, 
${\cal P}^{30}_{211}$, 
${\cal P}^{21}_{211}$, 
${\cal P}^{12}_{211}$, 
${\cal P}^{03}_{211}$.
This set is a suitable choice for all renormalizable theories, but note that not
all of these ten functions are independent. 

The UV-divergent part of the ten
master functions can be evaluated analytically, while their finite parts can be
expressed in terms of one-dimensional integral representations \cite{ghin2}. For
example, introducing two Feynman parameters to combine the three propagators
into one, expanding in $\epsilon$, and integrating over one Feynman parameter,
one finds for the function ${\cal P}^{00}_{211}$ \cite{ghin1}:
\begin{align} 
{\cal P}^{00}_{211}(\tilde{m}_1,\tilde{m}_2,\tilde{m}_3;\tilde{p}^2) &= 
\begin{aligned}[t] &\frac{1}{2\epsilon}
+ \frac{1}{\epsilon}\biggl [ \frac{1}{2} - \gamma_{\rm E} - \log \tilde{m}_1^2
  \biggr ] \\
&+ \Bigl ( \frac{1}{2} - \gamma_{\rm E} - \log \tilde{m}_1^2 \Bigr )^2 +
\frac{\pi^2-9}{12} + g(\tilde{m}_1,\tilde{m}_2,\tilde{m}_3;\tilde{p}^2),
\end{aligned} \\[1ex]
g(\tilde{m}_1,\tilde{m}_2,\tilde{m}_3;\tilde{p}^2) &= \int_0^1 dx \, \biggl [
 {\rm Li}_2\Bigl (\frac{1}{1-y_+}\Bigr ) + 
 {\rm Li}_2\Bigl (\frac{1}{1-y_-}\Bigr ) +
 y_+ \log \frac{y_+}{y_+-1} +
 y_- \log \frac{y_-}{y_--1} \biggr ], \label{eq:ghin_g} \\[1ex]
y_\pm &= -\frac{1}{2\tilde{p}^2}\bigl [\tilde{m}_1^2 - \mu^2 - \tilde{p}^2 
 \pm \lambda^{1/2}(\tilde{m}_1^2,\mu^2,\tilde{p}^2) \bigr],\\
\mu^2 &= \frac{\tilde{m}_2^2\, x + \tilde{m}_3^2(1-x)}{x(1-x)},
\end{align}
where $\gamma_{\rm E} \approx 0.577216$ is the Euler number, ${\rm Li}_2(z)$ is the
dilogarithm or Spence's function, and
\begin{equation}
\lambda(a,b,c) = a^2+b^2+c^2-2(ab+ac+bc) \label{eq:kallen}
\end{equation}
is the K\"all\'en function. Here it is understood that the usual Feynman
$i\varepsilon$ prescription is applied, $i.\,e.$ $\tilde{p}^2 \to \tilde{p}^2+i\varepsilon$.

The integration over the last Feynman parameter $x$, as well as the Feynman
parameters introduced in eq.~\eqref{eq:twol_fp}, can then be carried out
numerically, resulting in a $(N-2)$-dimensional numerical integration. 

For $\tilde{p}^2 > (\tilde{m}_1+\tilde{m}_2+\tilde{m}_3)^2$ the integrand of
\eqref{eq:ghin_g} (and similarly for the other ${\cal
P}^{ab}_{\nu_1\nu_2\nu_3}$ functions) develops singularities at 
\begin{align}
x &=
\frac{1}{2r^2}\bigl [-m_2^2+m_3^2+r^2\pm \lambda^{1/2}(m_2^2,m_3^2,r^2)\bigr ],
&
r^2 &= m_1^2+\tilde{p}^2-2\sqrt{\tilde{p}^2}.
\end{align}
These points must be circumvented by deforming the $x$-integration into the
complex plane. Similarly, values of the Feynman parameters $x_{1,2,\dots}$ in
\eqref{eq:twol_fp} where $\tilde{m}^2_{1,2,3}$ become zero should also be
avoided by choosing a complex integration path for these parameters. After this,
the integrand is reasonably smooth if all masses and external momentum
invariants are of similar order of magnitude, and the numerical integration can be
performed with standard discrete (for low dimensionality) or Monte-Carlo (for
high dimensionality) integration routines.

A suitable
integration path for $x$ is described in Ref.~\cite{ghin1}, but the choice of
complex contour for the other integration variables may depend on the topology
of the loop diagram and on the pattern of masses appearing inside it, and thus
it requires some case-by-case adaptation.

With this qualification in mind, the technique discussed in this section works
for fairly generic two-loop contributions, including UV-divergences. On the
other hand, it cannot handle IR divergences within dimensional regularization.
Instead one needs to use a mass regulator, leading to difficulties with
higher-order QCD corrections and to potential numerical
instabilities in the integration region where the mass-regulated
propagator becomes almost on-shell.

The method has
been used for the calculation of Higgs self-energy corrections \cite{ghin1} and 
corrections to the decays of Higgs bosons \cite{hdec2}, top quarks \cite{tdec2},
$Z$ bosons \cite{zdec2}, and rare $B$-meson decays \cite{bdec2}.


\subsection{Sector decomposition}
\label{sec:secdec}

Sector decomposition \cite{sdec} is an approach that is also based on Feynman parameter
integrals, but it provides a more systematic treatment of 
divergences in dimensional regularization.
It is based on the idea of iteratively dividing the Feynman parameter space into
sectors to disentangle overlapping soft, collinear and UV divergences \cite{hepp}. 
Each
singularity then becomes associated with a single Feynman parameter variable and
can be extracted with a suitable counterterm. The remaining non-singular
integrals, both for the coefficients of $1/\epsilon$ poles and the finite parts,
can then be evaluated numerically.

The starting point is a Feynman parameter integral as in eq.~\eqref{eq:feynp2},
and for simplicity only the case $\nu_1 = \nu_2 = ... = 1$ is considered here,
although the method also works for different propagator powers.
The vanishing of the function $\cal U$ is associated with UV (sub)divergences,
which may be identified and then subtracted in this way. On the other hand, IR
poles originate from regions where the function $\cal V$ vanishes, which happens
if some Feynman parameters are approaching zero. These singular regions
are in general overlapping in Feynman parameter space, but they can be separated
with the help of sector decomposition.

A ``primary'' sector decomposition eliminates the $\delta$-function and divides
the integral into $N$ integrals, where each integration variables runs from 0 to
1:
\begin{align}
\int_0^1 d^Nx &= \sum_{i=1}^N \int_0^1 d^Nx \prod_{j\neq i} \theta(x_i - x_j),
\end{align}
where $\theta$ is the Heaviside theta function. In the $i$th term of the sum one
then applies the variable substitution
\begin{align}
x_j &= \left\{ \begin{array}{ll}
 x_it_j, & j < i,\\
 x_i, & j=i,\\
 x_it_{j-1}, & j>i.
\end{array} \right.
\end{align}
Since $\cal U$ and $\cal V$ are homogeneous functions, the dependence on $x_i$
factorizes, ${\cal U}(\vec{x}) = x_i^L {\cal U}_i(\vec{t})$ and  ${\cal
V}(\vec{x}) = x_i^L {\cal V}_i(\vec{t})$. Performing the $x_i$-integral against
the $\delta$-function one then obtains
\begin{align}
I &= (-1)^N\,\Gamma(N-LD/2)\; \sum_{i=1}^N 
 \int_0^1 d^{N-1}t \; 
 \frac{{\cal U}_i^{N-(L+1)D/2}}{{\cal V}_i^{N-LD/2}}.
\label{eq:sec1}
\end{align}

Subsequent sector decompositions are performed iteratively until all
singularities are disentangled. For each term in the sum of integrals, a small
set of parameters ${\cal S} = \{t_{a_1}, \dots , t_{a_r}\}$ is chosen such that
either ${\cal U}_i$ or ${\cal V}_i$ vanishes if the elements of $\cal S$ are set
to zero. Then the integration region of $\cal S$ is subdivided into sectors
according to
\begin{align}
\int_0^1 d^rt &= \sum_{j=1}^r \int_0^1 d^rt \prod_{k\neq j} \theta(t_{a_j} -
t_{a_k}),
\end{align}
and in each new subsector the following variable substitution is performed,
\begin{align}
t_{a_k} &= \left\{ \begin{array}{ll}
 t_{a_j}t_{a_k}, & k \neq j,\\
 t_{a_j}, & k=j.
\end{array} \right.
\end{align}
This variable mapping ensures that the singularities are still located at the
lower limit (rather than the upper limit) of some of the new variable integrals.
Since either ${\cal U}_i$ or ${\cal V}_i$ vanishes for $t_{a_j} \to 0$, one can
factor out some power of $t_{a_j}$. Thus the subsector integrals have the form
\begin{align}
I_{ij} &= 
 \int_0^1 d^{N-1}t \;\Biggl (\prod_{k=1}^{N-1} t_k^{A_k-B_k\epsilon} \Biggr )
 \frac{{\cal U}_{ij}^{N-(L+1)D/2}}{{\cal V}_{ij}^{N-LD/2}}.
\label{eq:sec2}
\end{align}
These steps are repeated until no set ${\cal S}$ for any of the subsector
integrals can be found anymore.

Now the singularities can be extracted from the $t_k$ integrals of the form
\begin{equation}
I_k = \int_0^1 dt_k \, t_k^{A_k-B_k\epsilon} {\cal I}(t_k; \epsilon).
\end{equation}
For $A_k \geq 0$, this expression is finite and one can set $\epsilon \to 0$.
For $A_k < 0$, one performs the Taylor expansion
\begin{equation}
{\cal I}(t_k; \epsilon) = \sum_{n=0}^{|A_k|-1} \frac{{\cal I}^{(n)}(0;
\epsilon)}{n!} t_k^n + R(t_k; \epsilon),
\end{equation}
where ${\cal I}^{(n)}$ is the $n$th derivative of ${\cal I}$. Then
\begin{equation}
I_k = \sum_{n=0}^{|A_k|-1} \frac{{\cal I}^{(n)}(0;\epsilon)}{n!}
 \, \frac{1}{A_k+n+1-B_k\epsilon} +
\int_0^1 dt_k \, t_k^{A_k-B_k\epsilon} R(t_k; \epsilon).
\label{eq:secexp}
\end{equation}
A $1/\epsilon$ pole is contained in the highest term of the sum in
\eqref{eq:secexp}. By carrying out this step for all $N-1$ variables, one
obtains a series of $1/\epsilon^m$ poles whose coefficients are
$(N-1-m)$-dimensional integrals. These can be integrated numerically or, in
simple cases, also analytically.

This algorithm is straightforward to implement in a computer program, which can
be used for the evaluation of integrals with complicated singularity structures
\cite{sdec,Anastasiou:2005qj,sdec2}. The idea of sector decomposition has also been
extended to the case of phase-space integrals \cite{sdecph,Anastasiou:2005qj}.

A difficulty of sector decomposition is the exponential proliferation of
subsectors with increased number of iterations, leading to very large
expressions. From a naive application of the algorithm described above, one
typically generates many more subsectors than are needed for the disentanglement
of all singularities. However, the procedure can be optimized by making
intelligent choices for the subset $\cal S$ at every step, for which there is in
general no unique choice. Furthermore, in some cases with massive propagators
these improvements are  also necessary to ensure that the algorithm actually
terminates \cite{sector_decomp,sdec_term}.

Several computer codes have been developed that implement optimized
strategies for choosing the subsectors and perform the numerical integration:
{\sc sector\_decomposition/CSectors} \cite{sector_decomp, csectors}, \mbox{\sc Fiesta}
\cite{fiesta}, {\sc SecDec} \cite{secdec1,secdec2} and an unnamed implementation
\cite{sdec_form} in FORM \cite{form}.

The integrals obtained after sector decomposition may still
have points where the denominator of the integrand becomes zero, either at the
boundary or in the interior of the integration region. While these
singularities are formally integrable (and thus the integral will be finite),
they nevertheless can cause a numerical integration routine to converge slowly
or not at all. The numerical stability at the boundaries can be improved by
carrying out successive integrations by parts, until the exponent of the
denominator has been reduced sufficiently to ensure robust numerical evaluation
\cite{secdec1}\footnote{See section \ref{sec:bt} for related techniques that
apply to a wider range of integrable singularities.}.

The denominator zeros in the interior of the integration region are the result
of internal thresholds of the corresponding loop diagram. If a threshold is
crossed, the polynomial ${\cal V}_{ij}$ in the subsector integrals
changes sign. Such a
singular point can be avoided by choosing a complex integration path for the
Feynman parameter integrals. A convenient choice is realized by the variable
transformation \cite{nagysoper1,Binoth:2005ff}
\begin{equation}
t_k = z_k - i\lambda z_k(1-z_k)\frac{\partial {\cal V}_{ij}(\vec{t})}{\partial
t_k}, \qquad
0 \leq z_k \leq 1. \label{eq:cd}
\end{equation}
Here $0<\lambda<1$ is a constant parameter. To leading order in $\lambda$, this
generates an imaginary part in ${\cal V}_{ij}$ consistent with the Feynman
$i\varepsilon$ prescription:
\begin{equation}
{\cal V}_{ij}(\vec{t}) = {\cal V}_{ij}(\vec{z}) - i\lambda \sum_k z_k(1-z_k)
 \Bigl (\frac{\partial {\cal V}_{ij}}{\partial z_k}\Bigr )^2 + {\cal O}(\lambda^2).
 \label{eq:dencd}
\end{equation}
Thus, with a suitable choice of $\lambda$ the integrand becomes finite and
relatively smooth everywhere., except for the occurrence of a pinch singularity,
which will be discussed below.
There are two competing factors to consider for
picking the value of $\lambda$: Larger values of $\lambda$ move the integration
contour further away from the threshold singularities, thus improving the
smoothness of the integrand. However, if $\lambda$ becomes too large, the terms
of ${\cal O}(\lambda^2)$ and higher in \eqref{eq:dencd} cannot be neglected
anymore, and they may change the sign of the imaginary part of \eqref{eq:dencd}.
Typically, $\lambda \sim 0.5$ is a reasonable choice.
A complex contour deformation of this kind has been incorporated into several of
the public computer programs mentioned above \cite{secdec2,fiesta}.

Under certain circumstances, there are singular points in the integration volume
that cannot be avoided by a complex contour deformation \cite{nagysoper1}. A
singularity of this kind is called a ``pinch'' singularity. It arises, for
instance, from physical soft and collinear divergences. However, these have
already been removed by the counterterms in eq.~\eqref{eq:secexp} and thus are
they are no source for concern. In addition, pinch singularities can also occur
for special kinematical configurations of the loop momentum (or momenta), where
several loop propagators go on-shell simultaneously, see $e.\,g.$ chapter 13 of
Ref.~\cite{sterman}. In the event that a pinch singularity is encountered the
precision of the numerical integration will be negatively affected and one
should try to put a higher density of integration points near the singular
surface.

Numerical integration based on sector decomposition has been used for the
calculation of a variety of physics processes, including two-loop corrections to
Higgs production at hadron colliders \cite{sdhig} and one-loop
corrections to multi-particle production processes \cite{sdmult}.


\subsection{Mellin-Barnes representations}
\label{sec:mb}

Another powerful tool for disentangling overlapping singularities is the use of
Mellin-Barnes (MB) representations \cite{2lbox,mb,mb2}. It is based on the
replacement of a sum of terms in the $\cal V$ function of a Feynman parameter
integral, see expression in square brackets in eq.~\eqref{eq:feynpuv}, by a MB
integral of a product of terms. The $1/\epsilon$ poles can then be extracted
through analytical continuation and complex contour deformation.

\begin{figure}[tb]
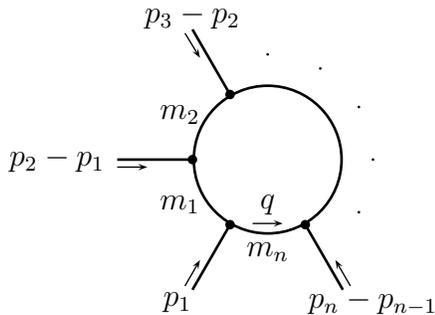

\centering
\pscircle[linewidth=1pt](0,-2){1}
\psline[linewidth=1pt](-2,-2)(-1,-2)\psdot(-1,-2)
\psline[linewidth=1pt](-1,-0.268)(-0.5,-1.134)\psdot(-0.5,-1.134)
\psline[linewidth=1pt](-1,-3.732)(-0.5,-2.866)\psdot(-0.5,-2.866)
\psline[linewidth=1pt](+1,-3.732)(+0.5,-2.866)\psdot(+0.5,-2.866)
\rput(-1.2,-3.9){$p_1$}\psline[linewidth=.5pt]{->}(-1.1,-3.682)(-0.9,-3.3356)
\rput(-2.8,-2){$p_2-p_1$}\psline[linewidth=.5pt]{->}(-2,-2.1)(-1.6,-2.1)
\rput(-1.0,-0.1){$p_3-p_2$}\psline[linewidth=.5pt]{->}(-1.1,-0.318)(-0.9,-0.6464)
\psdot[dotsize=1pt](0,-0.6)
\psdot[dotsize=1pt](0.7,-0.7876)
\psdot[dotsize=1pt](1.2124,-1.3)
\psdot[dotsize=1pt](1.4,-2)
\psdot[dotsize=1pt](1.2124,-2.7)
\rput(+1.4,-3.9){$p_n-p_{n-1}$}\psline[linewidth=.5pt]{->}(+1.1,-3.682)(+0.9,-3.3356)
\rput[t](0,-3.1){$m_n$}
\rput[r](-0.9,-2.6){$m_1$}
\rput[r](-0.9,-1.4){$m_2$}
\rput(0,-2.6){$q$}\psline[linewidth=.5pt]{->}(-0.2,-2.85)(0.2,-2.85)
\rule{0mm}{0mm}\\[4cm]
\caption{Topology of a general one-loop integral.
\label{fig:oneloop}}
\end{figure}

\begin{figure}
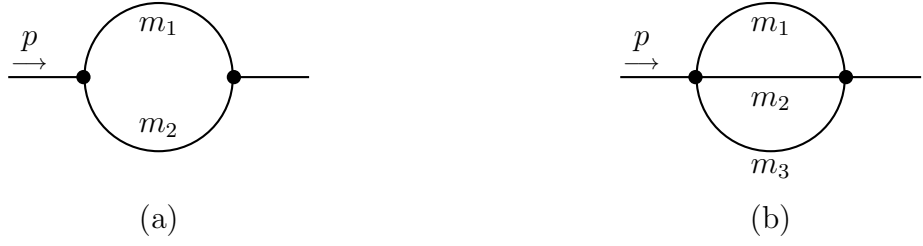

\vspace{1ex}
\begin{center}
\rule{0mm}{0mm}%
\psline(-2,-1)(-1,-1)
\psline(1,-1)(2,-1)
\pscircle(0,-1){1}
\psdot[dotscale=1.5](-1,-1)
\psdot[dotscale=1.5](1,-1)
\rput[t](0,-0.2){$m_1$}
\rput[b](0,-1.8){$m_2$}
\rput[lb](-2,-0.9){$\displaystyle p\atop\longrightarrow$}
\rput(0,-2.9){(a)}
\hspace{8cm}
\psline(-2,-1)(2,-1)
\pscircle(0,-1){1}
\psdot[dotscale=1.5](-1,-1)
\psdot[dotscale=1.5](1,-1)
\rput[t](0,-0.2){$m_1$}
\rput[t](0,-2.1){$m_3$}
\rput[t](0,-1.2){$m_2$}
\rput[lb](-2,-0.9){$\displaystyle p\atop\longrightarrow$}
\rput(0,-2.9){(b)}
\end{center}
\vspace{2.5cm}
\caption{Basic one-loop self-energy diagram (a); and ``sunset'' two-loop
diagram (b). All propagators are assumed to be scalars.
\label{fig:mb1}} 
\end{figure}

To avoid having to deal with the $\det(A)$ term in \eqref{eq:feynpuv}, it is
convenient to split a multi-loop integral into recursive subloop insertions.
Therefore, let us begin with a one-loop integral of the form
\begin{align}
I^{(1)} &= \int {\mathfrak D}q \; \frac{1}{D_1^{\nu_1} \cdots D_n^{\nu_n}}, &
D_j &= k_j^2-m_j^2 = (q-p_j)^2-m_j^2,
\label{eq:onel}
\end{align}
see also Fig.~\ref{fig:oneloop}.
Its Feynman parametrization is given by
\begin{align}
I^{(1)} &= (-1)^N \frac{\Gamma(N-D/2)}{\Gamma(\nu_1) \cdots \Gamma(\nu_n)}
\int _0^1 dx_1 \cdots dx_n \; 
\frac{\delta\bigl(1-\sum_{i=1}^n x_i\bigr) \; x_1^{\nu_1-1} \cdots x_n^{\nu_n-1}}{\Bigl [
 \sum_{i,j=1}^n x_i x_j \, p_i \cdot p_j 
 - \sum_{i=1}^n x_i(p_i^2-m_i^2)
\Bigr ]^{N-D/2}},
\label{eq:feynp1}
\end{align}
where $N = \nu_1 + ... + \nu_n$. Now the denominator term with a sum of terms
involving momentum invariants and masses can be transformed into a MB
representation with the formula
\begin{equation}
\begin{aligned}
\frac{1}{(A_0+\ldots+A_m)^Z} = \frac{1}{(2\pi i)^m} \int_{{\cal C}_1} dz_1
\cdots &\int_{{\cal C}_m} dz_m \; A_1^{z_1} \cdots A_m^{z_m}
A_0^{-Z-z_1-\ldots-z_m} \\
&\times \frac{\Gamma(-z_1) \cdots \Gamma(-z_m)\Gamma(Z+z_1+\ldots+z_m)}{\Gamma(Z)}
\end{aligned}
\end{equation}
when the following conditions are met:
\begin{enumerate}
\item \label{mbreq}
The integration contours ${\cal C}_i$ are straight lines parallel to the
imaginary axis of $z_i$ such that all gamma functions have arguments with
positive real parts;
\item The $A_i$ may in general be complex with $|\arg(A_i)-\arg(A_j)| < \pi$ for
any $i,j$.
\end{enumerate}
The second condition is not automatically fulfilled 
when writing the Feynman parameter integral as in eq.~\eqref{eq:feynp1}, since
one can get terms proportional to the same momentum invariant or mass but with
opposite signs. This can be solved by changing some terms using the relation
$\sum_i x_i=1$. For example, eq.~\eqref{eq:feynp1} for the primitive one-loop
self-energy diagram in Fig.~\ref{fig:mb1}~(a) reads
\begin{align}
I^{(1)}_{\rm fig\ref{fig:mb1}a} &= \Gamma(2-D/2)
 \int_0^1 dx_1 dx_2 \;
 \frac{\delta(1-x_1-x_2)}{[x_2^2p^2-x_2p^2+x_1m_1^2+x_2m_2^2]^{2-D/2}}.
\end{align}
Here the terms with $p^2$ violate condition 2 above. Replacing $x_2$ by
$x_2(x_1+x_2)$ one instead obtains
\begin{align}
I^{(1)}_{\rm fig\ref{fig:mb1}a} &= \Gamma(2-D/2)
 \int_0^1 dx_1 dx_2 \;
 \frac{\delta(1-x_1-x_2)}{[-x_1x_2p^2+x_1m_1^2+x_2m_2^2]^{2-D/2}}.
\label{eq:mbex1}
\end{align}
Here it is understood that $p^2 \to p^2+i\varepsilon$ due to the usual Feynman
$i\varepsilon$ prescription, and as $\arg(p^2+i\varepsilon) < \pi$ all
conditions for the MB integral are satisfied.

After introducing the MB representation, the Feynman parameter integrals can be
carried out using
\begin{equation}
\int_0^1 dx_0 \cdots dx_n \; \delta(1-x_0 - \ldots -x_n) \,
x_0^{\alpha_0-1} \cdots x_n^{\alpha_n-1}
= \frac{\Gamma(\alpha_0) \cdots \Gamma(\alpha_n)}{\Gamma(\alpha_0+\ldots+\alpha_n)}
,
\end{equation}
assuming that the exponents satisfy
\begin{equation}
{\rm Re}(\alpha_i) > 0.
\end{equation}
For the example in eq.~\eqref{eq:mbex1} one thus obtains the MB representation
\begin{equation}
\begin{aligned}
I^{(1)}_{\rm fig\ref{fig:mb1}a} =
 \frac{1}{(2\pi i)^2} \int_{{\cal C}_1} dz_1 \int_{{\cal C}_2} dz_2 \;
 &(m_1^2)^{-\epsilon-z_1-z_2}(m_2^2)^{z_2}(-p^2)^{z_1}\,
 \Gamma(-z_1)\Gamma(-z_2)\Gamma(1+z_1+z_2) \\
&\times \frac{\Gamma(1-\epsilon-z_2)
\Gamma(\epsilon+z_1+z_2)}{\Gamma(2-\epsilon+z_1)}.
\end{aligned}
\label{eq:mbex1r}
\end{equation}
As mentioned above, a MB representation for a two-loop integral can be
constructed by first deriving a MB representation for one subloop and inserting
the result into the second loop. Subsequently, the same steps are
followed to transform the second loop integral into a MB integral. The MB
integral for the first subloop will introduce new terms that depend on
invariants of the second loop momentum, which can be interpreted as additional
propagators, raised to some powers, for the second loop integration. 

As an example, let us consider the ``sunset'' diagram in
Fig.~\ref{fig:mb1}~(b). For the upper subloop, the result \eqref{eq:mbex1r} can
be used with the replacement $p \to q_2$. Thus one obtains the following
expression for the sunset diagram:
\begin{equation}
\begin{aligned}
I^{(2)}_{\rm fig\ref{fig:mb1}b} &=
 \frac{1}{(2\pi i)^2} \int dz_1\,dz_2 \int {\mathfrak D}q_2 \;
 \frac{1}{[q_2^2]^{-z_1}[(q_2-p)^2-m_3^2]} \\
 &\quad\times (-1)^{z_1}(m_1^2)^{-\epsilon-z_1-z_2}(m_2^2)^{z_2}\,
 \Gamma(-z_1)\Gamma(-z_2)\Gamma(1+z_1+z_2) \frac{\Gamma(1-\epsilon-z_2)
\Gamma(\epsilon+z_1+z_2)}{\Gamma(2-\epsilon+z_1)}.
\end{aligned}
\label{eq:mbex2}
\end{equation}
Thus the $q_2^2$ term becomes a new propagator raised to the power $-z_1$,
so that the $q_2$ integral has the form of a self-energy one-loop integral.
Introducing Feynman parameters and the MB representation as before, the final
result is
\begin{align}
I^{(2)}_{\rm fig\ref{fig:mb1}b} = \frac{-1}{(2\pi i)^3} &\int dz_1\,dz_2\,dz_3\; 
(m_1^2)^{-\epsilon-z_1-z_2} (m_2^2)^{z_2} (m_3^2)^{1-\epsilon+z_1-z_3} 
(-p^2)^{z_3} \; \Gamma(-z_2) \Gamma(-z_3) 
\nonumber\\  &\;
\times \Gamma(1+z_1+z_2) \Gamma(z_3-z_1) 
\frac{\Gamma(1-\epsilon-z_2)\Gamma(\epsilon+z_1+z_2)\Gamma(\epsilon-1-z_1+z_3)}
{\Gamma(2-\epsilon+z_3)}.
\label{eq:mbex2r}
\end{align}
As above, the Feynman $i\varepsilon$ prescription is implicitly assumed, $p^2 \to
p^2+i\varepsilon$.

The requirement 1 on page~\pageref{mbreq}, that all gamma functions have a
positive real part, can in
general only be satisfied if $\epsilon$ is chosen to differ from zero by a
finite amount. For instance, the conditions ${\rm Re}\,z_2<0$, ${\rm
Re}\,z_3<0$, $\epsilon + {\rm Re}(z_1+z_2) > 0$ and $\epsilon-1+{\rm
Re}(z_3-z_1)>0$ from eq.~\eqref{eq:mbex2r} imply that $\epsilon > 1/2$.

\begin{figure}
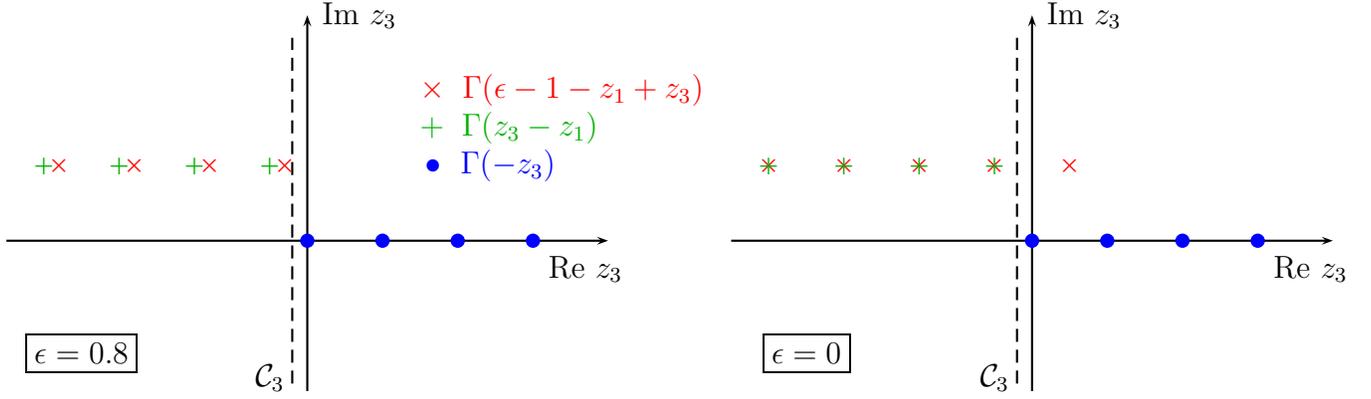

\vspace{1ex}
\begin{center}
\rule{0mm}{0mm}\\[3cm]
\psline{->}(-4,0)(4,0)
\psline{->}(0,-2)(0,3)
\rput[t](3.7,-0.2){Re $z_3$}
\rput[l](0.2,3){Im $z_3$}
\psline[linestyle=dashed](-0.2,-1.9)(-0.2,2.7)
\rput[rb](-0.3,-2){${\cal C}_3$}
\psdot[dotscale=1.5,linecolor=blue](0,0)
\psdot[dotscale=1.5,linecolor=blue](1,0)
\psdot[dotscale=1.5,linecolor=blue](2,0)
\psdot[dotscale=1.5,linecolor=blue](3,0)
\psdot[dotscale=1.5,dotstyle=x,linecolor=red](-0.3,1)
\psdot[dotscale=1.5,dotstyle=x,linecolor=red](-1.3,1)
\psdot[dotscale=1.5,dotstyle=x,linecolor=red](-2.3,1)
\psdot[dotscale=1.5,dotstyle=x,linecolor=red](-3.3,1)
\newrgbcolor{darkgreen}{0 0.7 0}%
\psdot[dotscale=1.5,dotstyle=+,linecolor=darkgreen](-0.5,1)
\psdot[dotscale=1.5,dotstyle=+,linecolor=darkgreen](-1.5,1)
\psdot[dotscale=1.5,dotstyle=+,linecolor=darkgreen](-2.5,1)
\psdot[dotscale=1.5,dotstyle=+,linecolor=darkgreen](-3.5,1)
\rput[l](1.5,1){\blue $\,\bullet\,\,\;\Gamma(-z_3)$}
\rput[l](1.5,1.5){\darkgreen $+\;\;\Gamma(z_3-z_1)$}
\rput[l](1.5,2){\red $\times\;\;\Gamma(\epsilon-1-z_1+z_3)$}
\rput(-3,-1.5){\fbox{$\epsilon = 0.8$}}
\hspace{9.5cm}
\psline{->}(-4,0)(4,0)
\psline{->}(0,-2)(0,3)
\rput[t](3.7,-0.2){Re $z_3$}
\rput[l](0.2,3){Im $z_3$}
\psline[linestyle=dashed](-0.2,-1.9)(-0.2,2.7)
\rput[rb](-0.3,-2){${\cal C}_3$}
\psdot[dotscale=1.5,linecolor=blue](0,0)
\psdot[dotscale=1.5,linecolor=blue](1,0)
\psdot[dotscale=1.5,linecolor=blue](2,0)
\psdot[dotscale=1.5,linecolor=blue](3,0)
\psdot[dotscale=1.5,dotstyle=x,linecolor=red](0.5,1)
\psdot[dotscale=1.5,dotstyle=x,linecolor=red](-0.5,1)
\psdot[dotscale=1.5,dotstyle=x,linecolor=red](-1.5,1)
\psdot[dotscale=1.5,dotstyle=x,linecolor=red](-2.5,1)
\psdot[dotscale=1.5,dotstyle=x,linecolor=red](-3.5,1)
\newrgbcolor{darkgreen}{0 0.7 0}%
\psdot[dotscale=1.5,dotstyle=+,linecolor=darkgreen](-0.5,1)
\psdot[dotscale=1.5,dotstyle=+,linecolor=darkgreen](-1.5,1)
\psdot[dotscale=1.5,dotstyle=+,linecolor=darkgreen](-2.5,1)
\psdot[dotscale=1.5,dotstyle=+,linecolor=darkgreen](-3.5,1)
\rput(-3,-1.5){\fbox{$\epsilon = 0$}}
\end{center}
\vspace{1.5cm}
\caption{Analytic continuation $\epsilon \to 0$ for the $z_3$-dependent 
gamma functions in the numerator of \eqref{eq:mbex2r}. For concreteness, the
following values have been chosen: Re$\,z_1=-0.5$, $\epsilon_{\rm initial} =
0.8$, and the contour ${\cal C}_3$ intersects the real $z_3$ axis at $z_3 =
-0.2$. The left plot is for $\epsilon = \epsilon_{\rm initial}$, while the right
plot depicts the limit $\epsilon \to 0$, where one of the poles of $\Gamma(\epsilon-1-z_1+z_3)$
has crossed $C_3$ from left to right (see red $\times$).
\label{fig:mbcontour}} 
\end{figure}
When taking the limit $\epsilon\to 0$, the poles of some gamma functions may
move across some of the integration contours. This is illustrated in
Fig.~\ref{fig:mbcontour} for the gamma functions
$\Gamma(-z_3)$, $\Gamma(z_3-z_1)$ and $\Gamma(\epsilon-1-z_1+z_3)$ from
\eqref{eq:mbex2r} in the $z_3$-plane.
In such a case, the residue for each crossed pole of a gamma function in the
numerator needs to be added back to the integral. As a
result, for $\epsilon\to 0$ one obtains the original MB integral plus a sum of
lower-dimensional MB integrals stemming from these residue contributions. An 
algorithm for performing this analytic continuation is described in detail in
Refs.~\cite{mb,mb2}.

If the contours have been placed properly to avoid touching any poles of the
gamma functions, all MB integrals in the finite expression are regular and
finite. Thus the $1/\epsilon$ singularities are contained in the coefficients of
these integrals, and they can be obtained explicitly when expanding the gamma
functions for $\epsilon\to 0$. 

For instance, for the pole crossing shown in
Fig.~\ref{fig:mbcontour}, one gets the residue contribution\linebreak
$\text{Res}_{z_3=1+z_1-\epsilon}^{}\,
\Gamma(-z_3)\Gamma(z_3-z_1)\Gamma(\epsilon-1-z_1+z_3) = 
\Gamma(\epsilon-1-z_1)\Gamma(1-\epsilon)$. When continuing to decrease
the value of $\epsilon$, the leading pole of $\Gamma(\epsilon-1-z_1)$ will cross the ${\cal
C}_1$ contour, so that one obtains
$\text{Res}_{z_1=\epsilon-1}\,\Gamma(-z_1)\Gamma(\epsilon-1-z_1)\cdots =
\Gamma(\epsilon-1)\cdots$, where the dots indicate other $z_1$-dependent terms
in the integrand of \eqref{eq:mbex2r} that are unimportant for the argument. For
$\epsilon \to 0$ one then has $\Gamma(\epsilon-1) = -1/\epsilon + \gamma_{\rm E}
-1 + {\cal O}(\epsilon)$.

Automated algorithms for the construction of MB representations for Feynman
integrals have been implemented in the public computer programs AMBRE
\cite{ambre}, MB \cite{mb2} and MBresolve \cite{mbresolve}. The method can also
be extended to include tensor integrals with non-trivial numerator terms, thus
avoiding the need to perform a tensor reduction \cite{ambre,mbtensor}.

\vspace{\medskipamount}
In principle, the MB integrals can be solved analytically by using Barnes' first
and second lemma and the convolution theorem for Mellin transforms, or by
closing the integration contours in the complex plane and summing up the
enclosed residues of the gamma functions. See $e.\,g.$
Refs.~\cite{Smirnov:2006ry,Freitas:2010nx,Ochman:2015fho} for more information on these
techniques. However, for complicated cases that depend on many different
mass scales, at least some of the MB integrals have to be performed numerically.
For this purpose one can introduce the simple parametrization
\begin{equation}
\int_{{\cal C}_i} dz_i \, f(z_i) = i \int_{-\infty}^\infty dy_i \, f(c_i+iy_i),
\label{eq:std}
\end{equation}
where the $c_i$ are real constants. For Euclidean external momenta, all mass and
momentum terms in the MB integrals are simple oscillating exponentials, such as
$(m_j^2)^{iy_i}$, which are bounded from above.
Returning to the example of the ``sunset''
diagram, see Fig.~\ref{fig:mb1}~(b) and eq.~\eqref{eq:mbex2r}, this corresponds
to the parameter region $p^2 < 0$. On the other hand, the gamma functions
rapidly decay to zero for increasing magnitude of the imaginary part of their
arguments. Therefore, numerical integration over a moderate finite integration
interval, $-{\cal O}(10) \lesssim y_i \lesssim {\cal O}(10)$, is adequate to
achieve a high-precision result.

\begin{figure}
\epsfig{figure=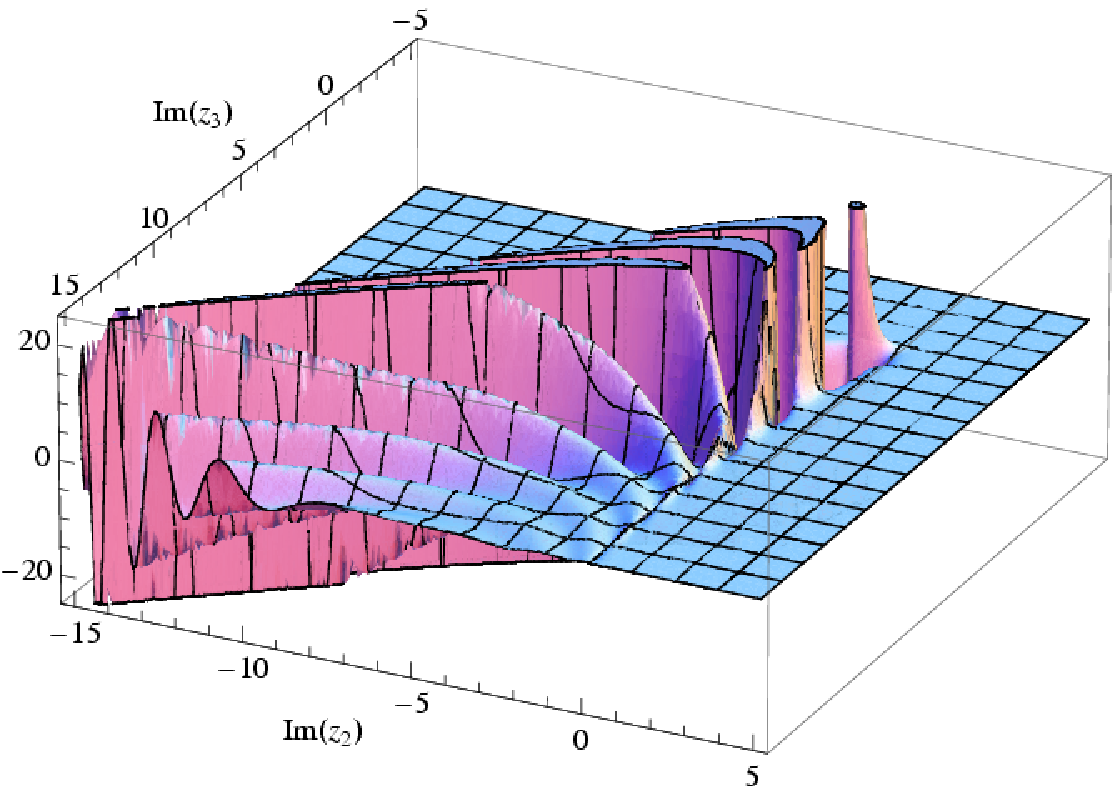, width=8cm}
\hfill
\epsfig{figure=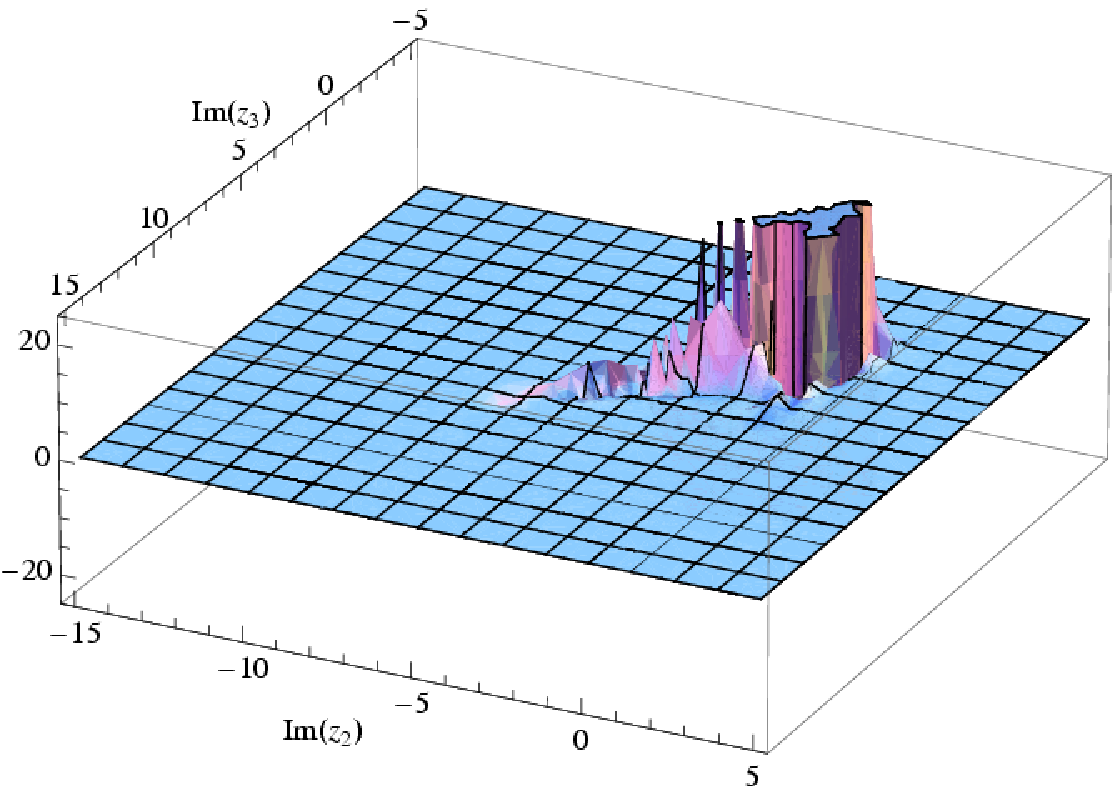, width=8cm}
\vspace{-1ex}
\caption{The real part of the integrand for the ``sunset'' diagram in
eq.~\eqref{eq:mbex2r}, for $p^2=\,$1, $m_1^2=\,$1, $m_2^2=\,$4, $m_3^2=\,$5. For the
left panel, the integration contours have been chosen as straight lines
parallel to the imaginary axis, and the $z_1$ integration has already been
carried out. For the right panel, the contours have been deformed by a
rotation in the complex plane, corresponding to $\theta=0.4$ in eq.~\eqref{eq:mbexpt}.
\label{fig:osc}} 
\end{figure}
However, for physical momenta, $p^2 > 0$, the integrand contains terms of the
form
\begin{equation} 
(-p^2)^{z_3} = (p^2)^{c_3+iy_3} (-1-i\epsilon)^{c_3+iy_3}
             = (p^2)^{c_3+iy_3} e^{-i\pi c_3} e^{\pi y_3}, \label{eq:mbexp}
\end{equation}
which grow exponentially for $y_3 \to\infty$. While the contribution of the
gamma functions is still dominant, so that the integral is formally finite,
there is a long oscillating tail for positive values of $y_3$, see
Fig.~\ref{fig:osc}~(left). As a result, numerical
integration routines often fail to converge in such a case.

This problem can be ameliorated by deforming the integration contours in the
complex plane. To ensure that no pole of the integrand is crossed, all MB
integrations need to be deformed in parallel. A possible choice is \cite{Freitas:2010nx}
\begin{equation}
c_i+iy_i \to c_i + (\theta+i)y_i. \label{eq:mbdeform}
\end{equation}
In eq.~\eqref{eq:mbexp} this leads to
\begin{equation}
(-p^2)^{z_3} = (p^2)^{c_3+iy_3} e^{-i\pi (c_3+\theta y_i)} 
               e^{(\pi+\theta\log p^2) y_3}. \label{eq:mbexpt}
\end{equation}
By choosing $\theta$ appropriate, one can in principle cancel the exponentially
growing term in \eqref{eq:mbexpt}, resulting in much improved numerical
convergence, see Fig.~\ref{fig:osc}~(right). The optimal value of $\theta$ for a given loop
integral can be found by numerically probing the behavior of the integrand for
large values $|y_i|$ of the integration variables.

A disadvantage of this method is that the parameter $\theta$ needs to be
adjusted individually for each type of integral in a given loop calculation, and
the choice depends on the values of the masses and external momenta.
Moreover, there is no guarantee, in particular for non-planar diagrams, that a
convergent result can always be obtained by varying the values of $\theta$.
However, for certain classes of physical two-loop diagrams, this technique
proved to be successful \cite{Freitas:2010nx} and it was applied to the calculation
of two-loop diagrams with triangle fermion subloops for the $Z \to b\bar{b}$
formfactor \cite{Freitas:2012sy}.

Further improvements of the numerical stability and convergence of the numerical
MB integrals can be achieved through other variable transformations. For recent
work in this direction, including possibilities for partial automatization, see
$e.\,g.$ Refs.~\cite{mbnumerics,mbtensor}.


\subsection{Subtraction terms}
\label{sec:sub}

In the previous two subsections, we discussed methods for the extraction of UV
and IR divergences in terms of explicit powers in $1/\epsilon$, whose
coefficients are finite multi-dimensional integrals that can be computed
numerically. This subsection, on the other hand, will outline an alternative
approach where the divergences are subtracted at the integrand level before any
non-trivial integration is performed. The subtracted loop integrals are finite
at every point in the integration region,
and thus they are directly suitable for numerical evaluation.
The subtraction terms are either simple
enough so that they can be computed analytically and added back to the final
expression, or they can be absorbed into
renormalization counterterms that are then evaluated numerically.

\paragraph{Subtractions for multi-loop QED corrections:}
In Ref.~\cite{kino_sub}, a subtraction scheme has been presented that can be
applied to theories with only Abelian gauge interactions, such as QED, and that
works to arbitrary loop order. It has been used for the calculation of 3-loop
\cite{amu3a,amu3}, 4-loop \cite{amu4} and 5-loop \cite{amu5} photonic corrections to
the lepton anomalous magnetic moment.

For the construction of subtraction terms for the UV divergences, the technique
of Kinoshita et al.\ starts with the Feynman parameter integral, see
eq.~\eqref{eq:feynp2}, and proceeds as follows:
\begin{enumerate}
\item For a given subdiagram $\cal S$, its UV limit is taken by retaining the
terms with the smallest number of external momenta in the numerator and taking
the limit $x_i \sim {\cal O}(\delta) \to 0$ for all Feynman parameters $x_i$ associated with $\cal S$.
\label{step1}
\item The function $\cal V$ in the Feynman parameter integral is replaced by
$\cal V_S + V_{\bar{S}}$, where $\cal V_S$ is the $\cal V$ 
function of the subdiagram $\cal S$ and $V_{\bar{S}}$ is the corresponding
function for the residual diagram that is obtained by shrinking $\cal S$ to a
point. 
\item In all other terms in the Feynman parameter integral, the UV limit from
step \ref{step1} is taken, keeping only the leading terms in $\delta$.
\item It can be shown \cite{kino_sub,amu3a} that the Feynman parameter integral then
factorizes into two parts, one that contains the UV divergence of the subdiagram
$\cal S$, and the other containing the contribution for the residual diagram
$\cal\bar S$. The first factor can then be canceled against the UV-divergent
part of the vertex or mass renormalization constant for $\cal S$.
\label{step4}
\end{enumerate}
Nested UV singularities are treated recursively from the minimal subdiagrams to
larger subdiagrams, to the whole diagram, resulting in what is called a
``forest'' of subdiagrams \cite{zimmermann}.

Concerning the treatment of IR divergences, the situation for the computation
of the lepton magnetic moment is somewhat special, since the magnetic form
factor is free from any physical IR singularities, which would need to be
canceled against soft real emission contributions. However, individual loop
diagrams may still contain IR divergences, which cancel against the
non-UV-divergent parts of the renormalization constants. In particular, IR
divergences from self-energy subdiagrams cancel against the mass counterterm,
while those from vertex subdiagrams cancel against the vertex counterterm
\cite{Aoyama:2007bs}. 

To achieve this cancellation in practice, point-by-point in the Feynman
parameter space, the IR-divergent parts of the renormalization constants need to
be written in a form such that they can be combined with the Feynman parameter
integral of the unrenormalized loop diagram. This can achieved by essentially
inverting the factorization  in step \ref{step4} above.

The IR subtraction terms may contain UV subdivergences, which must be removed
as described above. Similarly, the IR divergences may appear within different
subloops, leading to a nested structure. Therefore one needs to sum over all
possibilities of recursively selecting IR-divergent subdiagrams
\cite{Aoyama:2007bs} (again called ``forests'').

These procedures can be cast into an algorithm and implemented in an automated
computer program \cite{Aoyama:2007bs,Aoyama:2005kf}, which can, in principle, tackle arbitrary high loop orders.
However, the techniques have been tailored for QED corrections to magnetic
moments of fermions, and they cannot be easily adapted to other problems. 

\paragraph{Subtractions for general one- and two-loop processes:}
For more general processes and interactions, it is more straightforward to apply
singularity subtractions directly in loop momentum space, see
eq.~\eqref{eq:genint}, rather than in Feynman parameter space. General
subtraction schemes have been formulated for arbitrary one-loop amplitudes
\cite{ns1,Dittmaier:2003bc,becker1,becker2,loosub}, but only partial extensions to the two-loop level are available
\cite{tka_as,loosub,kreimeruv,xloops,kreimerir}. 

Let us begin by reviewing the subtraction approach for one-loop integrals of the
form
\begin{align}
I^{(1)} &= \int {\mathfrak D}q \;
\frac{N(q)}
{D_1 \cdots D_n} \label{eq:genone},
&
D_j &= k_j^2-m_j^2 = (q-p_j)^2-m_j^2, 
\end{align}
where $N(q)$ is a polynomial in the loop momentum $q$, which may also depend on
the external momenta and propagator masses. See Fig.~\ref{fig:oneloop} for a
graphical representation.

A soft singularity occurs if a massless propagator $D_i$ is adjoined at both
ends by two propagators $D_{i-1}$ and $D_{i+1}$ that become on-shell in the
limit that the momentum of $D_i$ vanishes, while all other terms in the
integrand remain regular in that limit. In other words, the necessary
condition for a soft divergence 
is given by $m_i=0$ and $k_{i-1,i+1} \to 0$, $N(q) \not\to 0$ for $k_i
\to 0$. It can be removed with the subtraction
term \cite{ns1,Dittmaier:2003bc,becker1,becker2}
\begin{align}
{\cal G}_{\rm soft}^{(1)} = \frac{1}{D_{i-1}D_iD_{i+1}}\;\lim_{k_i \to 0} \biggl [
N \hspace{-.5em}\prod_{\substack{j\neq \\  i-1,\,i,\,i+1}}\hspace{-.5em}
 D^{-1}_j \biggr ].
\end{align}
This function can be easily integrated analytically in terms of the
well known basic triangle function $C_0$. For explicit expressions, see $e.\,g.$
Refs.~\cite{Dittmaier:2003bc,loosub}.

A collinear singularity is encountered if two massless propagators meet at an
external leg with vanishing invariant mass, $i.\,e.$ $m_i=m_{i-1}=0$ and
$(k_i-k_{i-1})^2=(p_i-p_{i-1})^2=0$. It originates from the integration region
where $k_{i-1}$ (and thus also $k_i$) becomes parallel to the external momentum
$p_i-p_{i-1}$,  $k_{i-1} = x(p_i-p_{i-1})$. At the same time, the numerator
function should be regular in this limit, $N(q) \not\to 0$ for $q \to p_i$,
which otherwise would be a fake singularity. 

Collinear singularities associated with a soft singularity have already been
removed by ${\cal G}_{\rm soft}^{(1)}$. For the remaining collinear
singularities, the simplest subtraction term is given by \cite{loosub}
\begin{align}
{\cal G}^{(1)}_{\rm coll} &= \frac{1}{D_{i-1}D_i} \;\lim_{k_i \to p_i} \biggl [
N \hspace{-.4em}\prod_{j\neq i-1,\,i}\hspace{-.4em}
 D^{-1}_j \biggr ].
\end{align}
In dimensional regularization, the integrated collinear subtraction term is
simply zero. One drawback of this choice, however, is that ${\cal G}^{(1)}_{\rm
coll}$ contains a UV divergence. This can be avoided by using the modified
subtraction term \cite{ns1,becker2}
\begin{align}
{\cal G}^{(1)}_{\rm coll} &= \frac{f_{\rm UV}(k_{i-1}^2,k_i^2)}{D_{i-1}D_i} \;\lim_{k_i \to p_i} \biggl [
N \hspace{-.4em}\prod_{j\neq i-1,\,i}\hspace{-.4em}
 D^{-1}_j \biggr ],
\end{align}
where $f_{\rm UV}$ vanishes for $k \to \infty$ but equals 1 in the collinear
limit. A good choice is \cite{becker2}
\begin{align}
f_{\rm UV}(k_{i-1}^2,k_i^2) &= 1 - \frac{k_{i-1}^2k_i^2}{[(q-Q)^2-\mu_{\rm
UV}^2]^2}.
\end{align}
Here $Q$ and $\mu_{\rm UV}$ are a constant four-vector and constant mass
parameter, respectively, which can be chosen to be complex to prevent the appearance of
additional singularities from ${\cal G}^{(1)}_{\rm
coll}$ inside the integration region. An analytical result for the integrated
collinear subtraction term can be found in Ref.~\cite{becker2}.

At this point, the loop amplitude reads
\begin{align}
I^{(1)} &= \int {\mathfrak D}q \; \biggl [ 
\frac{N(q)}{D_1 \cdots D_n} - \sum_{\rm soft} {\cal G}_{\rm soft}^{(1)}
 - \sum_{\rm coll.} {\cal G}_{\rm coll}^{(1)} \biggr ]
+ \sum_{\rm soft} G_{\rm soft}^{(1)}
+ \sum_{\rm coll.} G_{\rm coll}^{(1)}, \label{eq:irsub}
\end{align}
where the sums run over all soft and collinear singularities in $I^{(1)}$, and
$G_x^{(1)} =  \int {\mathfrak D}q \, {\cal G}^{(1)}_x$ are the integrated
subtraction terms. The first term in \eqref{eq:irsub} is free of IR
divergences, but may still contain UV divergences. These may be extracted with
the help of Feynman parameters \cite{loosub} or by introducing suitable UV
subtraction terms \cite{ns1,becker2}.

In order to do so, it is helpful to first write the integrand in
\eqref{eq:irsub} on a common denominator,
\begin{align}
\frac{\widetilde{N}(q)}{D_1 \cdots D_n} \equiv \frac{N(q)}{D_1 \cdots D_n} - \sum_{\rm soft} {\cal G}_{\rm soft}^{(1)}
 - \sum_{\rm coll.} {\cal G}_{\rm coll}^{(1)},
\end{align}
where $\widetilde{N}(q)$ is a polynomial in $q$ and a rational function in the
other parameters. Then a UV subtraction term for vertices and fermion
self-energies can be constructed as \cite{ns1}
\begin{align}
{\cal G}_{\rm UV}^{(1)} &= \frac{\widetilde{N}_{\rm UV}(q)}{(q^2-\mu^2_{\rm
UV})^n} + {\cal G}_{\rm UV,fin}^{(1)}, \label{eq:uvsub}
\end{align}
where $\widetilde{N}_{\rm UV}$ contains only terms of order $q^{2n-4}$ or higher
from $\widetilde{N}(q)$. It is always possible to add a UV-finite piece,
${\cal G}_{\rm UV,fin}^{(1)}$, to the UV subtraction term, which can be adjusted
according to some renormalization scheme, such as the $\overline{\text{MS}}$
scheme \cite{ns1,becker2}.

\begin{figure}[tb]
\centering
\psfig{figure=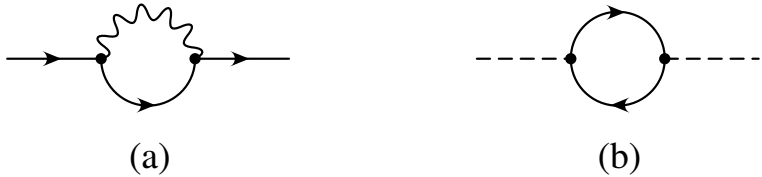, width=10cm}
\vspace{-1ex}
\caption{UV-divergent one-loop self-energy diagrams.
\label{fig:fse}}
\end{figure}

For instance, the UV subtraction term for a gauge-interaction correction to a
fermion self-energy, see Fig.~\ref{fig:fse}~(a), is given by
\begin{align}
\Sigma_{\rm UV}^{f(1)} &= \frac{g^2}{16\pi^2} \, \frac{\gamma^\mu (\qslash +
\frac{1}{2}\pslash + m_f) \gamma_\mu}{[q^2-\mu^2_{\rm UV}]^2} + 
\Sigma_{\rm UV,fin}^{f(1)}.
\end{align}
For boson self-energies, the form \eqref{eq:uvsub} is not adequate due to the
presence of quadratic divergences in the loop integral. Instead one needs to
perform an expansion of the loop propagators in addition to the numerator
$\widetilde{N}$. For example, one can construct in this way the following
subtraction terms for the scalar
self-energy as in Fig.~\ref{fig:fse}~(b):
\begin{align}
\Sigma_{\rm UV}^{\phi(1)} &= -\frac{g^2}{4\pi^2} \, \biggl [
 \frac{1}{q^2-\mu^2_{\rm UV}} + 
 \frac{p^2+q\cdot p + \mu_{\rm UV}^2 - 3m^2}{[q^2-\mu^2_{\rm UV}]^2} -
 \frac{2(q\cdot p)^2}{[q^2-\mu^2_{\rm UV}]^3} \biggr ] +
\Sigma_{\rm UV,fin}^{\phi(1)}.
\end{align}
Expressions for other UV subtraction terms can be found in Ref.~\cite{ns1} and,
including the finite part required for $\overline{\text{MS}}$ renormalization, 
in Ref.~\cite{becker2}.

Note that there is some ambiguity from the possibility of shifting $q$ by a
fixed amount in \eqref{eq:uvsub}, which can be exploited to make the UV
subtraction term better behaved for numerical evaluation.

\vspace{\medskipamount}

At the two-loop level, the situation becomes more involved. As long as a UV or
IR singularity is associated with one of the two subloops only, essentially the same
subtraction terms as above can be used \cite{loosub,kreimerir}. The only
difference is that the analytically integrated subtraction terms need to be
evaluated up to ${\cal O}(\epsilon)$.

In addition, subtraction terms for global UV divergences of both subloops have
been defined \cite{loosub,kreimeruv,xloops}. They are given in terms of simple
two-loop vacuum and self-energy integrals, which are known analytically
\cite{2lvac,Ford:1992pn,Scharf:1993ds}. However, no subtraction terms for overlapping IR
divergences between the two subloops have been constructed to date.

\paragraph{Contour deformation:}
After all IR and UV singularities have been subtracted, the loop integral is
ensured to have a finite result. However, the integrand may still contain
singularities associated with thresholds of the corresponding loop diagram,
which lead to problems for the numerical integration.  This situation is similar
to what has been discussed in section~\ref{sec:secdec}. One approach to
circumvent these singularities,  therefore, is to introduce Feynman parameters,
carry out the loop momentum integration, and then apply the complex contour
deformation in eq.~\eqref{eq:cd}. This method is implemented in the public
computer code {\sc Nicodemos} \cite{loosub}.

In some situations, it may be advantageous not to perform the loop momentum
integration analytically, but to evaluate the loop momentum and Feynman parameter
integrals together in a multi-dimensional loop integration
\cite{nagysoper1,becker2}. This offers more flexibility in choosing a contour
deformation that keeps the integrand fairly smooth everywhere and avoids large
cancellations between different integration regions.

Alternatively, the numerical integration, with a suitable complex contour
deformation, may be carried out directly in the $q$ momentum space 
\cite{soper,soper2,ns2,becker3,becker3a}. This has the advantage of keeping the
integration dimensionality low for multi-leg amplitudes, as well as to avoid a
denominator raised to a high power, which can cause numerical convergence
problems. It may also be beneficial for combining the virtual loop corrections
with contributions from real emission of extra partons
\cite{soper,soper2,rodrigo,kilian}.
On the other hand, the choice of the contour is more complicated for
the $q$-momentum integral.

In the following, the contour deformation for the case of massless loop
propagators will be sketched. A detailed description can be found in
Ref.~\cite{ns2}. The integrand has singularities for $(q-p_j)^2=0$. The goal is
to avoid the singularities by shifting the loop momentum into the complex plane in
their vicinity,
\begin{equation}
q^\mu \to q'^\mu = q^\mu + i\lambda\kappa^\mu_0(q),
\end{equation}
where $\lambda$ is a small parameter. Then
\begin{equation}
(q'-p_j)^2 = (q-p_j)^2 + 2i\lambda \, (q-p_j)\cdot \kappa_0(q) + {\cal
O}(\lambda^2). \label{eq:qdeform}
\end{equation}
To stay compatible with the Feynman $i\varepsilon$ prescription, one must demand
$(q-p_j)\cdot \kappa_0(q) \geq 0$. This condition can be interpreted
geometrically as the requirement that $\kappa_0$ point to the interior (for the $>$ sign) or along
the boundary (for the $=$ sign) of a light cone from the point $p_j$.

\begin{figure}[tb]
\centering
\epsfig{figure=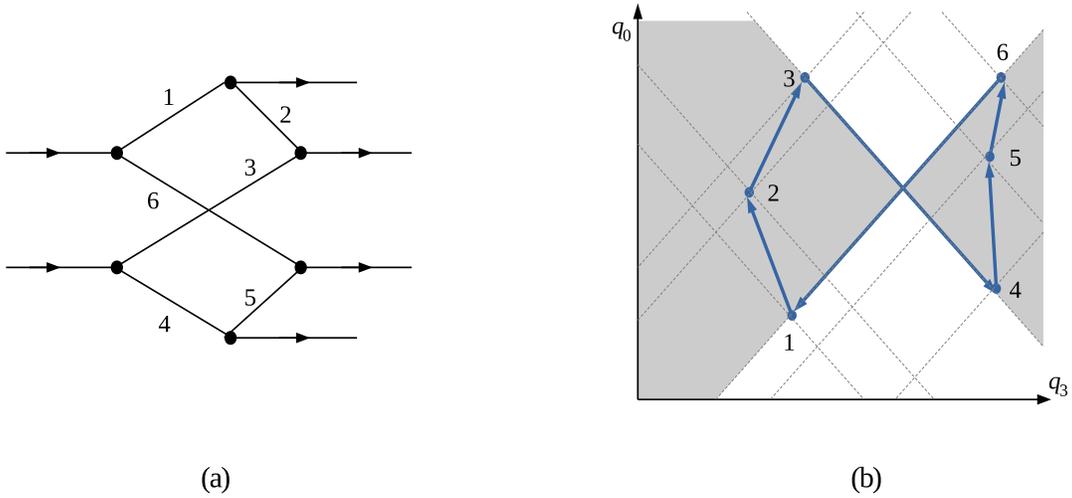, width=16cm, bb=0 40 612 288}
\caption{A sample $2\to 4$ diagram topology (a); and the projection of its
kinematics onto the $q_0$--$q_3$ plane (b).
\label{fig:ldeform}}
\end{figure}
To see how this can be achieved in practice, it is illustrative to consider a
concrete example, such as the diagram topology for a $2 \to 4$ process shown in
Fig.~\ref{fig:ldeform}~(a). Figure~\ref{fig:ldeform}~(b) shows the projection of
the four-dimensional $q$-space onto  the $q_0$--$q_3$ plane for a sample
kinematical configuration. Each dot corresponds to a possible point $q=p_j$,
while an
upward/downward line represents an outgoing/incoming external momentum. The
dashed lines indicate the light cones from the points $p_j$.

Now let us consider the deformation
\begin{align}
\kappa_0 &= -\sum_j c_j (q-p_j), \label{eq:kappa0}
\end{align}
where $c_j(q) \geq 0$ is a function to be specified below. For all $p_i$ lying
inside the backward light cone from $p_j$, one has $(q-p_j)\cdot(q-p_i) = (q-p_j)^2 +
(q-p_j)\cdot(p_j-p_i)>0$ when $q$ is on the surface of the backward light cone
($i.\,e.$ when $(q-p_j)^2=0$ and $q_0-p_{j,0} < 0$). Therefore, $c_i>0$ is a
good choice in these cases to ensure that $(q-p_j)\cdot \kappa_0(q) \geq 0$. On the other hand, when $q$ is on the forward light
cone, one has $(q-p_j)\cdot(q-p_i) < 0$ and thus we choose $c_i=0$.

Similarly, for all $p_i$ lying inside the forward light cone from $p_j$, one has
$(q-p_j)\cdot(q-p_i) > 0$ ($ <0$) when $q$ is on the surface of the forward
(backward) light cone, and thus we choose $c_i>0$ ($c_i=0$) in these cases. This
can be realized through
\begin{align}
c_j = h_+(q-p_{j+1}) \,h_-(q+p_{j+1}) \,g(q), \label{eq:cjcoeff}
\end{align}
where $h_\pm(k)$ are smooth functions with the properties $h_\pm(k) = 0$ for
$|\vec{k}| > \pm k_0$ and $h_\pm(k) \to 1$ for $k_0 \mp |\vec{k}| \to \infty$.
$g(k)$ is a smooth function that vanishes for $k^\mu \to \infty$, to ensure that
the contour deformation goes to zero at the integral boundaries. The precise
definition of $h_\pm$ and $g$ can be found in Ref.~\cite{ns2}.

The choice \eqref{eq:cjcoeff} ensures that $(q-p_j)\cdot \kappa_0(q) \geq 0$ if
$q$, $p_j$ and the $p_i$ are restricted to lie either inside the left or inside
the right shaded regions of Fig.~\ref{fig:ldeform}~(b). While some of the terms
in \eqref{eq:kappa0} are zero due to the $h_{\pm}$ functions, there is always at
least one non-zero term with the correct sign.
The only exceptions are
the points $q-p_j=0$ and the lines $q=xp_j-(1-x)p_i$ for $0<x<1$ and
$(p_j-p_i)^2-0$. In these two cases, which correspond to soft and collinear
singularities, $(q-p_j)\cdot \kappa_0(q) = 0$, $i.\,e.$ the contour is pinched.
However, the soft and collinear singularities have been subtracted already, so
these points do not require any special contour deformation.

The treatment of the unshaded regions in Fig.~\ref{fig:ldeform}~(b) requires the
consideration of special cases for the coefficients $c_j$ and the introduction
of additional terms in eq.~\eqref{eq:kappa0}. See Ref.~\cite{ns1} for more
information.

Finally, one has to define the scaling parameter $\lambda$ in
\eqref{eq:qdeform}. To improve the smoothness of the integrand and increase the
speed of convergence, it is advantageous to pick as large a value of $\lambda$
as possible. On the other hand, one must ensure that no other singularities are
crossed by the contour as $\lambda$ is increased. To balance these two
requirements, it is advantageous to define $\lambda$ as a function of $q$.
Suitable choices that avoid singularities from other propagators and from the UV
subtraction terms are given in Ref.~\cite{ns1} and Ref.~\cite{opti2},
respectively.

While this algorithm for the $q$-space contour deformation is rather involved,
it is straightforward to implement as a numerical computer program. Extensions
to handle massive loop propagators  \cite{becker3} and multi-loop integrals
\cite{becker3a} are also known. Techniques to improve the efficiency of the
numerical integration are discussed in Refs.~\cite{opti1,opti2}. One notable
application of the direct contour deformation methods is the calculation of
one-loop QCD corrections to
five-, six- and seven-jet production in $e^+e^-$ annihilation \cite{eenjet}.

\paragraph{Related methods:} Instead of carrying out all $4L$ dimensions of the
loop integral numerically, where $L$ is the number of loops, one can also try to
evaluate as many of the integrations analytically as possible, to arrive at a
low-dimensional numerical integral. This approach has been pioneered for
two-loop integrals in Ref.~\cite{Kreimer:1991jv}. For this purpose, the loop momenta $q_1$ and
$q_2$ are split into their energy components, $q_{10}$ and $q_{20}$, and their
momentum components parallel and transverse 
to an external momentum, $q_{1||}$, $q_{1\perp}$, $q_{2||}$, and $q_{2\perp}$.

For a general finite two-loop self-energy integral, it was shown that the
momentum-component integrations can be performed analytically, resulting in a
two-dimensional integral representation related to the $q_{10}$ and $q_{20}$
integrations \cite{Kreimer:1991jv,Czarnecki:1994td}. The finiteness of the
integral is assumed to have been achieved through suitable subtractions.
This method can be
straightforwardly extended to general $q_{1,2}$-dependent tensor structures in
the numerator of the integral, by splitting these into components in the same
fashion \cite{kreimeruv,xloops}.

Similarly, two-dimensional integral representations were obtained for
two-loop vertex-type master integrals, which have a trivial numerator function
$N(q_1,q_2)\equiv 1$ \cite{Czarnecki:1994td,Frink:1996ya}. Three-dimensional
integral representations for certain two-loop box master integrals were obtained
in Ref.~\cite{kreimerbox}. These techniques have
been applied towards the calculation of dominant two-loop corrections to
Higgs-boson decays in the limit of a large Higgs mass \cite{Frink:1996sv}.

An attractive feature of this method is the low dimension of the resulting
numerical integrals, which thus can be evaluated with deterministic integration
algorithms to high precision. However, it has certain shortcomings: No general
procedure
for numerator terms in three- and higher-point two-loop
integrals is known. In addition, loop diagrams with internal threshold have
singularities in the interior of the integration region where the integrand
denominator vanishes. In Refs.~\cite{xloops,xloops2}, these singular points are
circumvented by assigning a non-zero numerical value to the $\varepsilon$
parameter of the Feynman $i\varepsilon$ prescription. While in principle this
renders the integrand finite everywhere in the integration region, the numerical
integration will converge relatively slowly near such a point.


\subsection{Dispersion relations}
\label{sec:disp}

Dispersion relations are based on analytical properties of field theory
amplitudes, and they can be used to construct the value of a loop diagram from
its imaginary part. The latter is related to the discontinuity across the branch
cuts between different Riemann sheets. It can be constructed from cuts through
internal lines of the loop diagram using the Cutkosky rules
\cite{Cutkosky:1960sp}. See Ref.~\cite{Kniehl:1996rh} for a pedagogical introduction.

Generically, a dispersion relation has the form
\begin{align}
I(q^2) &= \frac{1}{2\pi i} \int_{s_0}^\infty \frac{\Delta
I(s)}{s-q^2-i\varepsilon}, \label{eq:disp}
\end{align}
where $q^2$ is a characteristic squared external momentum, and $\Delta I(s) =
\frac{1}{2i} \,{\rm Im}\,I(s)$ is the discontinuity of the loop integral $I(s)$.
The idea is that the imaginary part is relatively simple to determine from the
Cutkosky rules, and then one can use eq.~\eqref{eq:disp} to find the result for
the whole function $I(s)$. Note that the dispersion integral can be applied to
a multi-loop diagram itself or to some subloop, and either choice may be more
convenient for different types of Feynman diagrams.
If possible, one may try to perform the $s$-integral in \eqref{eq:disp}
analytically, see Ref.~\cite{dispana} for early applications. Here, we want to
focus on the numerical evaluation of the dispersion integral.

\begin{figure}
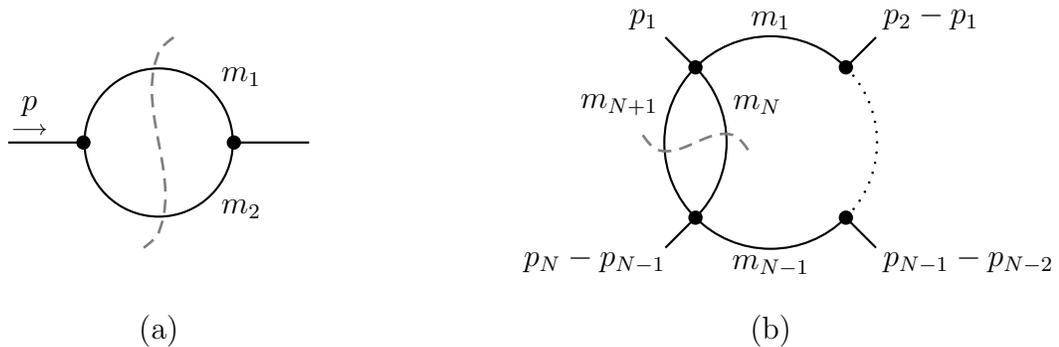

\vspace{5ex}
\begin{center}
\rule{0mm}{0mm}%
\psline(-2,-1)(-1,-1)
\psline(1,-1)(2,-1)
\pscircle(0,-1){1}
\psdot[dotscale=1.5](-1,-1)
\psdot[dotscale=1.5](1,-1)
\rput[t](1.1,0){$m_1$}
\rput[b](1.1,-2){$m_2$}
\rput[lb](-2,-0.9){$\displaystyle p\atop\longrightarrow$}
\pscurve[linestyle=dashed,linewidth=1pt,linecolor=gray](0.2,0.4)(0,0.2)(0,-2.2)(-0.2,-2.4)
\rput(0,-3.5){(a)}
\hspace{8cm}
\psarc(0,-1){1.414}{45}{-45}
\psarc(-2,-1){1.414}{-45}{45}
\psarc[linestyle=dotted,linewidth=1pt](0,-1){1.414}{-45}{45}
\psdot[dotscale=1.5](-1,0)
\psdot[dotscale=1.5](1,0)
\psdot[dotscale=1.5](-1,-2)
\psdot[dotscale=1.5](1,-2)
\psline(-1,0)(-1.4,0.4)
\psline(1,0)(1.4,0.4)
\psline(-1,-2)(-1.4,-2.4)
\psline(1,-2)(1.4,-2.4)
\pscurve[linestyle=dashed,linewidth=1pt,linecolor=gray](-1.7,-0.9)(-1.5,-1.1)(-0.5,-0.9)(-0.3,-1.1)
\rput[r](-1.5,-0.5){$m_{N+1}$}
\rput[l](-0.5,-0.5){$m_{N}$}
\rput[b](0,0.5){$m_1$}
\rput[t](0,-2.5){$m_{N-1}$}
\rput[rb](-1.5,0.5){$p_1$}
\rput[lb](1.5,0.5){$p_2-p_1$}
\rput[lt](1.5,-2.4){$p_{N-1}-p_{N-2}$}
\rput[rt](-1.4,-2.4){$p_{N}-p_{N-1}$}
\rput(0,-3.5){(b)}
\end{center}
\vspace{3.1cm}
\caption{One-loop scalar self-energy diagram with the cut contributing to its
discontinuity (a); and a general two-loop scalar diagram with a self-energy
subloop (b).
\label{fig:dispse}} 
\end{figure}
The simplest case is the one-loop scalar self-energy function, see
Fig.~\ref{fig:dispse}~(a), which will be called $B_0(p^2,m_1^2,m_2^2)$ in the
following. As a function of $p^2$, it exhibits a discontinuity along the
positive real axis for $p^2 > (m_1+m_2)^2$. The discontinuity can be calculated
by cutting the diagram through the $m_1$ and $m_2$ lines, resulting in a $1\to
2$ decay process. The result in dimensional regularization is
\begin{align}
B_0(p^2,m_1^2,m_2^2) &= \int_{(m_1+m_2)^2}^\infty {\rm d}s \,
  \frac{\Delta B_0(s,m_1^2,m_2^2)}{s - p^2-i\varepsilon}, \label{eq:dispb0} \\
\Delta B_0(s,m_1^2,m_2^2) &= 
  \frac{\Gamma(D/2-1)}{\Gamma(D-2)} \, \frac{\lambda^{(D-3)/2}(s,m_1^2,m_2^2)}%
  {s^{D/2-1}},
\end{align}
where $\lambda(a,b,c)$ is defined in \eqref{eq:kallen}. With this expression, a
scalar two-loop integral with a self-energy subloop, see
Fig.~\ref{fig:dispse}~(b), can be written as \cite{intnum}
\begin{equation}
\begin{aligned}
I^{(2)}_{\rm fig\ref{fig:dispse}b}(\{p_i\};\{m_i^2\}) = - &\int_{(m_N+m_{N+1})^2}^\infty {\rm d}s \;
  \Delta B_0(s,m_N^2,m_{N+1}^2) \\
  \times &\int {\mathfrak D}q  \,
  \frac{1}{q^2-s} \,
  \frac{1}{(q+p_1)^2 - m_1^2} \cdots \frac{1}{(q+p_{N-1})^2 -
  m_{N-1}^2}.
\end{aligned} \label{eq:subbubble}
\end{equation}
The integral in the second line is a $N$-point one-loop function, for which the
well-known analytical expression \cite{hv1,denner200} can be inserted. The remaining
integration over $s$ can then be carried out numerically.

This approach can be easily extended to deal with two-loop integrals with a
self-energy subloop and a non-trivial tensor structure ($i.\,e.$ with
non-trivial terms in the numerator of the integrand). For this purpose, one
first decomposes the self-energy subloop into a sum of Lorentz covariant
building blocks \cite{weiglein1,swucc1}. For example, a vector-boson and
a fermion self-energy can be written as
\begin{align}
\Sigma_{\mu\nu}^V(q) &= \Bigl ( g_{\mu\nu}-\frac{q_\mu q_\nu}{q^2} \Bigr )
\Sigma_{\rm T}^V(q^2) + \frac{q_\mu q_\nu}{q^2} \Sigma_{\rm L}^V(q^2), \\
\Sigma^f(q) &= \qslash P_{\rm L} \Sigma^f_{\rm L}(q^2)
 + \qslash P_{\rm R} \Sigma^f_{\rm R}(q^2) + m_f \Sigma^f_{\rm S}(q^2),
\end{align}
respectively. Here $P_{\rm L,R} = \frac{1}{2}(1\mp \gamma_5)$. Inserting these
expressions into the second loop, one obtains a dispersion integral similar to
\eqref{eq:subbubble}, except that the $q$-integral is in general a one-loop tensor
integral. The latter can be evaluated analytically with the standard
Passarino-Veltman decomposition \cite{pasvelt,denner200,Denner:2005nn}.

\begin{figure}
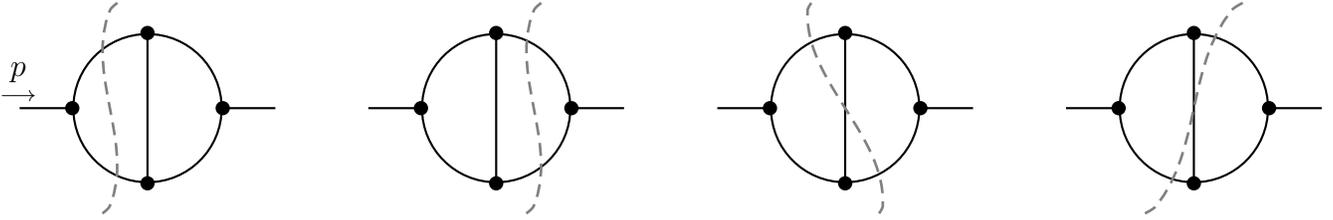

\vspace{3ex}
\begin{center}
\rule{0mm}{0mm}%
\psline(-1.7,-1)(-1,-1)
\psline(1,-1)(1.7,-1)
\psline(0,0)(0,-2)
\pscircle(0,-1){1}
\psdot[dotscale=1.5](-1,-1)
\psdot[dotscale=1.5](1,-1)
\psdot[dotscale=1.5](0,0)
\psdot[dotscale=1.5](0,-2)
\rput[lb](-2,-0.9){$\displaystyle p\atop\longrightarrow$}
\pscurve[linestyle=dashed,linewidth=1pt,linecolor=gray]
(-0.4,0.4)(-0.5,0.3)(-0.5,-2.3)(-0.6,-2.4)
\hspace{4.5cm}
\psline(-1.7,-1)(-1,-1)
\psline(1,-1)(1.7,-1)
\psline(0,0)(0,-2)
\pscircle(0,-1){1}
\psdot[dotscale=1.5](-1,-1)
\psdot[dotscale=1.5](1,-1)
\psdot[dotscale=1.5](0,0)
\psdot[dotscale=1.5](0,-2)
\pscurve[linestyle=dashed,linewidth=1pt,linecolor=gray]
(0.6,0.4)(0.5,0.3)(0.5,-2.3)(0.4,-2.4)
\hspace{4.5cm}
\psline(-1.7,-1)(-1,-1)
\psline(1,-1)(1.7,-1)
\psline(0,0)(0,-2)
\pscircle(0,-1){1}
\psdot[dotscale=1.5](-1,-1)
\psdot[dotscale=1.5](1,-1)
\psdot[dotscale=1.5](0,0)
\psdot[dotscale=1.5](0,-2)
\pscurve[linestyle=dashed,linewidth=1pt,linecolor=gray]
(-0.45,0.4)(-0.5,0.3)(0.5,-2.3)(0.45,-2.4)
\hspace{4.5cm}
\psline(-1.7,-1)(-1,-1)
\psline(1,-1)(1.7,-1)
\psline(0,0)(0,-2)
\pscircle(0,-1){1}
\psdot[dotscale=1.5](-1,-1)
\psdot[dotscale=1.5](1,-1)
\psdot[dotscale=1.5](0,0)
\psdot[dotscale=1.5](0,-2)
\pscurve[linestyle=dashed,linewidth=1pt,linecolor=gray]
(0.65,0.4)(0.5,0.3)(-0.5,-2.3)(-0.65,-2.4)
\end{center}
\vspace{2cm}
\caption{Cuts through the basic scalar two-loop self-energy diagram with
triangle subloops. \label{fig:master}} 
\end{figure}
In addition, the dispersion approach has been used to derive a one-dimensional
integral representation for the scalar self-energy integral in Fig.~\ref{fig:master}
\cite{master}.
The discontinuity of this diagram has been obtained by summing over the
contributions from all cuts shown in the figure, see Ref.~\cite{master} for more
details. The reduction of tensor integrals with the topology in
Fig.~\ref{fig:master} is described in Ref.~\cite{weiglein1}.

For the two-loop examples discussed in this section so far, the dispersion
relation method leads to one-dimensional integral expressions, which can be
evaluated to a very high precision with a deterministic integration algorithm.
The numerical integrals are free from problematic singularities in the interior
of the integration interval, even for loop diagrams with physical thresholds. In
some cases, the integrand may contain terms proportional to
$1/(s-p^2-i\varepsilon)$, which can be rendered smooth with the simple variable
transformation $s \to t = \log(s-p^2-i\varepsilon)$.

On the other hand, the method also has several drawbacks. Firstly, there is no
automated treatment of UV and IR divergences. These manifest themselves as
singularities at the lower or upper limit of the dispersion integral,
respectively, and they must be removed from the dispersion integral using
suitable subtraction terms. These terms have to be derived
by hand for each different class of diagram. 

For instance, a two-loop diagram of the form in Fig.~\ref{fig:dispse}~(b)
has a UV divergence stemming from the self-energy subloop, but no global UV
divergence if it has at least five propagators. The subloop UV divergence can be
removed by subtracting the term
\begin{equation}
B_0(M^2,m_N^2,m_{N+1}^2) \int {\mathfrak D}q  \,
  \frac{1}{(q+p_1)^2 - m_1^2} \cdots \frac{1}{(q+p_{N-1})^2 -
  m_{N-1}^2},
\end{equation}
which is a product of two one-loop functions. Here $M^2$ is an arbitrary mass
parameter.
The subtracted dispersion integral then reads
\begin{equation}
\begin{aligned}
I^{(2)}_{\rm fig\ref{fig:dispse}b,sub}(\{p_i\};\{m_i^2\}) = 
- &\int_{(m_N+m_{N+1})^2}^\infty {\rm d}s \;
  \frac{\Delta B_0(s,m_N^2,m_{N+1}^2)}{s-M^2} \\
  \times &\int {\mathfrak D}q  \,
  \frac{q^2-M^2}{[q^2-s][(q+p_1)^2 - m_1^2] \cdots [(q+p_{N-1})^2 -
  m_{N-1}^2]}.
\end{aligned} \label{eq:subUVsub}
\end{equation}
The integrand behaves like $s^{-2-\epsilon}$ for $s \to \infty$ and thus the
integral is UV-finite. 
Additional subtraction terms are needed for IR singularities.

\begin{figure}
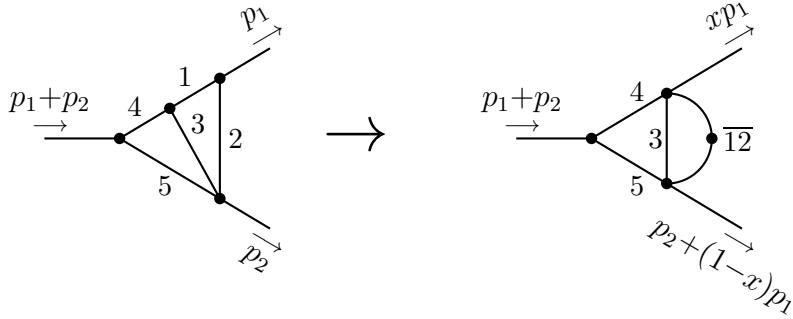

\vspace{2cm}
\begin{center}
\rule{0mm}{0mm}
\psline(-2,0)(-1,0)
\psline(-1,0)(1,1.2)
\psline(-1,0)(1,-1.2)
\psline(0.333,0.8)(0.333,-0.8)
\psline(-0.333,0.4)(0.333,-0.8)
\psdot[dotscale=1.2](-1,0)
\psdot[dotscale=1.2](0.333,0.8)
\psdot[dotscale=1.2](0.333,-0.8)
\psdot[dotscale=1.2](-0.333,0.4)
\rput[lb](-2.5,0.1){$\displaystyle p_1{+}p_2\atop\longrightarrow$}
\rput[b]{30}(1,1.3){$\displaystyle p_1\atop\longrightarrow$}
\rput[t]{-30}(1,-1.3){$\displaystyle \overset{\longrightarrow}{p_2}$}
\rput[rb](-0.7,0.3){\small 4}
\rput[rb](-0.033,0.7){\small 1}
\rput[l](0.45,0){\small 2}
\rput[l](-0.05,0.2){\small 3}
\rput[rt](-0.3,-0.5){\small 5}
\hspace{2cm}
\rput(0,0){\huge $\rightarrow$}
\hspace{4cm}
\psline(-2,0)(-1,0)
\psline(-1,0)(1,1.2)
\psline(-1,0)(1,-1.2)
\psarc(0,0){0.6}{-90}{90}
\psline(0,0.6)(0,-0.6)
\psdot[dotscale=1.2](-1,0)
\psdot[dotscale=1.2](0.6,0)
\psdot[dotscale=1.2](0,-0.6)
\psdot[dotscale=1.2](0,0.6)
\rput[lb](-2.5,0.1){$\displaystyle p_1{+}p_2\atop\longrightarrow$}
\rput[b]{30}(1,1.3){$\displaystyle xp_1\atop\longrightarrow$}
\rput[t]{-30}(1,-1.3){$\displaystyle \overset{\longrightarrow}{p_2{+}(1{-}x)p_1}$}
\rput[rb](-0.3,0.5){\small 4}
\rput[l](0.75,0){\small $\overline{12}$}
\rput[r](-0.05,0){\small 3}
\rput[rt](-0.3,-0.5){\small 5}
\end{center}
\vspace{1.8cm}
\caption{Reduction of triangle subloop to self-energy subloop by means of a
Feynman parameter. \label{fig:2ltri}} 
\end{figure}
Secondly, another limitation of the dispersion methods is the difficulty of
extending the elegant examples mentioned above to more complicated two-loop
topologies. One possibility is the introduction of Feynman parameters to reduce
triangle subloops to self-energy subloops, which then can be evaluated as in
\eqref{eq:subbubble} \cite{swlong}. For example, the propagators 1 and 2 of the
two-loop diagram in Fig.~\ref{fig:2ltri} can be combined using a Feynman
parameter $x$, resulting in a diagram with a self-energy subloop, where one
propagator is raised to the power two and has an $x$-dependent mass and
$x$-dependent external momenta:
\begin{equation}
\begin{gathered}
\ [(q+p_1)^2-m_1^2]^{-1} \; [q^2-m_2^2]^{-1} = \int_0^1 {\rm d}x
  \; [(q+xp_1)^2 - m_{\overline{12}}^2]^{-2} \\
m_{\overline{12}}^2 = x \, m_1^2 + (1-x) m_2^2 - x(1-x)p_1^2.
\end{gathered}
\end{equation}
The integration over Feynman parameters, as well as the dispersion integral, are
performed numerically. In this way, all basic two-loop vertex topologies can be
represented by at most two-dimensional numerical integrals.

However, the dispersion relation is only defined for non-negative masses $m_{1,2}
\geq 0$, whereas the Feynman-parameter dependent mass $m_{\overline{12}}$ can in
general also become negative. Thus, the combination of dispersion relations and
Feynman parameters can only be applied to restricted parameter regions.

Dispersion relations have been used for the calculation of two-loop corrections
to the prediction of the $W$-boson mass from muon decay in the full Standard
Model \cite{mw,mwlong}, for subsets of two-loop diagrams contributing to
the $Z \to f\bar{f}$ formfactors \cite{sweff,swucc1,swlong}, as well as for
two-loop quark loop corrections to Bhabha scattering \cite{bhabha}.


\subsection{Bernstein-Tkachov method}
\label{sec:bt}

The Bernstein-Tkachov method \cite{bernstein,tkachov} uses analytic properties of the integrand to render
singularity peaks into a smoother form that is suitable for numerical
integration without complex contour deformation. 

It can be applied most
straightforwardly for one-loop integrals, see eq.~\eqref{eq:onel}, whose Feynman parametrization can be
written as
\begin{align}
I^{(1)} &= (-1)\Gamma(n-D/2)\int_0^1 dx_1 \cdots dx_n \;
\delta\Bigl(1-\sum_{i=1}^n x_i\Bigr) \;{\cal Q}(\vec{x}_e)\,{\cal
V}^{-n+D/2}(\vec{x}_e),
\label{eq:feynpb}
\end{align}
where $\cal Q$ and $\cal V$ are polynomials in $\vec{x}_e \equiv (x_1, \dots,
x_n)^\top$, which also depend
on the internal masses and external momenta. 
A non-trivial function $Q \neq 1$
occurs as a result of a non-trivial numerator function $N(q)$ in
\eqref{eq:onel}. The $x_n$-integral can be evaluated to eliminate the
$\delta$-function, yielding
\begin{align}
I^{(1)} &= (-1)\Gamma(n-D/2)\int_0^1 dx_1 \int_0^{1-x_1} dx_2 \cdots
\int_0^{1-\sum_{i=1}^{n-2} x_i} dx_{n-1} \; {\cal Q}(\vec{x})\,{\cal
V}^{-\nu-\epsilon}(\vec{x}),
\label{eq:feynpa}
\end{align}
where $\vec{x} \equiv (x_1, \dots,
x_{n-1})^\top$, $\nu = n-2$ and $\epsilon = (4-D)/2$. 
$\cal V$ is a
quadratic form in the Feynman parameters,
\begin{align}
{\cal V}(\vec{x}) &= \vec{x}^\top H \vec{x} + 2\vec{K}^\top\vec{x} + L,
\end{align}
where $H$ is a $(n{-}1){\times}(n{-}1)$-matrix, $\vec{K}$ is a $(n{-}1)$-dimensional vector, and
$L$ is a scalar in Feynman-parameter space.
For
$\nu > 0$, the integrand in \eqref{eq:feynpa} exhibits singular behavior when
${\cal V}(\vec{x})$ becomes zero. This can occur for points $\vec{x}$ inside the
integration region if the loop diagram has internal thresholds. 

It was shown by Tkachov \cite{tkachov} that the following relation holds:
\begin{align}
{\cal
V}^{-\nu-\epsilon} = \frac{1}{B} \biggl [ 1- \frac{(\vec{x}+\vec{A})^\top
\vec{\partial}_x}{2(1-\nu-\epsilon)}\biggr ] {\cal V}^{-\nu-\epsilon+1},
\label{eq:bt}
\end{align}
where
\begin{align}
B &= L - \vec{K}^\top H^{-1} \vec{K}, &
\vec{A} &= H^{-1} \vec{K}, &
\vec{\partial}_x &= \bigl (\tfrac{\partial}{\partial x_1}, \dots,
\tfrac{\partial}{\partial x_{n-1}}\bigr )^\top.
\end{align}
This relation can be used to increase the power of the polynomial ${\cal
V}(\vec{x})$. For example, for the class of one-loop three-point functions with
two independent Feynman parameters, one finds
\begin{align}
&\int_0^1 dx_1 \int_0^{1-x_1} dx_2 \;{\cal Q}(\vec{x})\,
{\cal V}^{-\nu-\epsilon}(\vec{x}) \nonumber \\
&= \frac{1}{2(1-\nu-\epsilon)B} \,\biggl\{
\int_0^1 dx_1 \int_0^{1-x_1} dx_2 \; 
\biggl [(2-\nu-\epsilon){\cal Q}(\vec{x})+
\sum_{i=1}^2 A_k \frac{\partial \cal Q}{\partial x_k} \biggr ] 
{\cal V}^{-\nu-\epsilon+1}(\vec{x}) \nonumber \\
&\qquad\qquad\qquad\qquad + \int_0^1 dx_1 \; A_2{\cal Q}(\vec{x})\,
{\cal V}^{-\nu-\epsilon+1}(\vec{x})\Big|_{x_2=0} 
+ \int_0^1 dx_2 \; A_1{\cal Q}(\vec{x})\,
{\cal V}^{-\nu-\epsilon+1}(\vec{x})\Big|_{x_1=0} \nonumber \\
&\qquad\qquad\qquad\qquad - \int_0^1 dx_1 \; (1+A_1+A_2){\cal Q}(\vec{x})\,
{\cal V}^{-\nu-\epsilon+1}(\vec{x})\Big|_{x_2=1-x_1} \biggr\}.
\end{align}
Here the one-dimensional integrals stem from performing an integration-by-parts
operation on the derivative term in eq.~\eqref{eq:bt}. This operation can be
performed repeatedly until the power of $\cal V$ is reduced to ${\cal
V}^{-\epsilon}$. Then one can simply perform a Laurent expansion about
$\epsilon=0$ to extract the UV divergent contributions, while the finite terms
contain only logarithms of $\cal V$ as the worst singularities. These can be
integrated efficiently with standard numerical algorithms.

IR divergent configurations can be either evaluated by suitable rearrangements
of the Feynman parameter integral such that one Feynman parameter integration
can be performed analytically \cite{Passarino:2001wv}. With a subsequent Laurent expansion about
$\epsilon=0$, the IR divergent terms are obtained explicitly. Alternatively, the
IR divergent cases can be handled with the help of sector
decomposition or Mellin-Barnes representations \cite{fppu1}. See sections
\ref{sec:secdec} and \ref{sec:mb} for the definition of these methods.

\vspace{\medskipamount}
For two-loop integrals, the Bernstein-Tkachov relation \eqref{eq:bt} cannot
be employed straightforwardly, since in general the polynomial $\cal V$ in
\eqref{eq:feynpuv} is not a quadratic form in the Feynman parameters.
Instead, one may apply eq.~\eqref{eq:bt} to the one-loop subdiagram with the largest
number of internal lines \cite{Passarino:2001wv}. In other words, one introduces
two sets of Feynman parameters, one set $\vec{x}$ for the subloop with most
propagators, and another set $\vec{y}$ for the remainder of the two-loop
diagram. The Bernstein-Tkachov relation \eqref{eq:bt} is then applied to the
variables $\vec{x}$, but now the coefficients $A_{ij}$, $P_i$ and $M$ are
dependent on $\vec{y}$. Repeated application of \eqref{eq:bt} can then be used
to raise the power of the denominator function ${\cal V}(\vec{x},\vec{y})$.
Additional variable transformations can be used to ensure that one encounters at
most logarithmic behavior of the integrand near singularities in the interior
of the integration region \cite{Passarino:2001jd,fppu2}. Furthermore, oftentimes some of the
Feynman parameter integrations can be carried out analytically, thus reducing
the dimensionality of the numerical integral \cite{fppu2}.

Integrals with non-trivial tensor structures in the numerator can be handled
with essentially the same approach, since the contribution of the numerator
terms can be absorbed into the function $\cal Q$ in eq.~\eqref{eq:feynpb}. See
Ref.~\cite{afppu} for more details.

UV and IR divergences occurring in one subloop of a two-loop diagram can be
extracted by performing the Feynman parameter integrations associated with this
subloop analytically and then expanding the result in powers of $\epsilon$
\cite{Passarino:2006gv}. Sector decomposition can be used to simplify this procedure in more
complicated cases.
This approach leads to compact results that can be efficiently integrated
numerically, but it requires a separate derivation for each different loop
topology. Alternatively, the singularities can also be removed with
suitable subtraction terms before the Bernstein-Tkachov relation is
applied \cite{swucc1}. The subtraction methods are described in section~\ref{sec:sub}
in more detail.

\vspace{\medskipamount}
One difficulty of the original Bernstein-Tkachov formula \eqref{eq:bt} is the
appearance of the factor $B$ in the denominator. It may vanish for certain
configurations, some of which correspond to physical singularities. However, one
can also find $B \sim 0$ even in cases where the loop integral is regular. In
this situation, different terms in the numerator of \eqref{eq:bt} cancel each
other, leading to potential numerical instabilities. For a subloop within a
two-loop diagram, $B$ depends on the Feynman parameters $\vec{y}$
of the outer loop, and thus one can encounter $B \sim 0$ for particular values of
$\vec{y}$ inside the integration region, again resulting in numerical
instabilities.

Therefore, it is necessary to have a special treatment for regular integrals
with $B \sim 0$. For instance, a Taylor expansion about $B=0$ leads to a
well-defined and numerically stable expression \cite{fppu1,fppu2}. In
Ref.~\cite{uccirati}, a modified Bernstein-Tkachov-like relation was proposed
that avoids the appearance of the $1/B$ factors altogether. Writing
\begin{align}
{\cal V}(\vec{x}) &= Q(\vec{x})+B, & 
Q(\vec{x}) &= (\vec{x}+\vec{A})^\top H \; (\vec{x}+\vec{A}),
\end{align}
it can be shown that the following relation holds \cite{uccirati}:
\begin{align}
{\cal V}(\vec{x})^{-\nu-\epsilon} &= 
\bigl [\beta + (\vec{x}+\vec{A})^\top \vec{\partial}_x\bigr ]
\int_0^1 dy \, y^{\beta-1} [Q(\vec{x})y+B]^{-\nu-\epsilon},
\label{eq:ubt}
\end{align}
where $\beta>0$ is an arbitrary constant. The integral over $y$ can be evaluated
analytically in terms of the hypergeometric function ${}_2F_1$.  The
$\vec{\partial}_x$ can be eliminated by performed integration by parts in the
$\vec{x}$-integral. Expanding the
latter around $\epsilon=0$ then leads to an expression with a lower power of
$Q(\vec{x})+B$ than on the left-hand side of eq.~\eqref{eq:ubt}. This procedure
can be applied iteratively with suitable values of $\beta$ until sufficiently
smooth integrals are obtained.

\vspace{\medskipamount}
The advantage of the Bernstein-Tkachov method and related techniques is the fact
that it leads to relatively simple and smooth numerical integrals that converge
quickly and reliably. For most two-loop configurations with up to three external
legs, one can derive numerical integrals with just two or three dimensions.
UV and IR divergent configurations can also be handled. However, each different
diagram topology and IR singularity configuration requires a different
derivation of the final integral representation, so that this step cannot be
automatized easily.

Applications of the Bernstein-Tkachov method include two-loop electroweak corrections to
$Z\to f\bar{f}$ formfactors \cite{swucc1,hmu2,swbb} and next-to-leading order
electroweak corrections to the Higgs-boson couplings to photons and gluons
\cite{hgg}.


\subsection{Differential equations}
\label{sec:de}

Differential equations \cite{deq} are a well-known tool for the analytical
evaluation of loop integrals. However, they may also be integrated numerically.
For concreteness, let us begin by illustrating this approach for the
example for two-loop self-energy integrals, based on the work in
Refs.~\cite{de1,de2,martin1}.

\begin{figure}
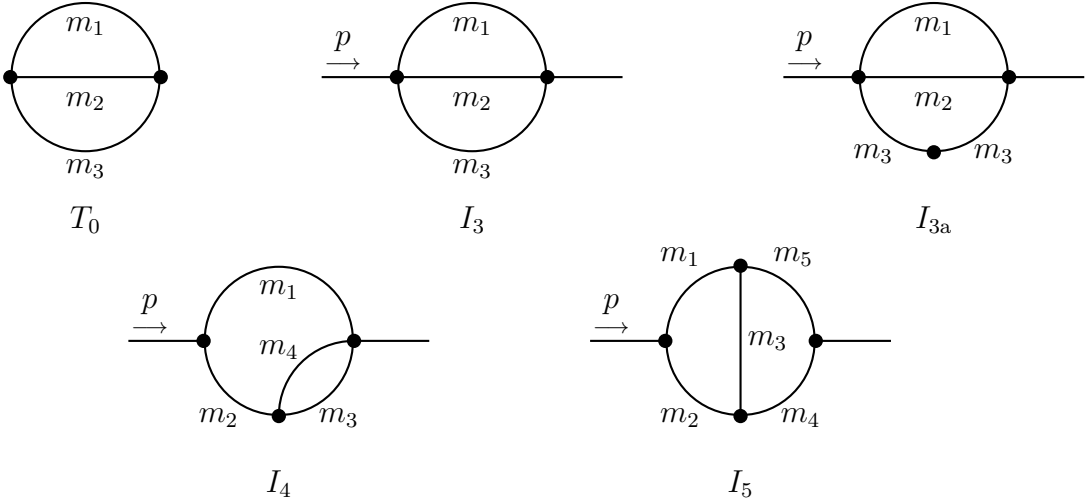

\vspace{1ex}
\begin{center}
\rule{0mm}{0mm}%
\psline(-1,-1)(1,-1)
\pscircle(0,-1){1}
\psdot[dotscale=1.5](-1,-1)
\psdot[dotscale=1.5](1,-1)
\rput[t](0,-0.2){$m_1$}
\rput[t](0,-2.1){$m_3$}
\rput[t](0,-1.2){$m_2$}
\rput(0,-2.9){$T_0$}
\hspace{5cm}
\psline(-2,-1)(2,-1)
\pscircle(0,-1){1}
\psdot[dotscale=1.5](-1,-1)
\psdot[dotscale=1.5](1,-1)
\rput[t](0,-0.2){$m_1$}
\rput[t](0,-2.1){$m_3$}
\rput[t](0,-1.2){$m_2$}
\rput[lb](-2,-0.9){$\displaystyle p\atop\longrightarrow$}
\rput(0,-2.9){$I_3$}
\hspace{6cm}
\psline(-2,-1)(2,-1)
\pscircle(0,-1){1}
\psdot[dotscale=1.5](-1,-1)
\psdot[dotscale=1.5](1,-1)
\psdot[dotscale=1.5](0,-2)
\rput[t](0,-0.2){$m_1$}
\rput[t](-0.8,-1.9){$m_3$}
\rput[t](0.8,-1.9){$m_3$}
\rput[t](0,-1.2){$m_2$}
\rput[lb](-2,-0.9){$\displaystyle p\atop\longrightarrow$}
\rput(0,-2.9){$I_{3\rm a}$}
\\[3cm]
\rule{0mm}{0mm}%
\psline(-2,-1)(-1,-1)
\psline(1,-1)(2,-1)
\pscircle(0,-1){1}
\psarc(1,-2){1}{90}{180}
\psdot[dotscale=1.5](-1,-1)
\psdot[dotscale=1.5](1,-1)
\psdot[dotscale=1.5](0,-2)
\rput[t](0,-0.2){$m_1$}
\rput[t](-0.8,-1.9){$m_2$}
\rput[t](0.8,-1.9){$m_3$}
\rput[t](0,-1){$m_4$}
\rput[lb](-2,-0.9){$\displaystyle p\atop\longrightarrow$}
\rput(0,-2.9){$I_4$}
\hspace{6cm}
\psline(-2,-1)(-1,-1)
\psline(1,-1)(2,-1)
\psline(0,0)(0,-2)
\pscircle(0,-1){1}
\psdot[dotscale=1.5](-1,-1)
\psdot[dotscale=1.5](1,-1)
\psdot[dotscale=1.5](0,0)
\psdot[dotscale=1.5](0,-2)
\rput[b](-0.8,0){$m_1$}
\rput[t](-0.8,-1.9){$m_2$}
\rput[t](0.8,-1.9){$m_4$}
\rput[b](0.7,0){$m_5$}
\rput[l](0.1,-1){$m_3$}
\rput[lb](-2,-0.9){$\displaystyle p\atop\longrightarrow$}
\rput(0,-2.9){$I_5$}
\end{center}
\vspace{2.5cm}
\caption{A convenient basis of two-loop self-energy master integrals. All propagators
are assumed to be scalars.
\label{fig:2lbas}} 
\end{figure}

Using integration by parts or other reduction methods, an arbitrary two-loop
self-energy integral can be written as a linear combination of five two-loop
master integrals shown in Fig.~\ref{fig:2lbas} and terms involving one-loop integrals.
The vacuum integral $T_0$ is known analytically in terms of dilogarithms
\cite{2lvac,Ford:1992pn,Scharf:1993ds}. The remaining integrals
$I_k(p^2;\{m_n^2\})$ depend on
the external invariant momentum, $p^2$. The derivative of $I_k$ with
respect to $p^2$ can be computed at the integrand level by using
\begin{align}
\frac{\partial}{\partial(p^2)}I_k &= \frac{1}{2p^2}\,p^\mu
\frac{\partial}{\partial(p^\mu)}I_k. \label{eq:deriv}
\end{align}
From the expression on the right-hand side of \eqref{eq:deriv} one obtains self-energy integrals with additional propagators
and/or numerator terms. These again can be reduced to a linear combination of
the $I_k$, $T_0$ and one-loop integrals. Thus one arrives at a differential
equation of the form
\begin{align}
\frac{\partial}{\partial(p^2)}I_k &= 
 \sum_l f_{kl} \, I_l + g_k.  \label{eq:deqp}
\end{align}
The coefficients $f_{kl}$ are rational functions of the masses $m_n$, the
momentum invariant $p^2$ and the dimension $D$. The inhomogeneous part involves
the two-loop vacuum function $T_0$ as well as one-loop functions, all of which
are known analytically. 

To eliminate the dependence of the space-time dimension $D=4-2\epsilon$, all terms in
\eqref{eq:deqp} can be expanded in powers of $\epsilon$. In the absence of IR
divergences, the master integrals $I_k$ may contain UV $1/\epsilon$ and
$1/\epsilon^2$ poles, so that one can write
\begin{align}
I_k &= I_k^{[-2]}\,\epsilon^{-2} + I_k^{[-1]}\,\epsilon^{-1} +
 I_k^{[0]} + {\cal O}(\epsilon), \nonumber \\
f_{kl} &= f^{[0]}_{kl} + f^{[1]}_{kl}\,\epsilon + f^{[2]}_{kl}\,\epsilon^2
 + {\cal O}(\epsilon^3), \label{eq:fexp} \\
g_k &= g_k^{[-2]}\,\epsilon^{-2} + g_k^{[-1]}\,\epsilon^{-1} +
 g_k^{[0]} + {\cal O}(\epsilon). \nonumber 
\end{align}
For general situations, the coefficient functions $f_{kl}$ may have singular
terms in $\epsilon$, so that the integrals $I_k$ need to be expanded to
higher powers $\epsilon$ beyond $\epsilon^0$. However,
for the class of two-loop self-energy integrals this is not needed.
Inserting \eqref{eq:fexp} into \eqref{eq:deqp} one obtains
\begin{align}
\frac{\partial}{\partial(p^2)}I_k^{[-2]} &= \sum_l f_{kl}^{[0]} \, I_l^{[-2]} + 
 g_k^{[-2]}, \label{eq:deqe2} \\
\frac{\partial}{\partial(p^2)}I_k^{[-1]} &= \sum_l \bigl (
 f_{kl}^{[0]} \, I_l^{[-1]} + f_{kl}^{[1]} \, I_l^{[-2]} \bigr )
 g_k^{[-1]}, \label{eq:deqe1} \\
\frac{\partial}{\partial(p^2)}I_k^{[0]} &= \sum_l \bigl (
 f_{kl}^{[0]} \, I_l^{[0]} + f_{kl}^{[1]} \, I_l^{[-1]} +
 f_{kl}^{[2]} \, I_l^{[-2]}\bigr )
 g_k^{[0]}. \label{eq:deqe0}
\end{align}
The differential equations 
\eqref{eq:deqe2} and \eqref{eq:deqe1} are simple enough such that they can
be solved analytically \cite{de1,de2}. On the other hand,
the system \eqref{eq:deqe0} of first-order linear differential
equations can be integrated numerically from an initial value $p_0^2$, 
for example using the Runge-Kutta
algorithm. A simple choice for the boundary value is $p_0^2=0$, where all
integrals $I_k(p_0^2)$ reduce to vacuum integrals and thus can be evaluated
analytically.

Similar to other methods discussed in the previous subsections, difficulties can
arise from singular points of the integrand associated with thresholds. While
these singularities are formally integrable by virtue of the Feynman
$i\varepsilon$ prescription, they can lead to numerical instabilities and
convergence problems. This problem can be avoided by using a complex integration
contour. A practical complex contour was suggested in Ref.~\cite{de2}: $0 \to
i\delta \to p^2 + i\delta \to p^2$, $i.\,e.$ the integration initially moves
along the imaginary axis to a fixed value $\delta>0$, then parallel to the real
axis, and finally back to the real axis.

Special cases occur if the integration endpoint $p^2$ is itself near a
threshold. In this case, one could use this threshold as the initial value
$p_0^2$ for the numerical integration, which requires the analytical evaluation
of the integrals at the threshold value to obtain the boundary value
\cite{de2,deexp}. In this context, it is worth mentioning that the differential
equations themselves  can be used as a tool to derive expansions about various
singular points, which then supply the necessary information for the boundary
condition \cite{de1,deexp}.
Alternatively, one can use a variable transformation to
improve the singular behavior of the integrand near the threshold point \cite{tsil}.

The techniques described above for the evaluation of two-loop self-energy master
integrals have been implemented in the public programs TSIL \cite{tsil} and {\sc
BoKaSun} \cite{Caffo:2008aw}.

More generally, a differential equation system can be built based on derivatives
with respect to momentum invariants, as in eq.~\eqref{eq:deqp} above, or with
respect to masses. In Ref.~\cite{ttbar1} this idea was extended to construct
differential equation systems that depend on two variables simultaneously. This
was done for the purpose of computing the master integrals required for the
evaluation of $t\bar{t}$ production at hadron colliders. The amplitudes for this
process can be expressed in terms of two independent variables,  $x \equiv -t/s$
and $y \equiv \mt^2/s$, where $s$ and $t$ are the usual Mandelstam variables
\cite{mandelstam}. The overall dimensionful scale $s$ can be factored out of the
problem, leaving only the dimensionless variables $x$ and $y$. Thus the
differential equation system takes the form
\begin{align}
\frac{\partial}{\partial x}I_k &= 
 \sum_l f^x_{kl} \, I_l + g^x_k, \\
\frac{\partial}{\partial y}I_k &= 
 \sum_l f^y_{kl} \, I_l + g^y_k,
\end{align}
where the master integrals $I_k$ and the coefficients $f^x_{kl}$, $f^y_{kl}$,
$g^x_k$, $g^y_k$ depend on $x$, $y$ and $D$.

For the boundary condition, it is convenient to choose a point in the high-energy
regime ($i.\,e.$ with a small value of $y$) \cite{ttbar1}. 
The boundary value
can be obtained from a small-mass expansion. By choosing the high-energy
boundary condition, no physical threshold is crossed when integrating the
differential equation system from the boundary to a physical kinematical point.
Nevertheless, there are spurious singularities from points that are regular but
involve large numerical cancellations, which should be avoided by means of a
complex contour deformation. The solution of the system is then obtained by
choosing a path in the complex $x$--$y$ plane, which corresponds to a
four-dimensional real space \cite{ttbar1}.

\vspace{\medskipamount}
Differential equations are a useful framework to determine numerical solutions
to multi-loop integrals with many beneficial properties: \emph{(i)} UV and IR singularities
can be systematically dealt with through an expansion in powers of $\epsilon$;
\emph{(ii)} the numerical integrals are of low dimensionality, which can be evaluated
efficiently and with high precision; and \emph{(iii)} in principle there is no
limit to the complexity of the integrals that can be handled with this method.
However, for each new class of loop integrals, several steps have to be worked
out to make the differential equation approach viable: \emph{(i)} a basis of
master integrals needs to be identified and the full 
amplitude needs to be algebraically reduced to this basis; \emph{(ii)} a suitable boundary
condition for the differential equation is required; and \emph{(iii)} the boundary terms
must be evaluated analytically or numerically using a different method
with very high precision. Computer algebra programs can assist in the execution
of these steps, but the entire procedure is difficult to be fully automated and
usually requires
substantial manual work\footnote{For recent work towards more complete
automatization, see $e.\,g.$ Ref.~\cite{Ablinger:2015tua}.}. 
Moreover, the reduction to master integrals may become
impractical for problems with many mass and momentum scales.

The numerical differential equation method has been used in a variety of
phenomenological applications, including the calculation of two-loop QCD
corrections to the production of $t\bar{t}$ pairs at hadron colliders
\cite{ttbar2}, of two-loop corrections to Higgs-boson masses in supersymmetric
theories \cite{susyhmass}, and of two-loop corrections to rare $B$-meson
decays \cite{bsg}.


\subsection{Comparison of numerical methods}

In section \ref{sec:numi}, three main challenges for numerical integration
methods were named: \emph{(i)} providing an algorithm for extracting UV and IR
singularities; \emph{(ii)} ensuring numerical stability and robust convergence;
and \emph{(iii)} being applicable to a large class of processes with different
numbers of loops and external legs and different configurations of massive
propagators. Tab.~\ref{tab:comp} provides a qualitative evaluation of the
strengths and weaknesses of the techniques discussed in this chapter according
to these criteria. It should be emphasized that this assessment is based on the
author's subjective opinion and does not claim to fully consider all pertinent
aspects.

\begin{table}[p]
\centering
\renewcommand{\arraystretch}{1.2}
\begin{tabular}{>{\RaggedRight\arraybackslash}p{4cm}>{\RaggedRight\arraybackslash}p{4cm}>{\RaggedRight\arraybackslash}p{4.1cm}>{\RaggedRight\arraybackslash}p{4.4cm}}
\hline
 & Treatment of\newline singularities & Stability and\newline convergence & Generality \\
\hline\hline
Feynman parameter integration of massive two-loop integrals\newline (section
\ref{sec:ghin}) & 
Provides general procedure for UV singularities, but IR singularities
require mass regulator & 
Good convergence and stability for massive two-loop
amplitudes with up to four external legs &
Applicable up to two-loop level; complex contour deformation requires
case-by-case adaptation \\
\hline
Sector decomposition\newline (section \ref{sec:secdec}) &
General algorithm for arbitrary UV and IR singularities &
Generates large expression which may slow down numerical evaluation; 
numerical stability deteriorates in presence of thresholds and pinch
singularities &
Applicable for any number of loops and legs; but convergence suffers for large
mass hierarchies \\
\hline
Mellin-Barnes representations\newline (section \ref{sec:mb}) &
General algorithm for arbitrary UV and IR singularities &
Improvement through contour
      deformation and variable transformations in semi-automatic
      way; need manual adaption to new diagram classes &
Applicable for any number of loops and legs; but convergence suffers for large
mass hierarchies \\
\hline
Subtraction terms\newline (section \ref{sec:sub}) &
Complete at one-loop; partial solutions at two-loop; only results for QED beyond
two-loop &
Numerical stability deteriorates in presence of thresholds and pinch
singularities &
Applicable for general one-loop and subset of general two-loop cases,
or for multi-loop QED amplitudes \\
\hline
Dispersion relations\newline (section \ref{sec:disp}) &
Removal of UV and IR singularities requires case-by-case treatment &
Excellent stability and convergence for simple two-loop topologies &
Applicable for two-loop cases with few legs; no general method for tensor
integrals; restrictions on the
occurrence of thresholds \\
\hline
Bernstein-Tkachov method\newline (section \ref{sec:bt}) &
Removal of UV and IR singularities for arbitrary one- and two-loop amplitudes,
but requires case-by-case treatment &
Very good numerical stability and convergence &
Applicable for general one-loop and two-loop amplitudes; algorithmic
automatization difficult \\
\hline
Differential equations\newline (section \ref{sec:de}) &
Systematic procedure for evaluation of UV and IR singularities, details depend
on integral basis and boundary terms &
Very good numerical convergence and precision except for special threshold cases
&
No fundamental limit on number of loops or legs, but requires choice of
(process-dependent) integral basis and boundary terms
\end{tabular}
\caption{Qualitative comparison of different numerical integration techniques.
\label{tab:comp}}
\end{table}


\section{Application to electroweak precision observables}
\label{sec:app}

One notable application of the numerical methods presented in the previous
chapter are electroweak two-loop corrections to electroweak precision
observables (EWPOs). These contributions typically depend on many independent mass and
momentum scales ($\MZ$, $\MW$, $\MH$, $\mt$, ...), which are a challenge for
analytical techniques.

In the following subsections, the phenomenology of electroweak precision
observables will be discussed in more detail. Particular emphasis is paid to
contributions beyond the one-loop order, including issues related to
renormalization, resummation of leading contributions, and the evaluation of
theory uncertainties.


\subsection{Precision observables}

Some of the observables most sensitive to quantum effects of physics beyond the
Standard Model are:
\begin{itemize}

\item
The $W$-boson mass as predicted from the muon decay rate. At low energies, muon
decay can be described through an effective four-fermion interaction with the
coupling strength given by the Fermi constant, $G_{\rm F} = 1.1663787(6) \times
10^{-5}\gev^{-2}$ \cite{pdg},
\begin{align}
\Gamma_\mu &= \frac{G_{\rm F}^2m_\mu^5}{192\pi^3} F\Bigl(\frac{m_{\rm
e}^2}{m_\mu^2}\Bigr) (1+\Delta q), \\
F(\rho) &= 1-8\rho+8\rho^3-\rho^4-12\rho^2\ln \rho,
\label{eq:gf}
\end{align}
where $\Delta q$ captures QED radiative
corrections, which will be covered in more detail in section~\ref{sec:qed}.
Within the Standard Model, $G_{\rm F}$ can be expressed as
\begin{align}
G_{\rm F} &= \frac{\pi\alpha}{\sqrt{2}\sw^2\MW^2}(1+\Delta r),
\label{eq:mw}
\end{align}
where $\Delta r$ summarizes the contribution from loop effects. Here $\sw^2 \equiv
\sin^2\theta_{\rm W} = 1-\MW^2/\MZ^2$ is the sine squared of the Weinberg angle, as defined
through the $W$- and $Z$-boson masses\footnote{In general, the Weinberg angle
may be defined through the weak boson masses or their couplings, both of which
are equivalent at tree-level, but differ at higher orders, see $e.\,g.$
section 10 in Ref.~\cite{pdg}.}. Eq.~\eqref{eq:mw} can be solved for $M_{\rm W}$ iteratively,
since $\Delta r$ also depends on $M_{\rm W}$.

\item
Observables related to the cross-section of $e^+e^- \to f\bar{f}$ near the
$Z$ pole, $i.\,e.$ $\sqrt{s}\approx \MZ$. These include \emph{(i)} the cross-sections for
different final states $f\bar{f}$ on the
$Z$ peak, $\sigma^0_f \equiv \sigma_f(s=\MZ^2)$, \emph{(ii)}  the total width of the
$Z$ boson, $\GZ$, extracted from measuring the shape of $\sigma_f(s)$ at several
values of $s$, \emph{(iii)} and branching ratios of different final states,
$\sigma_f/\sigma_{f'}$, see Fig.~\ref{fig:linez}.
\begin{figure}
\centering
\psfig{figure=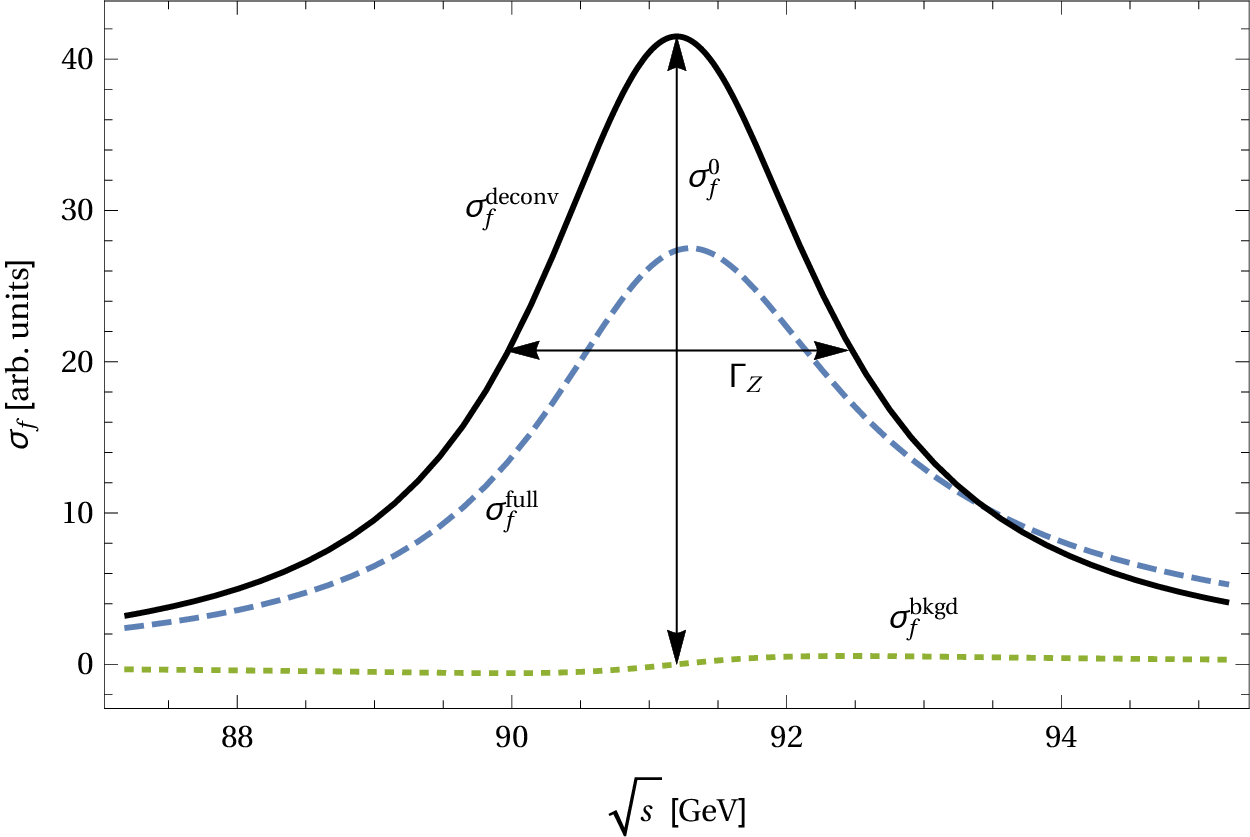, width=12cm}
\vspace{-1ex}
\caption{Illustration of $Z$-pole line-shape observables (not to scale). The dashed line
depicts the true line-shape, while the solid line indicates the line-shape after
deconvolution of initial-state QED radiation. The dotted line shows the
contribution of backgrounds from photon exchange and box contributions to the
deconvoluted cross-section.
\label{fig:linez}} 
\end{figure}

It is customary to select the following independent set of quantities to
describe this class of observables:
\begin{align}
\sigma^0_{\rm had} &= \sigma[e^+e^- \to \text{hadrons}]_{s=\MZ^2},\label{eq:sig0} \\[1ex]
\Gamma_Z &= \sum_f \Gamma[Z \to f\bar{f}], \\
R_\ell &= \Gamma[Z \to \text{hadrons}]/\Gamma[Z \to \ell^+\ell^-], &&
 (\ell = e,\mu,\tau)\\[1ex]
R_q &= \Gamma[Z \to q\bar{q}]/\Gamma[Z \to \text{hadrons}],
 && (q = b,c,s,d,u). \label{eq:rq}
\end{align}
Here it is implicitly assumed that an $f\bar{f}$ final state always includes
contributions from additional real photon and gluon radiation. In this sense,
$\sigma[e^+e^- \to \text{hadrons}] = \sum_q \sigma_q$.

Note that $\sigma_f(s)$ is not the observable cross-section. The latter, denoted
$\sigma^{\rm full}_f(s)$ in the following, receives additional contributions
from s-channel photon exchange, box contributions and 
initial-state QED radiation. These have an impact both on the magnitude and the
shape of the cross-section as a function of $s$, see Fig.~\ref{fig:linez} for
illustration. To a very good approximation, initial-state QED radiation can be
described through a convolution of $\sigma_f$ with a radiator function $H(x)$
\cite{ylep1},
\begin{align}
\sigma^{\rm full}_f(s) &= \int_0^{1-4m_f^2/s} dx \; H(x) \, \sigma^{\rm
deconv}_f (s'), \qquad
s' = s(1-x). \label{eq:qedconv}
\end{align}
Including resummed soft-photon and exact ${\cal O}(\alpha^2)$ contributions,
$H(x)$ reads
\begin{align}
H(x) &= \beta x^{\beta-1}(1+\delta_1^{\rm V+S}+\delta_2^{\rm V+S})
 + \delta_1^{\rm H} + \delta_2^{\rm H}, \label{eq:hfunc} \\[1ex]
\beta &= \frac{2\alpha}{\pi}(L-1), \qquad L = \log \frac{s}{m_{\rm e}^2}, \\
\delta_1^{\rm V+S} &= \frac{\alpha}{\pi}\biggl [
 \frac{3}{2}L + \frac{\pi^2}{3}-2\biggr ], \\
\delta_2^{\rm V+S} &= \Bigl (\frac{\alpha}{\pi}\Bigr )^2 \biggl [
 \biggl (\frac{9}{8}-\frac{\pi^2}{3}\biggr )L^2 + s_{21}L + s_{20}\biggr ], \\
\delta_1^{\rm H} &= \frac{\alpha}{\pi}(L-1)(x-2), \\
\delta_2^{\rm H} &= \Bigl (\frac{\alpha}{\pi}\Bigr )^2 \bigl [
h_{22}L^2 + h_{21}L + h_{20} \bigr ].
\end{align}
The coefficients $s_{21}$, $s_{20}$, $h_{22}$, $h_{21}$ and $h_{20}$ can be
found in Ref.~\cite{Berends:1987ab}.

The deconvoluted cross-section, $\sigma^{\rm deconv}_f(s')$ contains
contributions from s-channel $Z$-boson exchange, s-channel photon exchange,
photon-$Z$ interference, and
box diagrams. To extract the first part, the remaining pieces are subtracted:
\begin{equation}
\sigma_f(s') = \sigma^{\rm deconv}_f(s') - \sigma^\gamma_f(s')
 - \sigma^{\gamma \rm Z}_f(s')- \sigma^{\rm box}_f(s'). \label{eq:sigz}
\end{equation}

\item
Parity-violating asymmetries measured at the $Z$ pole. The forward-backward
asymmetry is defined as
\begin{align}
A^f_{\rm FB} &= \frac{\sigma_f(\theta<\frac{\pi}{2})-
 \sigma_f(\theta>\frac{\pi}{2})}{\sigma_f(\theta<\frac{\pi}{2})+
 \sigma_f(\theta>\frac{\pi}{2})}, \label{eq:afb}
\end{align}
where $\theta$ is the scattering angle between the incoming $e^-$ and the
outgoing $f$. It can be written as a product of two terms,
\begin{align}
A^f_{\rm FB} &= \tfrac{3}{4}{\cal A}_e{\cal A}_f, \\
{\cal A}_f &= \frac{1-4|Q_f|\seff{f}}{1-4|Q_f|\seff{f}+8 (Q_f\seff{f})^2},
\end{align}
where $\seff{f}$ is called the effective weak mixing angle. In the presence of
polarized electron beams, one can also measure the left-right asymmetry
\begin{align}
A^f_{\rm LR} &= \frac{\sigma_f(P_e<0)-
 \sigma_f(P_e>0)}{\sigma_f(P_e<0)+
 \sigma_f(P_e>0)}={\cal A}_e|P_e|.
\end{align}
Here $P_e$ is the polarization degree of the incident electrons, with $P_e<0$ ($P_e>0$)
referring to left-handed (right-handed) polarization. Finally, angular and
polarization information can be combined to construct
\begin{align}
A^f_{\rm LR,FB} &= \frac{\sigma_{f,\rm LF}-
 \sigma_{f,\rm LB}-\sigma_{f,\rm RF}+\sigma_{f,\rm RB}}{\sigma_{f,\rm LF}+
 \sigma_{f,\rm LB}+\sigma_{f,\rm RF}+\sigma_{f,\rm RB}}=\tfrac{3}{4}{\cal A}_f,
\end{align}
where $\sigma_{f,\rm LF}=\sigma_f(P_e<0,\theta<\frac{\pi}{2})$, etc.

\end{itemize}
The quantities introduced above can be computed within the Standard Model. At
tree-level they read, in the limit $m_f \ll \MZ$,
\begin{align}
\sigma^0_{\rm had} &= \sum_q \frac{12\pi}{\MZ^2} \,
\frac{\Gamma_e\Gamma_q}{\GZ^2}, & \GZ &= \sum_f \Gamma_f, &
\Gamma_f &= \frac{N_c^f\alpha\MZ}{24\sw^2\cw^2}(1-4\sw^2|Q_f|+8\sw^4|Q_f|^2), \\
\Delta r &= 0, & \seff{f} &= \sw^2,
\end{align}
where $N_c^f = 3(1)$ for quarks (leptons) and $\sw^2=\sin^2\theta_{\rm W}$,
$\cw^2=\cos^2\theta_{\rm W}$. However, they receive sizable corrections from
loop contributions that must be included to match the experimental precision.
The loop contributions depend on other Standard Model parameters, most notably
$\MW$, $\MH$, $\mt$ and $\as$, so that fits to the electroweak precision data can be used
to derive indirect bounds on these parameters.

For a proper theoretical definition of the EWPOs beyond tree-level, one needs to
ensure that gauge invariance is preserved at every loop order. For the $Z$-pole
observables, this can be achieved by expanding the matrix element for $e^+e^- \to
f\bar{f}$ about the complex pole, $s_0\equiv \mz^2-i\mz\gz$ of the
$Z$ propagator,
\begin{align}
{\cal M}[e^+e^- \to f\bar{f}] &= \frac{R}{s-s_0}+S+(s-s_0)S' + ...
\label{eq:poleexp}
\end{align}
Here $\mz$ and $\gz$ are the on-shell mass and width of the $Z$ boson. The
difference to the unbarred quantities ($\MZ$ and $\GZ$) will be explained below.
Since the pole is an analytical property of the $S$ matrix, it is
gauge-invariant to all orders and can be identified with an observable \cite{pole1}. It can
also be shown explicitly, with the help of Ward or Nielsen identities, that $s_0$ as
well as the coefficients $R,S,S',\dots$ are gauge-parameter independent
\cite{hveltman,nielsen}.

When ignoring the non-resonant contributions in \eqref{eq:poleexp}, 
it follows that the $s$-dependence of the
cross-section near the $Z$ pole is given by the Breit-Wigner function
\begin{align}
\sigma_f &\propto \frac{1}{(s-\mz^2)^2+\mz^2\gz^2}. \label{eq:bwth}
\end{align}
However, in experimental analyses, a different form of the Breit-Wigner
line-shape is being used,
\begin{align}
\sigma_f &\propto \frac{1}{(s-\MZ^2)^2+s^2\GZ^2/\MZ^2}. \label{eq:bwexp}
\end{align}
These two
Breit-Wigner forms can be translated into each other with the transformation
\cite{Bardin:1988xt}
\begin{align}
\mz &= \MZ\big/ \sqrt{1+\GZ^2/\MZ^2}, &
\gz &= \GZ\big/ \sqrt{1+\GZ^2/\MZ^2}. \label{eq:masstrans}
\end{align}
Similar relations hold for the $W$ mass. Numerically, they amount to
\begin{align}
\mz &\approx \MZ - 34\mev, &
\gz &\approx \GZ -0.9\mev, \\
\mw &\approx \MW -27\mev, &
\gw &\approx \GW -0.7\mev. &
\end{align}
As mentioned above, the non-resonant terms were disregarded in deriving these
relations. 

\vspace{\medskipamount} Quantities like $G_{\rm F}$, $\sigma^0_{\rm had}$,
$\GZ$, $R_\ell$, $R_{c,b}$, $A^f_{\rm FB}$, $A^f_{\rm LR}$, and $A^f_{\rm
LR,FB}$ are frequently used in precision tests of the Standard Model and to
derive constraints on new physics. As their definition requires the removal of
some photonic corrections, they are called ``pseudo-observables,'' in contrast
to the ``true observables'' (such as $\Gamma_\mu$, $\sigma^{\rm full}_f$, etc.).
For $G_{\rm F}$, the subtraction of QED corrections is straightforward since
these effects can be summarized in the single term $\Delta q$, see
eq.~\eqref{eq:gf}. On the other hand, for the $Z$-pole observables, the
deconvolution of soft and collinear initial-state radiation, see
eq.~\eqref{eq:qedconv}, and the subtraction of s-channel photon exchange and box
contributions, see eq.~\eqref{eq:sigz}, is more involved.

There are several public computer codes that carry out this procedure to extract
the pseudo-observables from the true observables. {\sc ZFitter}
\cite{zfitterold,zfitter}
and TOPAZ0 \cite{topaz0} use analytical expressions for the radiator function
$H(x)$, as in eqs.~\eqref{eq:hfunc}~ff. They also permit the implementation of
experimental cuts on the invariant mass, angular acceptance or acollinearity of
the outgoing fermions. 

A potential problem is the fact that these programs do not use a consistent
treatment of the complex pole expansion in \eqref{eq:poleexp} when performing
the subtraction of photon-exchange and box contributions. Nevertheless, it was
analyzed in Ref.~\cite{swlong} that the impact of this mismatch is
negligible compared to the current experimental uncertainties. It may, however,
become relevant for the increased precision of a future high-luminosity $e^+e^-$
machine (see section \ref{sec:future}). One should keep in mind that this
inconsistency also affects the validity of the translation between the two
Breit-Wigner forms in \eqref{eq:masstrans}. Thus it may be necessary to abandon
the form \eqref{eq:bwexp} in the future, since only the alternative form
\eqref{eq:bwth} has been shown to follow from
a self-consistent definition of resonant and non-resonant terms that can be
systematically extended to higher precision.

Several other computer codes use Monte-Carlo techniques for the
evaluation of external QED radiation. Recent examples are {\sc KoralZ}
\cite{koralz} and KK-MC \cite{kkmc}.
To improve the precision beyond full ${\cal O}(\alpha)$ and resummed 
leading-logarithmic order, they use exclusive exponentiation
\cite{ceex} based on the Yennie-Frautschi-Suura theory \cite{yfs} instead of
traditional parton showering. See also Ref.~\cite{babayaga} for similar
considerations of QED radiation in the context of Bhabha scattering.


\subsection{Renormalization and input parameters}

The loop corrections to the EWPOs contain UV divergences, which must be removed
with the help of suitable counterterms. Within the Standard Model, one may
choose the masses of all elementary particles, the wave function normalization
for asymptotic external states, and the electromagnetic coupling constant as
independent quantities, for which counterterms are introduced:
\begin{align}
&\text{Mass renormalization:} & 
&m_{f,0} = m_f + \delta m_f, &
&M^2_{\rm X,0} = M^2_{\rm X} + \delta M^2_{\rm X} \quad (X = W,Z,H), \nonumber \\
&\text{Wave function renormalization:} & 
&f^{\rm L}_0 = \sqrt{1+\delta Z^{f\rm L}}\, f^{\rm L}, & 
&f^{\rm R}_0 = \sqrt{1+\delta Z^{f\rm R}}\, f^{\rm R}, \\
&\text{Coupling renormalization:} & &e_0 = (1+\delta Z_e) e.\nonumber 
\end{align}
Here the symbols with (without) subscript 0 denote the bare (renormalized)
quantities, and superscripts L,R refer to left- and right-handed fermion fields,
respectively. Note that promptly decaying particles, such as the $W$ and $Z$
bosons, cannot be asymptotic external states, and thus no wave function
renormalization\footnote{In the literature, the term ``field renormalization''
is sometimes used instead of ``wave function renormalization.'' Both refer to
the same physical principle of properly normalizing the external legs of an
amplitude.} for these can be defined consistently to all orders in perturbation
theory.

One of most commonly used renormalization scheme for the calculation of EWPOs is
the ``on-shell scheme''. Within this scheme, the electromagnetic coupling constant
is defined through the $\gamma f \bar{f}$ vertex in the Thomson limit, where the
photon and fermions are on their mass shell. Furthermore, the masses of all
elementary Standard Model particles are defined through the pole of their
propagator. Note that the masses defined in this way correspond to the barred
quantities, $\overline{M}_{\rm X}$, introduced in the previous subsection.

Using these on-shell renormalization conditions, all counterterms can be
determined in terms of the one-particle irreducible self-energies
\begin{align}
&\Sigma_{\mu\nu}^{V_1V_2}(q) = \Bigl ( g_{\mu\nu}-\frac{q_\mu q_\nu}{q^2} \Bigr )
\Sigma_{\rm T}^{V_1V_2}(q^2) + \frac{q_\mu q_\nu}{q^2} \Sigma_{\rm L}^{V_1V_2}(q^2), 
\qquad (V_i=\rm \gamma,Z,W) \label{eq:sigmavv} \\
&\Sigma^{\rm H}(q^2),  \\
&\Sigma^f(q) = \qslash P_{\rm L} \Sigma^f_{\rm L(1)}(q^2)
 + \qslash P_{\rm R} \Sigma^f_{\rm R(1)}(q^2) + m_f \Sigma^f_{\rm S(1)}(q^2).
\end{align}
At one-loop order, the relevant on-shell counterterms read
\begin{align}
\delta \mw^2{}_{(1)} &= \text{Re}\bigl\{ \Sigma_{\rm T(1)}^{\rm WW}(\mw^2)\bigr\},
&
\delta Z_{e(1)} &= \frac{1}{2}\Sigma^{\gamma\gamma\,\prime}_{\rm T(1)}(0)
 - \frac{\sw}{\cw}\,\frac{\Sigma^{\gamma\rm Z}_{\rm T(1)}(0)}{\mz^2},
\\
\delta \mz^2{}_{(1)} &= \text{Re}\bigl\{ \Sigma_{\rm T(1)}^{\rm ZZ}(\mz^2)\bigr\},
&
\delta Z^{f\rm L}_{(1)} &= -\Sigma^{f\rm L}_{(1)}(0),
\\
\delta \MH^2{}_{(1)} &= \text{Re}\bigl\{ \Sigma_{\rm (1)}^{\rm H}(\MH^2)\bigr\},
&
\delta Z^{f\rm R}_{(1)} &= -\Sigma^{f\rm R}_{(1)}(0),
\\
\delta \mt{}_{(1)} &= \frac{\mt^2}{2} \,
 \text{Re}\bigl\{ \Sigma_{\rm L}^f(\mt^2) + \Sigma_{\rm R}^f(\mt^2) +
 2\Sigma_{\rm S}^f(\mt^2) \bigr\},
\end{align}
Here the numbers in brackets denote the loop orders, while
$\Sigma^\prime(p^2)$ stands for the derivative $\frac{\partial}{\partial
p^2}\Sigma(p^2)$. For the EWPOs introduced above, only light fermions ($i.\,e.$
all fermions except the top quark) can appear as external states. Their 
wave function renormalization can be computed in the limit of vanishing
fermion mass, since $m_f \ll \MW,\MZ$ for $f=e,\mu\,\tau,u,d,s,c,b$.

To compute two-loop electroweak corrections to the EWPOs introduced above, 
the following counterterms are also needed at order ${\cal O}(\alpha^2)$
\cite{mwlong}:
\begin{align}
&\begin{aligned} 
\delta \mw^2{}_{(2)} = \text{Re}\bigl\{ \Sigma^{\rm WW}_{\rm T(2)}(\mw^2)\bigr\} &+ 
\text{Im}\bigl\{\Sigma^{\rm WW}_{\rm T(1)}(\mw^2)\bigr\}
\text{Im}\bigl\{\Sigma^{\rm WW\,\prime}_{\rm T(1)}(\mw^2)\bigr\},
\end{aligned} \\[1ex]
&\begin{aligned}[b]
\delta \mz^2{}_{(2)} = \text{Re}\bigl\{ \Sigma^{\rm ZZ}_{\rm T(2)}(\mz^2)\bigr\}&+ 
\text{Im}\bigl\{\Sigma^{\rm ZZ}_{\rm T(1)}(\mz^2)\bigr\}
\text{Im}\bigl\{\Sigma^{\rm ZZ\,\prime}_{\rm T(1)}(\mz^2)\bigr\} \\ &
+ \frac{1}{\mz^2}\Bigl (\text{Re}\bigl\{\Sigma^{\rm \gamma Z}_{\rm T(1)}(\mz^2)\bigr\}\Bigr )^2 
+ \frac{1}{\mz^2}\Bigl (\text{Im}\bigl\{\Sigma^{\rm \gamma Z}_{\rm T(1)}(\mz^2)\bigr\}\Bigr )^2, 
\end{aligned} \hspace{-1em} \\[1ex]
&\delta Z^{f\rm L}_{(2)} = -\Sigma^{f\rm L}_{(2)}(0), \qquad
\delta Z^{f\rm R}_{(2)} = -\Sigma^{f\rm R}_{(2)}(0), 
\label{eq:dzf} \\
&\begin{aligned} 
\delta Z_{e(2)} = \,&\frac{1}{2}\Sigma^{\gamma\,\prime}_{\rm T(2)}(0)
- \frac{\sw}{\cw} \, \frac{\Sigma^{\gamma\rm Z}_{\rm T(2)}(0)}{\mz^2} 
+(\delta Z_{e(1)})^2 
+ \frac{1}{8}\bigl (\Sigma^{\gamma\,\prime}_{\rm T(1)}(0)\bigr )^2 \\
&+\frac{\Sigma^{\gamma\rm Z}_{\rm T(1)}(0)}{2\mz^2} \Biggl [
 \frac{1}{\sw\cw}\,\frac{\delta\mw^2{}_{(1)}}{\mw^2}
 + \frac{(\sw^2-\cw^2)}{2\sw\cw}\, \frac{\delta\mz^2{}_{(1)}}{\mz^2}
 + \frac{\Sigma^{\gamma\rm Z}_{\rm T(1)}(0)}{\mz^2} \Biggr ].
\end{aligned} 
\end{align}
Here the one-loop formula $\gz{}_{(1)} = \frac{1}{\mz}\text{Im}\{\Sigma^{\rm
Z}_{\rm T(1)}(\mz^2)\}$ and its equivalent form for the $W$-boson has been used.

The weak gauge couplings can be written as
\begin{align}
g_0 = \frac{e_0}{\sw{}_{,0}} = \frac{e(1+\delta Z_e)}{\sw + \delta \sw}, \qquad
g'_0 = \frac{e_0}{\cw{}_{,0}} = \frac{e(1+\delta Z_e)}{\cw + \delta \cw},
\end{align}
where
\begin{align}
\frac{\delta \sw}{\sw} = -\frac{\delta \cw}{\cw}
= \sqrt{1-\frac{\mw+\delta\mw}{\mz+\delta\mz}}-\sqrt{1-\frac{\mw}{\mz}}.
\end{align}
Thus the renormalization of $g$ and $g'$ is determined through the
renormalization of the electromagnetic couplings and the $W$ and $Z$ masses.

\vspace{\medskipamount}
Some of the counterterms contain contributions that are enhanced relative to the
remaining electroweak corrections. For instance, the charge counterterm can be
written as
\begin{align}
1+\delta Z_e &= \frac{1}{\sqrt{1-\Delta\alpha}} + \delta Z_{e,\rm rem},
&
\Delta\alpha &\equiv -\Sigma^{\gamma\gamma\,\prime}_{\rm T,lf}(0) + 
 \frac{\text{Re}\bigl\{\Sigma^{\gamma\gamma}_{\rm T,lf}(\MZ^2)\bigr\}}{\MZ^2},
\end{align}
where $\Sigma^{\gamma\gamma}_{\rm T,lf}(q^2)$ is the contribution to the photon
self-energy from light fermions ($e,\mu\,\tau,u,d,s,c,b$). The $\Delta\alpha$
term can be interpreted as a shift of the electromagnetic coupling due to
renormalization group running from the scale 0 to $\mz^2$:
\begin{equation}
\alpha(\MZ^2) = \frac{\alpha}{1-\Delta\alpha}.
\end{equation}
It is enhanced by logarithms of the light-fermion masses. For the leptonic
part one finds at three-loop order \cite{Steinhauser:1998rq}
\begin{equation}
\Delta\alpha_{\rm lept} = 0.0314976.
\end{equation}
The quark contribution to $\Delta\alpha$, however, receives large
non-perturbative QCD corrections, which are difficult to compute from first
principles. Instead, this part is typically extracted from experimental data for
the cross-section of $e^+e^- \to \text{hadrons}$ or from tau-lepton decay
distributions. Some recent evaluations of $\Delta\alpha_{\rm had}$ are listed in
Tab.~\ref{tab:dahad}.
\begin{table}[tb]
\centering
\renewcommand{\arraystretch}{1.1}
\begin{tabular}{lc}
\hline
Result & Reference \\
\hline
$0.02762\pm0.00011$ & \cite{Davier:2010nc} \\
$0.02750\pm0.00033$ & \cite{Burkhardt:2011ur} \\
$0.02764\pm0.00014$ & \cite{Hagiwara:2011af} \\
$0.02766\pm0.00018$ & \cite{Jegerlehner:2015stw} \\
\hline
\end{tabular}
\caption{Recent evaluations of $\Delta\alpha_{\rm had}$ from $e^+e^-$ or $\tau$
decay data.
\label{tab:dahad}}
\end{table}

A second important enhanced contribution is contained in the weak mixing angle
counterterm, which at one-loop order can be written as
\begin{align}
\delta \sw{}_{(1)} &= \frac{\cw^2}{2\sw} \Delta\rho^\alpha_{(1)} + \delta
\sw{}_{(1),\rm rem}, &
\Delta\rho^\alpha_{(1)} &= \frac{\Sigma^{\rm ZZ}_{\rm T(1)}(0)}{\MZ^2} -
 \frac{\Sigma^{\rm WW}_{\rm T(1)}(0)}{\MW^2} =
\frac{3\alpha}{16\pi\sw^2} \; \frac{\mt^2}{\MW^2} + \Delta\rho^\alpha_{(1),\rm
rem}.
\end{align}
The quantity $\Delta\rho^\alpha$, called the ``$\rho$ parameter,'' was
first studied in Ref.~\cite{drho1}. It captures the leading custodial-symmetry
violating effect due to the Standard Model Yukawa couplings, which in our limit
of vanishing light-fermion masses involves only the top Yukawa coupling.
When going beyond the Standard Model, the $\rho$ parameter similarly measures
the leading custodial-symmetry violating effects of the new particles.
Higher-order Standard Model contributions to the $\rho$ parameter were computed
in Refs.~\cite{drhoqcd2,drho2,drhoqcd3,drho3,drhoqcd4}. Numerically $\Delta\rho
\sim 1\%$, but since it enters with a prefactor $\propto \sw^{-2}$ in many
EWPOs, it can dominate over other loop contributions.

\vspace{\medskipamount}
It can be argued that by preforming the perturbation series in powers of $G_{\rm
F}/\MW^2$ rather than in powers of $\alpha$, part of the large
$(\Delta\alpha)^m(\Delta\rho)^{n-m}$ terms at $n$-loop order are included
automatically, thus leading to better convergence of the perturbation series.
Furthermore, it was shown \cite{Consoli:1989fg,ylep2} that
\begin{align}
1+\Delta r &=
\frac{1}{(1-\Delta\alpha)\bigl(1+\frac{\cw^2}{\sw^2}\Delta\rho^G\bigr)-\Delta
r_{\rm rem}}, \\
\seff{f} &= \sw^2\Bigl (1+\frac{\cw^2}{\sw^2}\Delta\rho^G + \Delta\kappa_{\rm
rem}\Bigr ),
\end{align}
correctly reproduce the $(\Delta\alpha)^m(\Delta\rho)^{n-m}$ terms for
$n$=2
and for $n$=3, $m{\geq}2$ for these quantities. Here $\Delta\rho^G = \frac{3G_{\rm
F}\mt^2}{8\sqrt{2}\pi^2}$ and $\Delta r_{\rm rem}$ and $\Delta\kappa_{\rm
rem}$ denote the remaining corrections.

However, the resummation of $\Delta\alpha$ and $\Delta\rho$ through these
prescriptions is not guaranteed to provide a
numerically dominant contribution to the $n$-loop corrections. The reason is
that there are large numerical cancellations between different
$(\Delta\alpha)^m(\Delta\rho)^{n-m}$ terms for fixed $n$. For example, for $n$=3
one obtains
\begin{align}
\Delta r_{\rm resum,(3)}  
&\begin{array}[t]{l@{\;-\;}c@{\;+\;}c@{\;-\;}c@{}l} 
= (\Delta\alpha)^3 \
 &3(\Delta\alpha)^2\bigl(\tfrac{\cw^2}{\sw^2}\Delta\rho^\alpha\bigr)
 &6(\Delta\alpha)\bigl(\tfrac{\cw^2}{\sw^2}\Delta\rho^\alpha\bigr)^2
 &5\bigl(\tfrac{\cw^2}{\sw^2}\Delta\rho^\alpha\bigr)^3 \\[1ex]
\approx (\;2.05 &3.40 &3.74 &1.72&)\times 10^{-4} \\[1ex]
\multicolumn{2}{l}{= 0.68 \times 10^{-4}.} \end{array}
\end{align}
The sum is comparable in size to some other three-loop contributions.

\vspace{\medskipamount}
Another frequently used renormalization scheme is the ``\msbar scheme''
\cite{msbar}. In this scheme, the counterterm consists only of a divergent
piece together with some universal parameters. Specifically, the counterterms have the form
\begin{align}
\delta_\msbar X &= C_X \bigl (\mu^2 e^{\gamma_{\rm E}}\bigr
)^{4-D} \, \frac{1}{\epsilon},
\end{align}
where the coefficients $C_X$ are to be chosen appropriately to render all physical
quantities finite. $\mu$ is a free parameter, which will drop out of physical
observables when all orders of perturbation theory are included, but a
fixed-order perturbative result has a residual dependence on $\mu$ from
missing higher orders. In the calculation of EWPOs, one typically chooses
$\mu=\MZ$.

As before, one could choose the elementary particle masses and the
electromagnetic coupling as the independent quantities $\{X\}$ for which \msbar
counterterms are introduced. Instead, the following set of independent
quantities is used more commonly \cite{sirlinDR,dfs1,dgv}:
\begin{enumerate}
\item Fermion masses $\hat{m}_f$; \label{msbar1}
\item Higgs mass $\hMH$; \label{msbar2}
\item The couplings $\hat g$ and $\hat g'$ or, equivalently, $\hat{s}^2 =
\frac{\hat{g}'^2}{\hat{g}^2 + \hat{g}'^2}$ and $\hat e = 
\frac{\hat{g}}{\hat{s}}$.
\end{enumerate}
Here and in the following, the hat ($\hat{X}$) denotes \msbar quantities.
Note that the \msbar masses in points \ref{msbar1} and \ref{msbar2} depend on
the unphysical scale $\mu$ (at finite order in perturbation theory) and do not
directly correspond to any observable. In practice, one needs to compute the
translation between the \msbar masses and, say, the on-shell masses at the required
order before the results can be compared to experiment. Two-loop results and
higher-order QCD contributions for
the translation of the top-quark mass were presented in
Refs.~\cite{topmaa,master3,topmqcd}.

The on-shell masses of the $W$ and $Z$ bosons can be computed from $G_{\rm F}$,
$\hat{s}^2$ and $\hat{e}$ using the relations \cite{sirlinDR}
\begin{align}
\mw^2 &= \frac{\hat{e}^2}{4\sqrt{2}G_{\rm F}\hat{s}^2}\;\frac{1}{1+\Delta
\hat{r}_{\rm W}}, &
\mz^2 &= \frac{\hat{e}^2}{4\sqrt{2}G_{\rm
F}\hat{s}^2(1-\hat{s}^2)}\;\frac{1}{1+\Delta \hat{r}}, 
\end{align}
where $\hat{r}_{\rm W}$ and $\delta \hat{r}$ contain the required radiative
corrections.

Two-loop results for the \msbar renormalization of $\hat{e}$, $\hat{s}^2$ and
$\hmt$, as well as for $\hat{r}_{\rm W}$ were computed in
Refs.~\cite{fks1,topmaa,dems,dgg}.

The \msbar scheme has both advantages and disadvantages compared to the on-shell
scheme. For instance, the \msbar scheme is convenient for drawing connections to
low-energy physics or to GUT physics through renormalization group running.
Furthermore, when employing the \msbar top mass for computing ${\cal
O}(\alpha\as^n)$ corrections to the $\rho$ parameter, the coefficients for
increasing $n$ are much smaller compared to the case when the on-shell mass is
used. Specifically, one finds \cite{drhoqcd2,drhoqcd3,drhoqcd4}
\begin{align}
\Delta\rho^\msbar_{\rm QCD} &= \frac{3G_{\rm F}\hmt}{8\sqrt{2}\pi^2} \biggl[
 -0.193 \Bigl(\frac{\as}{\pi}\Bigr)
 -2.860 \Bigl(\frac{\as}{\pi}\Bigr)^2
 -1.680 \Bigl(\frac{\as}{\pi}\Bigr)^3 \biggr], \\
\Delta\rho^{\rm OS}_{\rm QCD} &= \frac{3G_{\rm F}\mt}{8\sqrt{2}\pi^2} \biggl[
 -3.970 \Bigl(\frac{\as}{\pi}\Bigr)
 -14.59 \Bigl(\frac{\as}{\pi}\Bigr)^2
 -93.15 \Bigl(\frac{\as}{\pi}\Bigr)^3 \biggr].
\end{align}
On the other hand, the \msbar scheme requires additional translations between
\msbar quantities and actual observables. Furthermore, the \msbar quantities are
in general not guaranteed to be gauge-invariant, although gauge invariance of
the \msbar masses can be achieved in the full Standard Model ($i.\,e.$ without
integrating out any heavy particles) through a suitable renormalization of the
electroweak vacuum \cite{Kniehl:2015nwa}.


\subsection{QED/QCD and short-distance corrections}
\label{sec:qed}

When computing radiative corrections to EWPOs, it is convenient to separate them
into two categories:
\begin{enumerate}
\item[a)]
Virtual and real QED and QCD radiation from the incoming and outgoing fermions.
These contributions contain physical soft and collinear IR divergences that
cancel between the virtual and real emission diagrams. They are, furthermore,
subject to experimental acceptances and cuts, so that it may be advantageous to
treat them with Monte-Carlo techniques.
\item[b)]
Electroweak corrections involving massive gauge or Higgs bosons. These
contributions are IR finite and can be absorbed into short-distance effective
couplings, such as $G_{\rm F}$ or the vector and axial-vector couplings of a
gauge-boson to a $f\bar{f}$ pair.
\end{enumerate}
It is desirable to factorize the complete set of all radiative corrections into
a product of the external radiation part (a) and the massive short-distance part
(b). For example, the muon decay rate in eqs.~\eqref{eq:gf} and \eqref{eq:mw} is
written as
\begin{align}
\Gamma_\mu &= \frac{\alpha^2m_\mu^5}{384\pi\sw^2\MW^2} F\Bigl(\frac{m_{\rm
e}^2}{m_\mu^2}\Bigr) (1+\Delta q)(1+\Delta r)^2,
\end{align}
where $\Delta q$ represents initial- and final-state QED radiation, while
$\Delta r$ denotes massive electroweak corrections. However, in general this factorization
is not exact, and there are additional non-factorizable contributions. These
first occur at the two-loop level and stem from diagrams with both massless and
massive boson exchange. Nevertheless, a factorized form can be recovered by
shifting the non-trivial non-factorizable terms into the short-distance part.
This can be illustrated by referring again to the example of muon decay:
\begin{align} 
&(1+\Delta q)(1+\Delta r_{\rm fact})^2 + \Delta x_{\rm nonfact}
\equiv (1+\Delta q)(1+\Delta r)^2, \\
&\Delta r = \Delta r_{\rm fact} + \frac{\Delta x_{\rm nonfact}}{2(1+\Delta
q)(1+\Delta r_{\rm fact})} + {\cal O}\bigl((\Delta x_{\rm nonfact})^2\bigr),
\end{align}
where the symbol $\Delta x_{\rm nonfact}$ has been introduced to denote the
non-factorizable contributions.

\vspace{\medskipamount}
For the QED corrections $\Delta q$ defined in this way, complete ${\cal
O}(\alpha)$ \cite{dq1} and ${\cal O}(\alpha^2)$ \cite{dq2} contributions are
known, including terms that are suppressed by powers of $m_{\rm e}/m_\mu$
\cite{dqme}. As a result, there is a very small relative uncertainty of $6 \times
10^{-7}$ for the extracted value of the Fermi constant $G_{\rm F}$.

As far as the short-distance corrections $\Delta r$ are concerned, the full
${\cal O}(\alpha)$ \cite{dr1}, ${\cal O}(\alpha\as)$ \cite{drhoqcd2,drqcd2},
${\cal O}(\alpha\as^2)$ \cite{drhoqcd3,drqcd3} and ${\cal O}(\alpha^2)$
\cite{mw,mwlong,dr2bos} have been computed. For the calculation of the two-loop
electroweak corrections in Refs.~\cite{mw,mwlong,dr2bos}, some of the numerical
techniques described in section \ref{sec:num} have been used. In addition, some
leading three- and four-loop corrections in the large-$\mt$ limit, that enter
through the $\rho$ parameter, are known. These include corrections of order
${\cal O}(\at^2\as)$, ${\cal O}(\at^3)$ \cite{drho3} and ${\cal O}(\at\as^3)$
\cite{drhoqcd4}, where $\at = y_{\rm t}^2/(4\pi)$ and $y_{\rm t}$ is the top
Yukawa coupling.

\vspace{\medskipamount}
Let us now move to the analysis of the $Z$-pole observables. The total $Z$
width, $\gz$, can be obtained from the requirement that $s_0 \equiv \mz^2 -
i\mz\gz$ is the pole of the $Z$-boson propagator,
\begin{equation}
(s_0-\mz^2) + \Sigma^{\rm Z}_{\rm T}(s_0) = 0, \label{eq:comppole}.
\end{equation}
Here $\Sigma^{\rm Z}_{\rm T}$, in contrast to $\Sigma^{\rm ZZ}_{\rm T}$ defined
in eq.~\eqref{eq:sigmavv}, contains contributions from $\gamma$--$Z$ mixing,
\begin{equation}
\Sigma^{\rm Z}_{\rm T}(s) = \Sigma^{\rm ZZ}_{\rm T}(s) 
 - \frac{[\Sigma^{\rm \gamma Z}_{\rm T}(s)]^2}{s+\Sigma^{\gamma\gamma}_{\rm
 T}(s)}.
\end{equation}
From the imaginary part of \eqref{eq:comppole} one finds
\begin{align}
\gz &= \frac{1}{\mz} \text{Im}\bigl \{ \Sigma^{\rm Z}_{\rm T}(s_0) \bigr\}
\nonumber \\
 &= \frac{1}{\mz}\Bigl [ \text{Im}\bigl \{ \Sigma^{\rm Z}_{\rm T}(\mz^2) \bigr\}
  - \mz\gz \, \text{Re}\bigl \{ \Sigma^{\rm Z\prime}_{\rm T}(\mz^2) \bigr\}
  - \tfrac{1}{2}\mz^2\gz^2 \text{Im}\bigl \{ \Sigma^{\rm Z\prime\prime}_{\rm T}(\mz^2) \bigr\}
  + {\cal O}(\gz^3) \Bigr ].
\label{eq:gz}
\end{align}
Eq.~\eqref{eq:gz} can be solved recursively for $\gz$. The imaginary part of the
self-energy can be related, with the help of the optical theorem, to the decay
process $Z \to f\bar{f}$:
\begin{equation}
\text{Im}\,\Sigma^{\rm Z}_{\rm T} = \frac{1}{3\mz} \sum_{f} \sum_{\rm
spins} \int d\Phi \;\bigl ({\cal R}_{\rm V}^f|v_f|^2 + {\cal R}_{\rm A}^f|a_f|^2\bigr ), 
\label{eq:opt}
\end{equation}
where $d\Phi$ is the integration measure for the $f\bar{f}$ phase-space
integration. In writing \eqref{eq:opt}, the radiative corrections have been split
up, as before, into final-state QED and QCD contributions, which are
denoted by the radiator functions ${\cal R}_{\rm V,A}$, and short-distance
contributions contained in the effective vector coupling $v_f$ and axial vector coupling
$a_f$ of the $Zf\bar{f}$ vertex. These effective couplings include massive
electroweak vertex corrections and $Z$--$\gamma$ mixing contributions. At
tree-level, they are given by
\begin{align}
v_{f(0)} &= \frac{e}{\sw\cw}\bigl (\tfrac{1}{2}I_f - Q_f\sw^2), &
a_{f(0)} &= \frac{eI_f}{2\sw\cw},
\end{align}
where $I_f$ and $Q_f$ are the weak isospin and electric charge quantum numbers,
respectively.

Inserting \eqref{eq:opt} into \eqref{eq:gz} and expanding the electroweak
contributions up to next-to-next-to-leading order, one obtains \cite{gz,gzlong}
\begin{align}
\gz &= \sum_f \overline{\Gamma}_f\,, \qquad
\overline{\Gamma}_f = \frac{N_c^f\mz}{12\pi} \Bigl [
 {\cal R}_{\rm V}^f F_{\rm V}^f + {\cal R}_{\rm A}^f F_{\rm A}^f \Bigr ]_{s=\mz^2} 
 \;, \label{Gz} \\
F_{\rm V}^f &= v_{f(0)}^2
 \bigl [1-\text{Re}\,\Sigma^{\rm Z\prime}_{\rm T(1)}-
 \text{Re}\,\Sigma^{\rm Z\prime}_{\rm T(2)}
  + (\text{Re}\,\Sigma^{\rm Z\prime}_{\rm T(1)})^2 \bigr ] 
 + 2 \,\text{Re}\, (v_{f(0)}v_{f(1)})\bigl [1-\text{Re}\,\Sigma^{\rm Z\prime}_{\rm T(1)} \bigr ] 
\nonumber \\[.5ex] 
 &\quad + 2 \,\text{Re}\, (v_{f(0)}v_{f(2)}) + |v_{f(1)}|^2
 - \tfrac{1}{2}\mz \gz v_{f(0)}^2
 \;\text{Im}\,\Sigma^{\rm Z\prime\prime}_{\rm T(1)}\,, \label{Fv} \\[1ex]
F_{\rm A}^f &= a_{f(0)}^2
 \bigl [1-\text{Re}\,\Sigma^{\rm Z\prime}_{\rm T(1)}-\text{Re}\,\Sigma^{\rm Z\prime}_{\rm T(2)}
  + (\text{Re}\,\Sigma^{\rm Z\prime}_{\rm T(1)})^2 \bigr ] 
 + 2 \,\text{Re}\, (a_{f(0)}a_{f(1)})\bigl [1-\text{Re}\,\Sigma^{\rm Z\prime}_{\rm T(1)} \bigr ] 
\nonumber \\[.5ex] 
 &\quad + 2 \,\text{Re}\, (a_{f(0)}a_{f(2)}) + |a_{f(1)}|^2
 - \tfrac{1}{2}\mz \gz a_{f(0)}^2
 \;\text{Im}\,\Sigma^{\rm Z\prime\prime}_{\rm T(1)}\,. \label{Fa}
\end{align}
The radiator functions ${\cal R}_{\rm V,A}$ are known with QCD corrections up to
${\cal O}(\as^4)$ \cite{rad,rad2}, QED corrections up to ${\cal O}(\alpha^2)$
\cite{Kataev:1992dg} and mixed QED-QCD corrections of ${\cal O}(\alpha\as)$
\cite{rad}. These have been computed in the limit of massless final-state
quarks. Additionally, terms that are suppressed by powers of $m_q^2/s$ are known
up to ${\cal O}(\as^3)$ \cite{rad}.

For the short-distance loop corrections in $F^f_{\rm V,A}$, entering through
$v_f$, $a_f$ and $\Sigma^{\rm Z}_{\rm T}$, complete electroweak one-loop and
fermionic two-loop results have been computed \cite{gz,gzlong}. Here
``fermionic'' stands for contributions from diagrams with closed fermions loops,
which are numerically dominant compared to the ``bosonic'' contributions. 
The two-loop electroweak corrections have been computed using numerical loop integration
techniques. In
addition, the two-loop ${\cal O}(\alpha\as)$ corrections are available
\cite{drhoqcd2,drqcd2,ck,aasbb}, as well as leading three- and four-loop contributions of
${\cal O}(\at\as^2)$ \cite{drhoqcd3}, ${\cal O}(\at^2\as)$, ${\cal O}(\at^3)$
\cite{drho3} and ${\cal O}(\at\as^3)$ \cite{drhoqcd4}.

The higher-order predictions for the partial widths $\overline{\Gamma}_f$ can be
used to derive predictions for other pseudo-observables in
eqs.~\eqref{eq:sig0}--\eqref{eq:rq}:
\begin{align}
\sigma^0_{\rm had} &= \sum_q\frac{12\pi}{\mz^2} \; 
 \frac{\overline{\Gamma}_e\overline{\Gamma}_q}{\gz^2}(1+\delta X), \\
R_\ell &= \frac{\sum_q \overline{\Gamma}_q}{\sum_\ell \overline{\Gamma}_\ell},
\qquad
R_c = \frac{\overline{\Gamma}_c}{\sum_q \overline{\Gamma}_q}, \qquad
R_b = \frac{\overline{\Gamma}_b}{\sum_q \overline{\Gamma}_q}.
\end{align}
The correction factor $\delta X$ is obtained by performing the systematic
expansion of the matrix element for $e^+e^- \to f\bar{f}$ about the complex pole
$s_0$, see eq.~\eqref{eq:poleexp}, and then collecting all terms for $s=\mz^2$ at the required
order\footnote{It is interesting to note that $\sigma_f(s=\MZ^2) = 
\sigma_f(s=\mz^2)$, so that the ``on-shell'' peak cross-section at $s=\mz^2$ is
consistent with the definition in eq.~\eqref{eq:sig0}.}. At two-loop order one
finds \cite{gz}
\begin{equation}
\delta X_{(2)} = -\Bigl [
 (\text{Im}\,\Sigma^{\rm Z\prime}_{\rm T(1)})^2 +
 2\mz \gz 
 \;\text{Im}\,\Sigma^{\rm Z\prime\prime}_{\rm T(1)} \Bigr]_{s=\mz^2}.
\end{equation}
The $Z$-pole asymmetries can also be defined in terms of the effective couplings
$v_f$ and $a_f$, leading to
\begin{align}
\seff{f} &= \frac{1}{4|Q_f|}\biggl (1-{\rm Re}\frac{v_f}{a_f}\biggr )
 = \biggl(1-\frac{\mw^2}{\mz^2}\biggr ) \, 
  \frac{1-{\rm Re}\{v_f/a_f\}}{1-v_{f(0)}/a_{f(0)}},
\end{align}
from which the left-right and forward-backward asymmetries can be constructed.
The left-right asymmetry, $A^f_{\rm LR}$, does not receive any contributions
from initial/final-state QED or QCD corrections. The effect of initial- and
final-state radiation on the forward-backward asymmetry, $A^f_{\rm FB}$,
strongly depends on the experimental cuts applied, but it is generally small
\cite{afbr}. Note, however, that the observable forward-backward asymmetry is
significantly affected by soft and collinear initial-state radiation, which is
already removed in the definition of the pseudo-observable $A^f_{\rm FB}$ in
terms of the deconvoluted cross-section $\sigma_f$ in \eqref{eq:afb}.

The complete electroweak one-loop and two-loop corrections to the leptonic
effective weak mixing angle $\seff{\ell}$ have been computed in
Refs.~\cite{dr1,sw1} and \cite{sweff,swlong,swucc1,hmu2}, respectively. As before, numerical loop
integration methods were used for the two-loop contributions. Furthermore, mixed
electroweak--QCD corrections of ${\cal O}(\alpha\as)$ \cite{drhoqcd2,drqcd2} and ${\cal
O}(\alpha\as^2)$ \cite{drhoqcd3,drqcd3} are known, as well as leading contributions entering
through the $\rho$ parameter, of order ${\cal O}(\at^2\as)$, ${\cal O}(\at^3)$
\cite{drho3} and ${\cal O}(\at\as^3)$ \cite{drhoqcd4}.

For the quark weak mixing angle $\seff{q},\;q=u,d,c,s,b$, the situation is
similar, except that only fermionic but no bosonic electroweak two-loop
corrections are available. Furthermore, the calculation of mixed
electroweak--QCD corrections  becomes more involved in this case. The
additional contributions at order ${\cal O}(\alpha\as)$ have been computed in
Ref.~\cite{ck,aasbb}, but
only the leading $\rho$-parameter contribution is known at the
next order ${\cal O}(\at\as^2)$ \cite{drhoqcd3}.

\vspace{\medskipamount}
The use of numerical loop integration techniques has enabled the computation of
precise results for the two-loop electroweak contributions to various EWPOs,
without needing to resort to any large-mass or small-momentum expansions.
However, the numerical evaluation is relatively slow and involves large
expressions for the integrands, so that it is difficult to implement these
results in this form into other computer codes. Instead they have been made
available in the form of convenient parametrization formulae, whose coefficients
have been fitted to reproduce the full numerical result over a wide range of
values for the input parameters \cite{swlong,swbb,gzlong,mwfinal}. They have
been implemented in the global fitting programs {\sc GFitter} \cite{gfitter}, {\sc
GAPP} \cite{gapp}, in Ref.~\cite{mishima} and partially in {\sc ZFitter 6.42}
\cite{zfit2}.


\subsection{Impact of corrections beyond one-loop}

Global fits to a set of EWPOs lead to important indirect constraints on physics
beyond the Standard Model. See Ref.~\cite{newphys} for recent examples
within the context of a variety of new physics models. One can also use a
model-independent effective field theory framework to parametrize any deviations
from the Standard Model (see Refs.~\cite{mishima,eftfit} for more information and recent
results). At the same time, a fit of the Standard Model to EWPOs tests the
theory at the quantum level and leads to indirect constraints on the masses of
heavy particles, such as the top-quark and Higgs-boson masses
\cite{mishima,gfitter,pdg}.

\begin{figure}[tb]
\centering
\psfig{figure=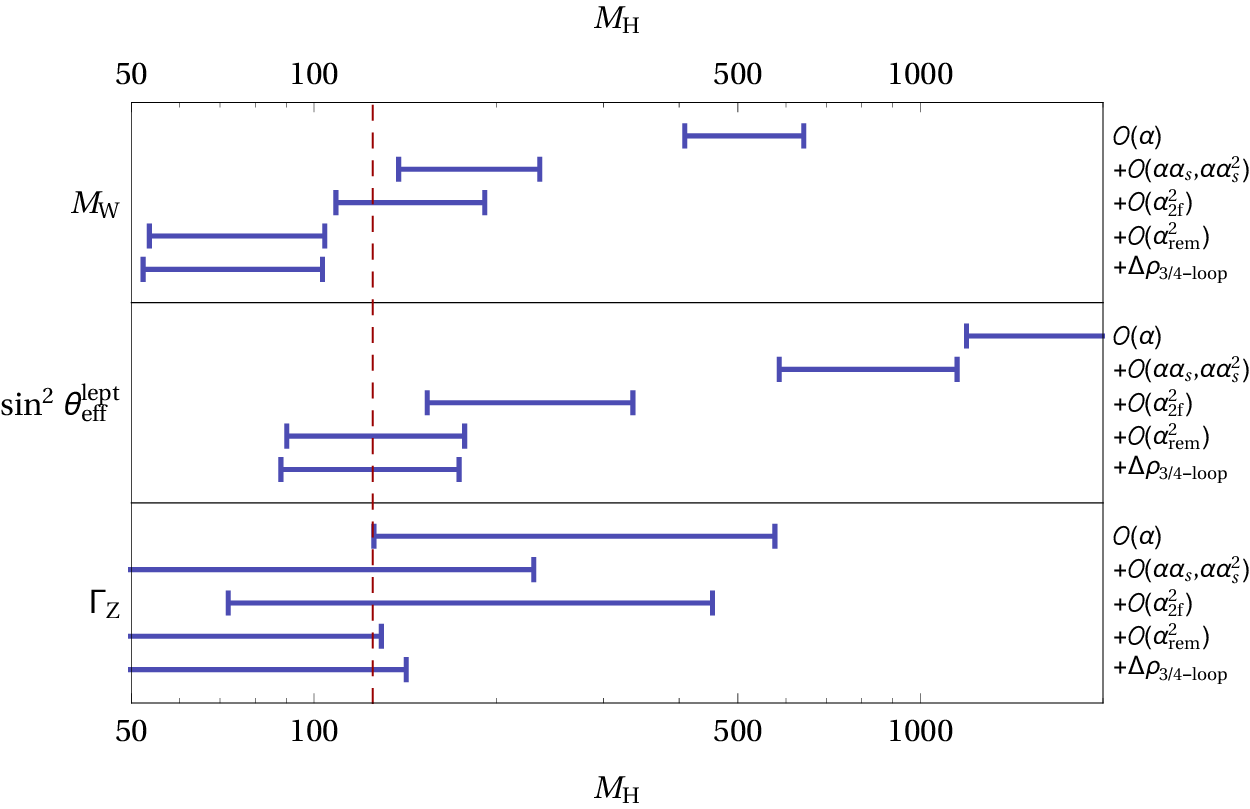, width=16cm}
\caption{Impact of higher-order corrections on the indirect determination of
$\MH$ from $\MW$ (predicted from $G_{\rm F}$), $\seff{\ell}$ and $\GZ$.
Corrections of different order are cumulatively added as indicated on the right.
Here ${\cal O}(\alpha^2_{\rm 2f})$ stands for electroweak two-loop contributions with
two closed fermion loops, ${\cal O}(\alpha^2_{\rm rem})$ denotes the remaining
two-loop contributions (with one or no closed fermion loop), and $\Delta
\rho_{\rm 3/4-loop}$ are leading 3- and 4-loop corrections of ${\cal O}(\at^2\as)$, ${\cal O}(\at^3)$ and ${\cal O}(\at\as^3)$.
The error bars reflect the parametric uncertainty from the input
parameters in Tab.~\ref{tab:smval}, but not the theory error from missing higher
orders. The dashed line indicates the value $\MH=125\gev$ from the direct measurement of
the Higgs mass.
\label{fig:limpact}}
\end{figure}

The inclusion of higher-order corrections in these fits is mandatory to ensure
that the results are not subject to systematic errors from missing theory
contributions. In Fig.~\ref{fig:limpact}, the impact of corrections of different
order on the indirect prediction of the Higgs mass, $\MH$, from various EWPOs is
shown. In each case, the Standard Model prediction of one EWPO is compared to
the experimental value in the left side of Tab.~\ref{tab:smval}, using the
values on the right side of Tab.~\ref{tab:smval} for the remaining parametric
inputs.
\begin{table}[tb]
\renewcommand{\arraystretch}{1.1}
\begin{tabular}[t]{llc}
\hline
Quantity & Value & Ref.\\
\hline
$\MW$ & $80.385 \pm 0.015\gev$ & \cite{pdg} \\
$\seff{\ell}$ & $0.23153 \pm 0.00016$ & \cite{lep1} \\
$\GZ$ & $2.4952 \pm 0.0023\gev$ & \cite{lep1} \\
\hline
\end{tabular}
\hfill
\begin{tabular}[t]{llc}
\hline
Quantity & Value & Ref. \\
\hline
$G_{\rm F}$ & $(11663787 \pm 6)\times 10^{-12}\gev^{-2}$ & \cite{pdg} \\
$\MZ$ & $91.1876 \pm 0.0021\gev$ & \cite{lep1} \\
$\mt$ & $173.24 \pm 0.95\gev$ & \cite{pdg} \\
$\as(\MZ^2)$ & $0.1185 \pm 0.0006$ & \cite{pdg} \\
$\Delta\alpha_{\rm had}$ & $0.02766\pm0.00018$ & \cite{Jegerlehner:2015stw} \\
\hline
\end{tabular}
\caption{Direct measurements of various electroweak precision observables (left);
and measurements of other important Standard Model inputs (right).}
\label{tab:smval}
\end{table}

To produce the data in Fig.~\ref{fig:limpact}, the calculations were performed
in the on-shell scheme, using an expansion of perturbative electroweak
corrections in powers of $\alpha$ (rather than $G_{\rm F}/\MW^2$). Higher-order
corrections are included step-by-step in the following order:
\begin{itemize}
\item One-loop corrections, ${\cal O}(\alpha)$;
\item Two- and three-loop QCD corrections, ${\cal O}(\alpha\as)$ and  
 ${\cal O}(\alpha\as^2)$;
\item Two-loop electroweak corrections from diagrams with two-closed fermion
loops, denoted ${\cal O}(\alpha_{\rm 2f})$, which include resummed terms
proportional to powers of $\Delta\alpha$ and $\Delta\rho$;
\item The remaining two-loop electroweak corrections, denoted ${\cal
O}(\alpha_{\rm rem})$, which include the two-loop terms that cannot be
obtained from resummation or scheme changes;
\item Leading three- and four-loop corrections in the large-$\mt$ limit
entering through the $\rho$
parameter, ${\cal O}(\at^2\as)$, ${\cal O}(\at^3)$ and ${\cal O}(\at\as^3)$.
\end{itemize}

As evident from the figure, radiative corrections beyond the one-loop level are
essential for a reliable test of the Standard Model at the current level of
experimental precision. In particular, the non-trivial two-loop electroweak
corrections with one or no closed fermion loop, ${\cal
O}(\alpha_{\rm rem})$, are sizable and shift the indirect value of $\MH$ by
about one standard deviation for the most sensitive observables $\MW$ and
$\seff{\ell}$. These are the corrections for which the use of
numerical loop integration methods has been instrumental.
The intrinsic uncertainty from the numerical integration is entirely negligible
on the scale shown in the figure.
The rather small numerical impact of the leading three- and four-loop
corrections, $\Delta \rho_{\rm 3/4-loop}$, is due to the fact the 
contributions of ${\cal O}(\at^2\as)$ and ${\cal O}(\at^3)$ are partially
canceled by the ${\cal O}(\at\as^3)$ terms.


\subsection{Theory uncertainties from missing higher-order contributions}
\label{sec:err}

As was demonstrated in the previous subsection, multi-loop radiative corrections
have a sizable impact on electroweak precision tests. Therefore it is
important to estimate the uncertainty from missing higher-order contributions.
Unfortunately there is no undisputable method for arriving at this estimate.
Some of the common approaches are:
\begin{itemize}
\item Collect all relevant prefactors of the most important missing
contributions, such as couplings, group factors, particle
multiplicities, and mass ratios. This method has been advocated, for example, in
Refs.~\cite{gapp,Baur:2002gp}.
\item When using the \msbar renormalization scheme, an estimate of missing
higher orders may be obtained from varying the renormalization scale, $\mu$, within
a certain range, such as $\MZ/2 < \mu < 2\MZ$. This idea has been widely used
for the evaluation of theory errors from missing QCD corrections, but its
application in the electroweak sector stands on less firm footing.
\item Another approach is to compare results between different renormalization
schemes, for instance between the on-shell and \msbar schemes. See for example
Ref.~\cite{kataev1,dgs}. 
\item When several orders of a certain quantity have already been computed, one
could try to extrapolate to higher orders by assuming that the coefficients of
the perturbative series approximately follow a geometric series. There is no
formal reason for the validity of this assumption, but in practice it has led to
reasonable theory error estimates.
This method has been used, for example, in
Refs.~\cite{mwlong,swlong,gzlong}.
\end{itemize}
In the following, the outcome of these methods
will be illustrated with a few examples,
and their advantages and disadvantages will be briefly discussed.

\vspace{\medskipamount}
Let us begin by examining the use of the prefactor method to estimate the theory
error in the prediction of $\GZ$. The most important missing higher order
corrections are the bosonic electroweak two-loop contributions, ${\cal
O}(\alpha_{\rm bos})$, three-loop contributions of orders $\alpha^3$,
$\alpha^2\as$ and $\alpha\as^2$, and four-loop contributions
of order $\alpha\as^3$. For all these contributions except the ${\cal
O}(\alpha_{\rm bos})$ terms, the leading terms in the large-$\mt$ limit are
already known, so the uncertainty pertains only to the remaining, non-enhanced
terms. The following estimates are obtained with the prefactor method:
\begin{align}
&{\cal O}(\alpha_{\rm bos}) \sim \GZ \alpha^2 \approx 0.13\mev, \nonumber\\
&{\cal O}(\alpha^3)-{\cal O}(\at^3) \sim \GZ \alpha \at^2 \approx 0.12\mev,\nonumber\\
&{\cal O}(\alpha^2\as)-{\cal O}(\at^2\as) \sim 
  \GZ \,\frac{\alpha \at n_q}{\pi} \as(\mt) \approx 0.23\mev, \label{eq:prefac}  \\
&{\cal O}(\alpha\as^2)-{\cal O}(\at\as^2) \sim 
  \GZ\,\frac{\alpha n_q}{\pi} \as^2(\mt) \approx 0.35\mev, \nonumber \\
&{\cal O}(\alpha\as^3)-{\cal O}(\at\as^3) \sim 
  \GZ\,\frac{\alpha n_q}{\pi} \as^3(\mt) \approx 0.04\mev.\nonumber
\end{align}
Alternatively, one may estimate the theory error for $\GZ$ using the geometric
series extrapolation method. As mentioned above, we are only interested in the
missing contributions of orders $\alpha^3$,
$\alpha^2\as$, $\alpha\as^2$ and $\alpha\as^3$ beyond the known leading
large-$\mt$ part. Consequently, the leading large-$\mt$ part must also be
excluded for the lower order which is used as reference for the extrapolation.
In this way, one obtains \cite{gzlong}\begin{align}
&{\cal O}(\alpha_{\rm bos}) \sim [{\cal O}(\alpha_{\rm bos})]^2 \approx 0.10\mev,
 \nonumber \\
&{\cal O}(\alpha^3)-{\cal O}(\at^3) \sim 
 \frac{{\cal O}(\alpha^2)}{{\cal O}(\alpha)}\,
 [{\cal O}(\alpha^2)-{\cal O}(\at^2)] \approx 0.26\mev, \nonumber \\
&{\cal O}(\alpha^2\as)-{\cal O}(\at^2\as) \sim
 \frac{{\cal O}(\alpha\as)}{{\cal O}(\alpha)}\,
 [{\cal O}(\alpha^2)-{\cal O}(\at^2)] \approx 0.30\mev, \label{eq:geom} \\
&{\cal O}(\alpha\as^2)-{\cal O}(\at\as^2) \sim
 \frac{{\cal O}(\alpha\as)}{{\cal O}(\alpha)}\,
 [{\cal O}(\alpha\as)-{\cal O}(\at\as)] \approx 0.23\mev, \nonumber \\
&{\cal O}(\alpha\as^3)-{\cal O}(\at\as^3) \sim
 \frac{{\cal O}(\alpha\as^2)}{{\cal O}(\alpha)}\,
 [{\cal O}(\alpha\as)-{\cal O}(\at\as)] \approx 0.035\mev. \nonumber 
\end{align}
By comparing \eqref{eq:prefac} and \eqref{eq:geom}, one can see that both
methods lead to comparable estimates, although there are sizable differences for
certain individual contributions. 
Adding all contributions in quadrature, the total error is estimated to be
$\delta_{\rm th}\GZ \sim 0.5\mev$.

The scale variation method has mostly been applied for the estimation of
higher-order QCD corrections to EWPOs. In Ref.~\cite{mwlong} it has been used to estimate
the ${\cal O}(\alpha^2\as)$ and ${\cal O}(\alpha\as^3)$ contributions to $\MW$
to amount to $\approx 3.8\mev$ and $\approx 0.7\mev$, respectively. Since then,
the leading large-$\mt$ contributions at order $\at^2\as$ and $\at\as^3$ have been
calculated \cite{drho3,drhoqcd4}, yielding about 2~MeV each when using the
on-shell definition for the top-quark mass. Thus the magnitude of the ${\cal
O}(\alpha^2\as)$ contributions has been properly estimated from the scale
variation, whereas the ${\cal O}(\alpha\as^3)$ contributions were somewhat
underestimated. This may be partially due to the use of the on-shell top-quark
mass, whereas better estimates may be obtained when consistently using \msbar
renormalization for all quantities that receive QCD corrections.

Finally, the analysis of scheme variation has also been used to evaluate the
impact of missing higher-order corrections. In Ref.~\cite{dgs,dgs2}, approximate results
for the electroweak two-loop corrections to $\MW$, $\seff{\ell}$ and
$\Gamma_\ell$ were computed in the \msbar and the on-shell scheme. These results
incorporated the leading, ${\cal O}(\alpha^2\mt^4/MW^4)$, and next-to-leading,
${\cal O}(\alpha^2\mt^2\MW^2)$, contributions in an expansion for
large values of $\mt$, besides resummed terms involving $\Delta\alpha$.
The \msbar and on-shell results can be compared to
estimate the remaining two-loop electroweak contributions. The observed
differences were $\delta\MW \approx 2\mev$, 
$\delta\seff{\ell} \approx 3 \times 10^{-5}$ and $\delta\Gamma_\ell \approx
0.001\mev$ for $\MH \sim 100\gev$. Later, explicit results for the remaining
${\cal O}(\alpha^2)$ contributions were obtained, resulting in significantly
larger numerical effects: $\delta \MW \approx 4\mev$ \cite{mw,mwlong},
$\delta\seff{\ell} \gtrsim 4\times 10^{-5}$ \cite{sweff,swlong} and
$\delta\Gamma_\ell \approx 0.02\mev$ \cite{gzlong}.

This underestimation of the theory error may not be entirely surprising since
the difference between renormalization
schemes is part of the missing higher-order corrections, but there is no strong
reason to assume that it is the numerically dominant part. More recently, the
complete two-loop corrections to the prediction of the $W$ mass have been
computed both in the on-shell and \msbar schemes. The difference between the
results in Ref.~\cite{dgg} and in the arXiv update of Ref.~\cite{mwfinal} 
(hep-ph/0311186v2) can be taken as an estimate of the missing three-loop
contributions. It amounts to 4--5~MeV, in reasonable agreement with other
estimates of the missing higher-order contributions \cite{mwfinal}.


\subsection{Future projections}
\label{sec:future}

Table~\ref{tab:curerr} gives a summary of the estimated theory uncertainty for
several important electroweak precision observables, compared with their current
experimental precision from measurements at LEP, SLC and Tevatron. In all cases,
the theory error is smaller than the experimental uncertainty by a factor of a
few, which is a desirable situation since it implies a subdominant impact from
ambiguities in defining and evaluating the theoretical uncertainty (see previous
subsection).

\begin{table}[tb]
\centering
\renewcommand{\arraystretch}{1.1}
\begin{tabular}[t]{lcc}
\hline
Quantity & Theory error & Exp.\ error \\
\hline
$\MW$ [MeV] & 4 & 15 \\
$\seff{\ell}$ [$10^{-5}$] & 4.5 & 16 \\
$\GZ$ [MeV] & 0.5 & 2.3 \\
$R_b$ [$10^{-5}$] & 15 & 66 \\
\hline
\end{tabular}
\caption{Estimated theory error of available predictions for several important
electroweak precision observables (from Refs.~\cite{mwfinal,swlong,gzlong}), compared to the
current experimental error (from Refs.~\cite{lep1,pdg}).}
\label{tab:curerr}
\end{table}

There are several proposals for future high-luminosity $e^+e^-$ colliders, which
are expected to measure electroweak precision observables, in particular
$Z$-pole observables and the $W$ mass, to significantly higher precision. The
first proposal, the International Linear Collider (ILC), is planned to be a
linear $e^+e^-$ machine with adjustable center-of-mass energy in the range
$\sqrt{s} \sim 90\dots 500\gev$, extendable to 1~TeV \cite{tesla,ilc}. It can
accommodate polarized $e^-$ and $e^+$ beams and is expected to collect more than
50~fb$^{-1}$ of data near the $Z$ pole and 100~fb$^{-1}$ near the $W$ pair
production threshold. An alternative proposal, the Future Circular Collider
(FCC-ee), is based on a 80--100~km circumference accelerator ring with $\sqrt{s}
\sim 90\dots 350\gev$ \cite{fccee}. It has the potential to generate several
ab$^{-1}$ of data near the $Z$ pole and a comparable amount at the $WW$
threshold. Finally, there is the Circular Electron-Positron Collider (CEPC)
proposal \cite{cepc}, which is also a ring collider with 50--70~km circumference
and $\sqrt{s} \sim 90\dots 250\gev$. Its target luminosities are 150~fb$^{-1}$
at the $Z$ pole and 100~fb$^{-1}$ near the $WW$ threshold.

\begin{table}[tb]
\centering
\renewcommand{\arraystretch}{1.1}
\begin{tabular}[t]{lcccc}
\hline
Quantity & ILC & FCC-ee & CEPC & Projected theory error \\
\hline
$\MW$ [MeV] & 3--4 & 1 & 3 & 1 \\
$\seff{\ell}$ [$10^{-5}$] & 1 & 0.6 & 2.3 & 1.5 \\
$\GZ$ [MeV] & 0.8 & 0.1 & 0.5 & 0.2 \\
$R_b$ [$10^{-5}$] & 14 & 6 & 17 & 5--10 \\
\hline
\end{tabular}
\caption{Estimated experimental precision for several important
electroweak precision observables at future $e^+e^-$ colliders
\cite{ilc,Hawkings:1999ac,fccee,d'Enterria:2016yqx,cepc} (no theory
uncertainties included, see text). In the last column,
the estimated error for the theoretical predictions of these quantities is
given, under the assumption that ${\cal
O}(\alpha\as^2)$, fermionic ${\cal O}(\alpha^2\as)$, fermionic ${\cal
O}(\alpha^3)$, and leading four-loop corrections entering through the
$\rho$-parameter will become available \cite{futth}.}
\label{tab:futerr}
\end{table}

All of these machines will significantly improve the experimental uncertainty
for the determination of electroweak precision observables (EWPOs), see
Tab.~\ref{tab:futerr} \cite{ilc,Hawkings:1999ac,fccee,d'Enterria:2016yqx,cepc}.
As a consequence, the experimental error for many quantities will become
comparable or even subdominant compared to the theory error, from missing
higher-order corrections, in the prediction of these quantities within the
Standard Model. Therefore it will be necessary to compute three-loop and even
leading four-loop corrections to be able to take full advantage of the potential
of these future accelerators. In Ref.~\cite{futth}, it has been estimated by how much
the theory error may be expected to be reduced if the complete ${\cal
O}(\alpha\as^2)$ corrections, the fermionic ${\cal O}(\alpha^2\as)$ and ${\cal
O}(\alpha^3)$ corrections\footnote{As above, the term ``fermionic'' refers to
diagrams with closed fermion loops.}, and the leading four-loop corrections in the
large-$\mt$ limit will become available. This estimate is based on the geometric 
series extrapolation method discussed in the previous subsection, and only
provides an order-of-magnitude projection. The results are shown in the last
column of Tab.~\ref{tab:futerr}.

On the technical side, the calculation of these corrections will include
three-loop self-energy and vertex integrals with many different masses in the
propagators. It appears highly unlikely that they can be tackled with analytical
methods. In some cases, in particular diagrams involving top-quark propagators,
asymptotic expansion techniques may be helpful to arrive at a result with
sufficient accuracy. Alternatively, and for the remaining cases, numerical
integration methods will need to be employed. None of the techniques described
in this review can be immediately deployed to this problem, but further
developments and improvements in numerical efficiency and convergence will be
required to carry out these calculations. 

In addition, theoretical improvements will be necessary for the processes that
are being used to extract the relevant Standard Model input parameters. These
include $\MW$, which can be determined from the $e^+e^- \to W^+W^-$
cross-section near threshold, and $\mt$, which can be determined from the
$e^+e^- \to t\bar{t}$ cross-section near threshold. A lot of effort has been
invested into precision calculations for the $t\bar{t}$ cross-section,
culminating in the recent completion of full ${\cal O}(\as^3)$ \cite{ttnnnlo}
and partial ${\cal O}(\alpha)$ electroweak \cite{ttew} corrections. Together
with future improvements in the determination of $\as$, this appears to enable a
determination of $\mt$ from $e^+e^- \to t\bar{t}$ with a theory error of about
50~MeV \cite{hoang}.

On the other hand, a much more ambitious precision target of 1~MeV is envisioned
for $\MW$ at FCC-ee, see Tab.~\ref{tab:futerr}. On the theory side, this will
require the inclusion of multi-loop electroweak corrections for the prediction
of $e^+e^- \to W^+W^-$ near threshold. The current state of the art includes
complete one-loop electroweak corrections for $W$-boson pair production and
decay with off-shell effects \cite{ee4f}, as well as two-loop contributions that
are enhanced by the Coulomb singularity from soft photon exchange
\cite{Beneke:2015lwa}. The theory uncertainty is estimated to be $\delta_{\rm
th}\MW \sim 3\mev$, which is not sufficient for the FCC-ee and CEPC precision
targets. The most important steps for improving the theoretical precision are
the calculation of full ${\cal O}(\alpha^2)$ electroweak corrections to on-shell
$W$-boson pair production ($e^+e^- \to W^+W^-$) and to on-shell $W$-boson decay
($W \to f\bar{f}'$), which are building blocks for the effective field theory
framework in Refs.~\cite{Beneke:2015lwa,Beneke:2007zg}. Since these
contributions depend on many independent mass and momentum scales, it again is
likely that numerical loop integration methods are the most promising avenue
towards dealing with these challenges.

Another important quantity is the shift in the electromagnetic fine structure
constant from hadronic contributions, $\Delta\alpha_{\rm had}$. It can be
extracted from data on $e^+e^- \to \text{hadrons}$, which may be improved in the
future by, $e.\,g.$, radiative return events measured at Belle-II \cite{belle}.
Alternatively, $\alpha(\MZ^2)$ could be measured directly at a
very-high-luminosity $e^+e^-$ machine, such as FCC-ee, by taking data at a few
GeV above and below the $Z$ peak \cite{janot}. This approach will require very
precise Standard Model predictions for $e^+e^- \to f\bar{f}$, including fermionic
three-loop vertex and box corrections. This is another
example where numerical integration techniques can play an
essential role.


\section{Summary}

Since the 1980s, results from particle physics experiments have reached a level
of precision that makes the consideration of radiative corrections a necessary
element of many experimental analyses and interpretations thereof. With
increasing number of loops and increasing number of mass and momentum scales, it
becomes more difficult to compute these corrections analytically. Thus, both
from a conceptual and practical point of view, one needs to resort to
non-analytical methods for the evaluation of multi-loop multi-scale problems.
Broadly speaking, these can be divided into two categories: 
asymptotic expansions and numerical integration techniques. 

The asymptotic expansion technique aims to write a loop amplitude as a series in
powers of a suitable small parameter, such as a small mass or momentum scale,
the inverse of a large mass or momentum scale, or a small mass difference. The
coefficients of this series are simpler loop integrals, which are more amenable
for analytical evaluation. An asymptotic expansion may also be performed in
terms of more than one expansion parameter, in order to simplify the analytics
even further. By computing ${\cal O}(10)$ terms for each expansion parameter, 
one can obtain sufficiently accurate results for many practical purposes.
Nevertheless, by its very nature, this method delivers only approximate results.
In principle, the quality of the approximation can be systematically improved by
including more terms of the series, but the computational complexity grows
significantly with each additional term.

Numerical integration techniques, on the other hand, aim to evaluate a given
loop amplitude without any specific approximation, by performing some subset or
all integrations of a multi-dimensional integral numerically. Depending on the
chosen technique, the numerical integration may be carried out directly in the
loop momentum space, or one may transform the integral to a different set of
variables, such as Feynman parameters. In any case, a successful numerical
integration technique will be subject to three general requirements: \emph{(i)}
a method for the removal or regularization of UV and IR singularities;
\emph{(ii)} sufficiently fast convergence of the numerical integration for a
range of values of the input parameters; and \emph{(iii)} applicability to a
variety of physical processes with different numbers of external legs and
different types of particles in the loops.

In section~\ref{sec:num}, a variety of different numerical loop integration
techniques have been reviewed, and their approaches to addressing the
aforementioned requirements have been discussed. Due to the contribution of many
researchers, tremendous progress in the development and improvement of numerical
integration techniques has been achieved over the last 20 years. Nevertheless,
no single technique is applicable to any arbitrary multi-loop multi-scale
problem. Instead, different techniques have advantages and disadvantages for
certain applications. 

For instance, sector decomposition and Mellin-Barnes representations provide a
general and fully automatizable algorithm for the extraction of UV and IR
divergences from arbitrary multi-loop integrals. However, they have problems
with reaching high numerical stability and precision in many physical
applications. At the other extreme are techniques like dispersion relations,
the Bernstein-Tkachov method, and differential equations, which are capable of
producing high-precision numerical results, but which require substantial work
to adapt them to a new class of processes. Here a class of processes is
characterized by the number of loops and external legs, as well as the types of
subloops that can appear in a given Feynman diagram.

The availability of a variety of different numerical loop integration techniques
has led to tremendous progress in the computation of difficult higher-loop
contributions. As a concrete example, the computation of electroweak two-loop
corrections to the most important electroweak precision observables has been
reviewed in section~\ref{sec:app}. These observables are the $W$-mass, which can
be predicted from the muon decay rate, and $Z$-pole observables, such as the
peak $Z$ production cross-section, the total $Z$ width and branching ratios, and
various parity-breaking asymmetries. 

These corrections can be computed in different renormalization schemes, which
differ in the higher-order terms that are implicitly resummed. Due to
cancellations between different terms, the resummed terms are generically
numerically small, so that there is no phenomenologically motivated advantage of
one renormalization scheme over another. For practical purposes, the radiative
corrections are factorized into initial- and final-state QED and QCD
corrections, which include real emission contributions, and massive electroweak
corrections, which are free of physical IR divergences. Due to the efforts of
many groups, high-precision results are available for both of these categories,
resulting in an estimated theory error that is safely below the experimental
uncertainties for all relevant electroweak precision observables.

The estimation of theory uncertainties has been discussed in some detail in
section~\ref{sec:err}, with the conclusion that no single method is fully
reliable, but instead it is advantageous to compare the outcomes of different
error estimation methods. These issues are important in the context of planned
future high-luminosity $e^+e^-$ colliders that are projected to obtain
electroweak precision data with substantially improved precision. To match the
projected experimental precision, new theory calculations will be necessary,
including three-loop electroweak corrections. Numerical methods will likely 
play an essential role in achieving this goal. However, none of the available
methods is immediately applicable to this task, but new developments and
improvements will be needed.


\section*{Acknowledgments}

The author is indebted to M.~Czakon, T.~Riemann and S.~Heinemeyer 
for many valuable comments on the manuscript.
This work was supported in part by
the U.S.\ National Science Foundation under grant PHY-1519175.


\end{document}